\documentclass[12pt]{article}
\usepackage{graphicx} 
\usepackage{authblk} 
\usepackage{setspace} 
\usepackage{natbib} 
\usepackage{amsmath}
\usepackage{verbatim} 
\usepackage[a4paper, total={6in, 8in}]{geometry} 
\usepackage{float}  
\usepackage{hyperref} 
\usepackage{amssymb} 
\usepackage{amsfonts}
\usepackage{dutchcal} 
\usepackage{booktabs} 
\usepackage{algorithm} 
\usepackage{algorithmic}
\usepackage{multirow} 
\usepackage{makecell}
\usepackage{subcaption} 
\usepackage{orcidlink}
\usepackage{rotating}

\usepackage{geometry}
\usepackage{xr}
\makeatletter

\newcommand*{\addFileDependency}[1]{
\typeout{(#1)}
\@addtofilelist{#1}
\IfFileExists{#1}{}{\typeout{No file #1.}}
}\makeatother

\newcommand*{\myexternaldocument}[1]{%
\externaldocument{#1}%
\addFileDependency{#1.tex}%
\addFileDependency{#1.aux}%
}

\myexternaldocument{supplementary}

\author{Anjana Wijayawardhana\thanks{Corresponding author: anjanaw@uow.edu.au} \orcidlink{0000-0003-3847-2671}}
\author{David Gunawan}
\author{Thomas Suesse}
\affil{School of Mathematics and Applied Statistics, University of Wollongong, Wollongong, NSW, Australia}
\date{}

\title{Variational Bayes Inference for Spatial Error Models with Missing Data}

\begin{document}
\doublespacing
\maketitle

\begin{abstract}

The spatial error model (SEM) is a type of simultaneous autoregressive (SAR) model for analysing spatially correlated data. Markov chain Monte Carlo (MCMC) is one of the most widely used Bayesian methods for estimating SEM, but it has significant limitations when it comes to handling missing data in the response variable due to its high computational cost. Variational Bayes (VB) approximation offers an alternative solution to this problem. Two VB-based algorithms employing Gaussian variational approximation with factor covariance structure are presented, joint VB (JVB) and hybrid VB (HVB), suitable for both missing at random and not at random inference. When dealing with many missing values, the JVB is inaccurate, and the standard HVB algorithm struggles to achieve accurate inferences. Our modified versions of HVB enable accurate inference within a reasonable computational time, thus improving its performance. The performance of the VB methods is evaluated using simulated and real datasets.

Keywords: Missing at random; Missing not at random; Selection model; Factor covariance structure; Stochastic gradient ascent

\end{abstract}

\section{Introduction}

The simultaneous autoregressive (SAR) models are ideal for analysing spatially correlated data, since they extend a linear regression model to take into account spatial correlations. There are three commonly used types of SAR models: spatial error models (SEMs), spatial autoregressive models (SAMs), and spatial Durbin models (SDMs). SAR models are applied in diverse applied research, including ecology~\citep{cassemiro2007spatial,ver2018spatial}, political science~\citep{di2021spatial}, social network analysis~\citep{leenders2002modeling,zhu2020multivariate}, and epidemiology~\citep{mollalo2020gis, wong2020spreading}.



An extensive literature explores sampling-based Bayesian Markov chain Monte Carlo (MCMC) methods for estimating SAR models \citep{gelfand1990illustration,lesage1997bayesian,de2008bayesian}. However, it is computationally expensive to estimate SAR models with many observations and missing data. 
Variational Bayes (VB) has recently emerged as a faster alternative to MCMC for estimating complex statistical models \citep{chappell2008variational,han2016variational,ong2018gaussian,gunawan2021variational,gunawan2023flexible}; see Section~\ref{sec:methods} for further details on VB methods.
There are several commonly used VB methods, including mean-field variational Bayes (MFVB)~\citep{ormerod2010explaining}, integrated non-factorised variational inference  (INFVB)~\citep{han2013integrated}, and Gaussian variational approximation~\citep{ong2018gaussian,tan2018gaussian}. 

Although VB methods are a promising alternative to MCMC methods, their use in estimating SAR models has been limited even where there are no missing values in the response variable.~\citet{wu2018fast} employed two variational Bayes methods, hybrid mean-field variational Bayes (MFVB) and integrated non-factorised variational Bayes (INFVB), to estimate the spatial autoregressive confused (SAC) and matrix exponential spatial specification (MESS) models, both belonging to the SAR family. In \citet{bansal2021fast}, spatial count data models were estimated using MFVB and INFVB, incorporating a MESS model to capture spatial dependence in error terms. 

Having missing values in the response variable is common in practice. When estimating SAR models, ignoring missing response values can lead to inconsistency and bias ~\citep{wang2013estimation, benedetti2020spatial}. Extensive literature has explored the estimation of SAR models under missing at random (MAR) mechanism ~\citep{lesage2004models,wang2013estimation,suesse2017computational,suesse2018marginal,wijayawardhana2024statistical}. 
There has been a limited exploration of estimating SEM under the missing not at random (MNAR) mechanism. \cite{flores2012estimation} introduced a Generalized Method of Moments (GMM) estimator that performs poorly with small sample sizes. \cite{Seya2021parameter} and \cite{Dougan2018bayesian} used Metropolis-Hastings (MH) algorithms, which are computationally expensive when the number of observations is large. A more recent study by ~\cite{rabovivc2023estimation} examined the estimation of the SEM using a partial maximum likelihood (ML) method.


Current VB methods have only been applied to estimate SAR models with full data \citep{wu2018fast,bansal2021fast}. The MFVB method is not suitable for estimating SAR models with missing data because it assumes posterior independence over the model parameters and missing values, resulting in underestimating posterior variance ~\citep{ 10.1214/06-BA104}. To address this issue, we employ the 
Gaussian variational approximation with a factor covariance structure proposed by~\cite{ong2018gaussian} in Section ~\ref{sec:methods}.

Our paper proposes two efficient VB algorithms, called joint VB (JVB) and hybrid VB (HVB), that are less computationally demanding than MCMC for estimating SEM under MAR and MNAR. The JVB method uses a Gaussian variational approximation with a factor covariance structure to approximate the joint posterior of the model parameters and the missing values. The HVB method significantly modifies the VB methods proposed by \cite{loaiza2022fast} and \cite{Dao:2013}, which combine VB optimisation with MCMC steps. A Gaussian variational approximation with a factor covariance structure is used for approximating the posterior distribution of the model parameters, and MCMC steps are used to sample the missing response values from their conditional posterior distribution. The conditional posterior distribution of missing response values is available in closed form under MAR. Under MNAR, however, the conditional posterior distribution is not available in closed form, making Bayesian inference more challenging. We propose several MCMC schemes for the HVB method under MNAR to address low acceptance percentages, especially for cases with many missing values.  


The performance of the VB methods is investigated using simulated and real datasets with different numbers of observations and missing data percentages. 
We compare the performance of the VB methods with Hamiltonian Monte Carlo (HMC)~\citep{duane1987hybrid, neal2011mcmc}, implemented using RStan, an interface to the Stan programming language~\citep{stan}. In particular, we use the HMC algorithm of~\cite{hoffman2014no}, called the No U-Turn Sampler (NUTS), which adaptively selects the number of leapfrogs and the step size. Section ~\ref{sec:HMC} of the online supplement provides detailed information on the HMC algorithm used.

The rest of this paper is organised as follows. Section~\ref{sec:Models} presents the spatial error models and discusses different missing data mechanisms.
In Section~\ref{sec:methods}, we present the variational Bayes methods to estimate the SEM with missing data. In Section~\ref{sec:SimulationStudy}, Simulation studies are conducted to evaluate the performance of the VB methods. Section~\ref{sec:RealWorldAnalysis} applies the VB methods to a real-world dataset. Section \ref{sec:Conclusion} discusses our major results and findings.   The paper also has an online supplement with additional technical details. 

\section{Spatial Error Models and Missing Data Mechanisms}
\label{sec:Models}
\subsection{Spatial Error Model}
\label{sec:SEM.model}


Let $\textbf{y}=(y_1,y_2,...,y_n)^\top$ be the vector of response variable observed at $n$ spatial locations $s_1, \hdots,s_n$, $\textbf{X}$ be the $n\times (r+1)$ design matrix containing the covariates, and $\textbf{W}$ be the $n\times n$ spatial weight matrix. The SEM is given by
\begin{align}
\label{eq:SEM}
\begin{split}
\textbf{y}=\textbf{X}\boldsymbol{\beta}+\textbf{v},\\
 \textbf{v}=\rho \textbf{W}\textbf{v}+\textbf{e} ,
\end{split}
\end{align}

\noindent where $\textbf{e}\sim N(0,\sigma_{\textbf{y}}^2\textbf{I}_n)$, $\textbf{I}_n$ denotes the $n\times n$ identity matrix, and $\sigma_{\textbf{y}}^2$ is a variance parameter. The vector $\boldsymbol{\beta}=(\beta_0, \beta_1, \hdots, \beta_r)^\top$ contains the fixed effects parameters, and $\rho$ is the spatial autocorrelation parameter which measures the strength and the direction of spatial dependence~\citep{anselin1988spatial,allison2001missing,lesage2009introduction}. 

Let $W_{ij}$ be the $i^{th}$ row and $j^{th}$ column entry of the spatial weight matrix $\textbf{W}$. The entry $W_{ij}$ is non-zero if the unit $i$ is a neighbour of the unit $j$. The diagonal of the spatial weight matrix $\textbf{W}$ is zero. There have been several strategies proposed for constructing $\textbf{W}$ in the literature  
(see ~\citet{ ord1975estimation,anselin1988spatial, kelley1997fast} for further details). The $\textbf{W}$ is commonly constructed to be sparse and symmetric.

For the SEM, when the error vector $\textbf{e}$ is normally distributed, the response variable $\textbf{y}$ is multivariate Gaussian with the mean vector $\boldsymbol{\mu}_{\textbf{y}}=\textbf{X}\boldsymbol{\beta}$ and covariance matrix $\boldsymbol{\Sigma}_{\textbf{y}}=\sigma^2_{\textbf{y}}(\textbf{A}^\top\textbf{A})^{-1}$. To ensure the validity of $\boldsymbol{\Sigma}_{\textbf{y}}$ as a proper covariance matrix, the parameter $\rho$ does not take on any of the values $\frac{1}{\lambda_{(1)}}, \frac{1}{\lambda_{(2)}}, \ldots, \frac{1}{\lambda_{(n)}}$, where $\lambda_{(1)}, \lambda_{(2)}, \ldots, \lambda_{(n)}$ represent the eigenvalues of the matrix $\textbf{W}$ sorted in ascending order~\citep{li2012one}. It is common practice to perform row or column normalisation (ensuring that the sum of the rows or columns is 1) on $\textbf{W}$, thus restricting $\rho$ to the range $\frac{1}{\lambda_{(1)}} < \rho < 1$~\citep{lesage2009introduction}. 



Table~\ref{tbl:properties_SAR-M} provides expressions for the mean vector, covariance matrix, and precision matrix for the distribution of $\textbf{y}$. Let $\boldsymbol{\phi} = (\boldsymbol{\beta}^\top, \rho, \sigma^2_{\textbf{y}})^\top$ be the vector of model parameters of the SEM. The log-likelihood of $\textbf{y}$ is given by
\begin{equation} 
\label{eq:log_like}
    \text{log}~p(\textbf{y} \mid \boldsymbol{\phi})=-\frac{n}{2}\textrm{log}(2\pi)-\frac{n}{2}\textrm{log}(\sigma^2_{\boldsymbol{y}})+\frac{1}{2}\textrm{log}|\textbf{M}_{\textbf{y}}|-\frac{1}{2\sigma^2_{\boldsymbol{y}}}\textbf{r}^\top\textbf{M}_{\textbf{y}}\textbf{r},
\end{equation} 

\noindent where $\textbf{r}=\textbf{y}-\boldsymbol{\mu}_{\textbf{y}}$.

\begin{table}
\caption {Expressions for mean vector ($\boldsymbol{\mu}_{\textbf{y}}$), covariance matrix ($\boldsymbol{\Sigma}_{\textbf{y}}$), $\textbf{V}_{\textbf{y}}$, and precision matrix ($\textbf{M}_{\textbf{y}}$) of SEM with $\textbf{A}=\textbf{I}_n-\rho\textbf{W}$}
\label{tbl:properties_SAR-M}
\centering

\begin{tabular}{c|c}
Term               & Expression \\ 
\cline{1-2}
$\boldsymbol{\mu}_{\textbf{y}}$              & $\textbf{X}\boldsymbol{\beta}$ \\
$\boldsymbol{\Sigma}_{\textbf{y}}$ & $\sigma^2_{\textbf{y}}(\textbf{A}^\top\textbf{A})^{-1}$     \\
$\textbf{V}_{\textbf{y}}$ &$(\textbf{A}^\top\textbf{A})^{-1}$  \\
$\textbf{M}_{y}=\textbf{V}_{y}^{-1}$ & $\textbf{A}^\top\textbf{A}$ 
\end{tabular} 
\end{table}

\subsection{Missing Data Mechanisms}
\label{sec:missmach}



Consider that the response vector $\textbf{y}$ of an SEM in Equation \eqref{eq:SEM} contains missing values. Let $\textbf{y}_o$ be the subset of $\textbf{y}$ with $n_o$ observed units, and $\textbf{y}_u$ be the subset of $\textbf{y}$ with $n_u$ unobserved units. The complete response vector is $\textbf{y} = (\textbf{y}_o^\top, \textbf{y}_u^\top)^\top$. A missing data indicator vector $\textbf{m}$ of length ${n}$ containing 1's and 0's is defined. If an element in $\textbf{y}$ is missing, then the corresponding element in $\textbf{m}$ is 1 and 0, otherwise. The missing data mechanism is characterised by the conditional distribution of $\textbf{m}$ given $\textbf{y}$, say $p(\textbf{m}|\textbf{y},\boldsymbol{\psi},\textbf{X}^*)$, where $\boldsymbol{\psi}$ is a vector of unknown parameters, and $\textbf{X}^{*}$ is an $n \times (q+1)$ design matrix containing the covariates of the missing data model. The covariates of the missing data model can be a subset of the covariates of the SEM. The main process of interest ($\textbf{y}$) and the missing data mechanism ($\textbf{m}$) should be jointly modeled in statistical modeling~\cite{rubin1976inference}. There are three missing data mechanisms \citep{rubin1976inference}.
The first mechanism is missing completely at random (MCAR). In MCAR, there is no relationship between the values of the vector $\textbf{y}$ (both observed and missing values) and the probability that they are missing,   $p(\textbf{m}|\textbf{y},\boldsymbol{\psi},\textbf{X}^{*})=p(\textbf{m}\mid  \boldsymbol{\psi},\textbf{X}^{*})$, for all $\textbf{y}$ and $\boldsymbol{\psi}$. 
    
The second mechanism is missing at random (MAR). In this case, the probability of missing an element depends only on the observed data $\textbf{y}_o$ and does not depend on the missing data themselves,  $p(\textbf{m}|\textbf{y},\boldsymbol{\psi}, \textbf{X}^{*})=p(\textbf{m} \mid \textbf{y}_o, \boldsymbol{\psi}, \textbf{X}^{*})$, for all $\textbf{y}_o$ and $\boldsymbol{\psi}$. As demonstrated in Section ~\ref{sec:methods}, under assumptions of the MAR missing data mechanism and distinct parameters of the missing data model and the SEM, Bayesian inference on the SEM parameters can be performed without explicitly considering the missing data model and its parameters.

The third mechanism is missing not at random (MNAR). The probability that an element is missing depends on both the observed data and the unobserved data, $p(\textbf{m}|\textbf{y},\boldsymbol{\psi}, \textbf{X}^{*})$. 
Under MNAR mechanism, we assume that the distribution of $\textbf{m}$ is independent given $\textbf{y}$, $\textbf{X}^{*}$, and $\boldsymbol{\psi}$. With this assumption, the density $p(\textbf{m} \mid \textbf{y}, \textbf{X}^{*}, \boldsymbol{\psi})$ is a product of $p(m_i \mid y_i, \textbf{x}^{*}_{i}, \boldsymbol{\psi})$ for $i = 1, \ldots, n$, where $m_i$ and $y_i$ denote the $i^{th}$ elements of $\textbf{m}$ and $\textbf{y}$, respectively, and $\textbf{x}_i^*$ is the $i$th row vector of $\textbf{X}^{*}$. The parameter vector $\boldsymbol{\psi}=(\boldsymbol{\psi}_\textbf{x}^\top,\psi_{\textbf{y}})^\top$ consists of the fixed effects vector associated with covariates $\textbf{X}^{*}$; $\boldsymbol{\psi}_\textbf{x} = (\psi_0, \psi_1, \psi_2, \hdots, \psi_{q})^\top$, and the fixed effect corresponding to $\textbf{y}$, denoted as $\psi_{\textbf{y}}$. A logistic regression model is  used to model $p(m_i \mid y_i, \textbf{x}^{*}_{i}, \boldsymbol{\psi})$, leading to:

\begin{equation}
 \label{eq:joint.logistic.SEM}
    p(\textbf{m} \mid \textbf{y}, \textbf{X}^{*},\boldsymbol{\psi}) = \prod_{i=1}^{n} \frac{e^{\textbf{x}^{*}_{i}\boldsymbol{\psi}_\textbf{x}+{y}_i\psi_{y}}}{1 + e^{\textbf{x}^{*}_{i}\boldsymbol{\psi}_\textbf{x}+{y}_i\psi_{y}}}.
\end{equation}

In the presence of missing responses, the matrices $\textbf{X}$, $\textbf{W}$, and $\textbf{M}_{y}$ are divided into distinct parts as follows:
\begin{equation}
\label{mat:portions_of_xwM}
\textbf{X}=
\begin{pmatrix}
    \textbf{X}_o\\
   \textbf{X}_u
\end{pmatrix},
~\textbf{W}=
\begin{pmatrix}
    \textbf{W}_{oo} &  \textbf{W}_{ou}\\
    \textbf{W}_{uo} & \textbf{W}_{uu}
\end{pmatrix},
~\textbf{M}_y=
\begin{pmatrix}
    \textbf{M}_{y,oo} &  \textbf{M}_{y,ou}\\
    \textbf{M}_{y,uo} & \textbf{M}_{y,uu}
\end{pmatrix},
\end{equation}

\noindent where $\textbf{X}_o$ and $\textbf{X}_u$ are the corresponding design matrices for the observed and unobserved responses, respectively, and $\textbf{W}_{oo}$, $\textbf{W}_{ou}$, $\textbf{W}_{uo}$, and $\textbf{W}_{uu}$ represent the sub-matrices of $\textbf{W}$, and  $\textbf{M}_{y,oo}$, $\textbf{M}_{y,ou}$, $\textbf{M}_{y,uo}$, and $\textbf{M}_{y,uu}$ are sub-matrices of $\textbf{M}_{y}$.

\section{Bayesian Inference}
\label{sec:methods}


Let $\boldsymbol{\phi}=(\boldsymbol{\beta}^\top,\sigma^2_{y},\rho)^\top$ and $\boldsymbol{\psi}=(\boldsymbol{\psi}_\textbf{x}^\top,\psi_{\textbf{y}})^\top$ be the vectors of parameters of the SEM in Equation~\eqref{eq:SEM} and missing data model described in Section~\ref{sec:missmach}, respectively. Consider Bayesian inference for the parameters $\boldsymbol{\phi}$, $\boldsymbol{\psi}$, and the missing values $\textbf{y}_u$, with a prior distribution $p(\textbf{y}_u \mid \boldsymbol{\phi})p\left(\boldsymbol{\phi},\boldsymbol{\psi}\right)$.  The term $p\left(\textbf{y}_o,\textbf{m}|\boldsymbol{\phi},\boldsymbol{\psi},\textbf{y}_u\right)$  denotes the joint density of $\textbf{y}_o$ and $\textbf{m}$ conditional on $\boldsymbol{\phi}$, $\boldsymbol{\psi}$, and $\textbf{y}_u$, and the term $p\left(\boldsymbol{\phi}, \boldsymbol{\psi}, \textbf{y}_u|\textbf{y}_o,\textbf{m}\right)$ is the joint posterior distribution of $\boldsymbol{\phi}$, $\boldsymbol{\psi}$ and $\textbf{y}_u$ and is given by
\begin{equation}
\label{eq:bay.posterior}
\begin{split}
    p\left(\boldsymbol{\phi}, \boldsymbol{\psi},\textbf{y}_u|\textbf{y}_o, \textbf{m}\right) &\propto p\left(\textbf{y}_o, \textbf{m}|\boldsymbol{\phi}, \boldsymbol{\psi},\textbf{y}_u\right)p(\textbf{y}_u \mid \boldsymbol{\phi})p\left(\boldsymbol{\phi}, \boldsymbol{\psi}\right)\\
    &\propto p\left(\textbf{y}, \textbf{m}|\boldsymbol{\phi}, \boldsymbol{\psi}\right)p\left(\boldsymbol{\phi}, \boldsymbol{\psi}\right).\\
\end{split}
\end{equation}
The first term in RHS of Equation~\eqref{eq:bay.posterior} is the joint distribution of $\textbf{y}$ and $\textbf{m}$. The selection model~\citep{little2019statistical} decomposes $p(\textbf{y},\textbf{m} \mid \boldsymbol{\phi}, \boldsymbol{\psi})$ into two factors as follows,
\begin{equation}
\label{eq:selection_model_ym}
        p(\textbf{y},\textbf{m}\mid \boldsymbol{\phi}, \boldsymbol{\psi})=p(\textbf{y}\mid \boldsymbol{\phi})p(\textbf{m}\mid \textbf{y},\boldsymbol{\psi}),
\end{equation}


\noindent where $p(\textbf{y}\mid \boldsymbol{\phi})$ denotes the density function of the SEM, which follows a multivariate Gaussian distribution with the mean vector $\boldsymbol{\mu_y}$ and covariance matrix $\boldsymbol{\Sigma}_{\textbf{y}}$ given in Table \ref{tbl:properties_SAR-M}. Additionally, $p(\textbf{m}\mid \textbf{y},\boldsymbol{\psi})$ is the conditional distribution of $\textbf{m}$ given $\textbf{y}$ and the parameter $\boldsymbol{\psi}$. By substituting the selection model factorisation in Equation~\eqref{eq:selection_model_ym} into the joint distribution of $\textbf{y}$ and $\textbf{m}$ in Equation~\eqref{eq:bay.posterior}, we obtain 
\begin{equation}
\label{eq:bay.posterior_2}
    p\left(\boldsymbol{\phi}, \boldsymbol{\psi},\textbf{y}_u|\textbf{y}_o, \textbf{m}\right)
     \propto p\left(\textbf{y}|\boldsymbol{\phi}\right) p\left( \textbf{m}|\textbf{y} , \boldsymbol{\psi}\right) p\left(\boldsymbol{\phi}, \boldsymbol{\psi}\right)\\
\end{equation}
We now consider Bayesian inference under the MAR mechanism. Assume that $\boldsymbol{\phi}$ and $\boldsymbol{\psi}$ are distinct and a priori independent, $p\left(\boldsymbol{\phi}, \boldsymbol{\psi}\right)=p\left(\boldsymbol{\phi}\right) p \left(\boldsymbol{\psi}\right)$. Section~\ref{sec:missmach} shows that under MAR $p\left( \textbf{m}|\textbf{y} , \boldsymbol{\psi}\right)=p\left( \textbf{m}|\textbf{y}_o , \boldsymbol{\psi}\right)$. Substituting these terms to Equation~\eqref{eq:bay.posterior_2}, we obtain
\begin{equation}
\label{eq:bay.MAR.posterior}
\begin{split}
    p\left(\boldsymbol{\phi}, \boldsymbol{\psi},\textbf{y}_u \mid \textbf{y}_o, \textbf{m}\right) 
    & \propto p\left(\textbf{y}\mid \boldsymbol{\phi}\right) p\left( \textbf{m}|\textbf{y}_o , \boldsymbol{\psi}\right) p\left(\boldsymbol{\phi}\right) p \left(\boldsymbol{\psi}\right)\\
    & \propto p\left(\textbf{y}_o\mid \textbf{y}_u, \boldsymbol{\phi}\right) 
 p\left(\textbf{y}_u \mid \boldsymbol{\phi}\right) p\left(\boldsymbol{\phi}\right) p\left( \textbf{m}\mid\textbf{y}_o , \boldsymbol{\psi}\right)  p\left(\boldsymbol{\psi}\right)\\
    & \propto p\left(\boldsymbol{\phi}, \textbf{y}_u \mid \textbf{y}_o\right) p\left(\boldsymbol{\psi} \mid \textbf{m}, \textbf{y}_o\right).\\
\end{split}
\end{equation}
The first term in Equation~\eqref{eq:bay.MAR.posterior}
is the posterior distribution of $\boldsymbol{\phi}$ and $\textbf{y}_u$, which does not contain $\boldsymbol{\psi}$. The second term is the posterior distribution of $\boldsymbol{\psi}$, which does not contain $\boldsymbol{\phi}$ and $\textbf{y}_u$. This suggests that Bayesian inference for $\boldsymbol{\phi}$ and $\textbf{y}_u$ is based only on
the posterior distribution $p\left(\boldsymbol{\phi}, \textbf{y}_u \mid \textbf{y}_o\right)$, ignoring the missing data model and its parameters, $\boldsymbol{\psi}$. Therefore, the joint posterior distribution of SEM parameters and missing data is 
\begin{equation}
    \begin{split}
    \label{eq:post.SEM.MAR}
            p(\boldsymbol{\phi},\textbf{y}_u \mid \textbf{y}_o)&\propto p(\textbf{y}_o \mid \boldsymbol{\phi},\textbf{y}_u)p(\textbf{y}_u \mid \boldsymbol{\phi})p(\boldsymbol{\phi})\\
            &\propto p(\textbf{y}\mid \boldsymbol{\phi})p(\boldsymbol{\phi}).
    \end{split}
\end{equation}
We now consider Bayesian inference under the MNAR mechanism. When making Bayesian inference on parameters $\boldsymbol{\phi}$ and missing values $\textbf{y}_u$, even if we assume that $\boldsymbol{\phi}$ and $\boldsymbol{\psi}$ are distinct and a priori independent, we cannot ignore the missing data model.

The following notations are used in the subsequent sections. Let $\boldsymbol{\theta}$ represent the vector of model parameters. The parameters are $\boldsymbol{\theta}=\boldsymbol{\phi}$ under MAR and $\boldsymbol{\theta}=(\boldsymbol{\phi}^\top,\boldsymbol{\psi}^\top)^\top$ under MNAR.  Let $\textbf{O}$ be the observed data. The observed data  $\textbf{O}=\textbf{y}_o$ under MAR and $\textbf{O}=(\textbf{y}_o,\textbf{m})$ under MNAR. Given the prior $p(\textbf{y}_u \mid \boldsymbol{\theta})p\left(\boldsymbol{\theta}\right)$, the joint posterior distribution of $\boldsymbol{\theta}$ and $\textbf{y}_u$ given $\textbf{O}$, denoted as $p(\boldsymbol{\theta},\textbf{y}_u\mid \textbf{O})$, is
\begin{equation} \label{eq:post_theta_yu}p(\boldsymbol{\theta},\textbf{y}_u\mid \textbf{O}) \propto p\left(\textbf{O}|\boldsymbol{\theta},\textbf{y}_u\right)p(\textbf{y}_u \mid \boldsymbol{\theta})p\left(\boldsymbol{\theta}\right),
\end{equation}
\noindent where $p\left(\textbf{O}|\boldsymbol{\theta},\textbf{y}_u\right)$ denotes the density of $\textbf{O}$ conditional on $\boldsymbol{\phi}$, $\boldsymbol{\psi}$, and $\textbf{y}_u$.
We also let $h(\boldsymbol{\theta},\textbf{y}_u)=p\left(\textbf{O}|\boldsymbol{\theta},\textbf{y}_u\right)p(\textbf{y}_u \mid \boldsymbol{\theta})p\left(\boldsymbol{\theta}\right)$. Table~\ref{tab:model_yusummary} summarises $\boldsymbol{\theta}$, $\textbf{O}$, the number of parameters $S$, and expressions for $h(\boldsymbol{\theta},\textbf{y}_u)$ under different missing data mechanisms. 

\begin{table}
    \centering
    \caption{The vector of parameters ($\boldsymbol{\theta}$), the set of observed data ($\textbf{O}$), $h(\boldsymbol{\theta},\textbf{y}_u)$, and the number of parameters to be estimated ($S$) under MAR and MNAR}
    \label{tab:model_yusummary}
    \begin{tabular}{ccc} 
        \toprule
          \makecell{Missing \\ mechanism}   & MAR & MNAR \\
          \hline
          $\boldsymbol{\theta}$ & $\boldsymbol{\phi}$ & $(\boldsymbol{\phi}^\top, \boldsymbol{\psi}^\top)^\top$ \\
          $\textbf{O}$ & $\textbf{y}_o$& $(\textbf{y}_o,\textbf{m})$\\
          $h(\boldsymbol{\theta},\textbf{y}_u)$& $p(\textbf{y}\mid \boldsymbol{\phi})p(\boldsymbol{\phi})$ & $p(\textbf{y}\mid \boldsymbol{\phi})$ $p(\textbf{m}\mid \textbf{y},\boldsymbol{\psi})p(\boldsymbol{\phi},\boldsymbol{\psi})$\\
          $S$ & $r+3$ & $r+q+5$ \\
        \bottomrule
    \end{tabular}
\end{table}



\subsection{Variational Bayes Inference}
\label{sec:stdvb}

Consider Bayesian inference for the parameters $\boldsymbol{\theta}$ and the missing values $\textbf{y}_u$ given the observed values $\textbf{O}$. Table ~\ref{tab:model_yusummary} gives the parameters $\boldsymbol{\theta}$ and the observed data $\textbf{O}$ for different missing data mechanisms.
We consider the variational approximation $q_{\boldsymbol{\lambda}}(\boldsymbol{\theta},\textbf{y}_u)$, indexed by the variational parameter $\boldsymbol{\lambda}$ to approximate the joint posterior $p\left(\boldsymbol{\theta},\textbf{y}_u|\textbf{O}\right)$. The VB approach approximates this posterior distribution by minimising the Kullback-Leibler (KL) divergence
between $q_{\boldsymbol{\lambda}}(\boldsymbol{\theta},\textbf{y}_u)$ and $p(\boldsymbol{\theta},\textbf{y}_u \mid \textbf{O})$. The KL divergence between these two distributions is
\begin{equation}
    \label{eq:ELBO}
    \begin{split}
     \text{KL}(\boldsymbol{\lambda})&=\text{KL}\left(q_{\boldsymbol{\lambda}}(\boldsymbol{\theta},\textbf{y}_u)\mid \mid p(\boldsymbol{\theta},\textbf{y}_u \mid \textbf{O})) \right)\\
           & =\int   \text{log} \left(\frac{q_{\boldsymbol{\lambda}}(\boldsymbol{\theta},\textbf{y}_u)}{p\left(\boldsymbol{\theta},\textbf{y}_u|\textbf{O}\right)} \right) q_{\boldsymbol{\lambda}}(\boldsymbol{\theta},\textbf{y}_u)d\boldsymbol{\theta}d\textbf{y}_u. \\
    \end{split}
\end{equation}
Minimising KL divergence between $q_{\boldsymbol{\lambda}}(\boldsymbol{\theta},\textbf{y}_u)$ and $p(\boldsymbol{\theta},\textbf{y}_u \mid \textbf{O})$ is equivalent to maximising evidence lower bound (ELBO) on the marginal likelihood, $\text{log}~p(\textbf{O})$, denoted by $\mathcal{L}(\boldsymbol{\lambda})$, with $p(\textbf{O})=\int ~p(\textbf{O}\mid \boldsymbol{\theta},  \textbf{y}_u)p(\textbf{y}_u \mid \boldsymbol{\theta})p\left(\boldsymbol{\theta}\right)d\boldsymbol{\theta}d\textbf{y}_u$~\citep{blei2017variational}. The ELBO is
\begin{equation}
    \label{eq:ELBO}
    \begin{split}
            \mathcal{L}(\boldsymbol{\lambda})& =\int   \text{log} \left(\frac{h(\boldsymbol{\theta},\textbf{y}_u)}{q_{\boldsymbol{\lambda}}(\boldsymbol{\theta},\textbf{y}_u)} \right) q_{\boldsymbol{\lambda}}(\boldsymbol{\theta},\textbf{y}_u) d\boldsymbol{\theta}d\textbf{y}_u, \\
    \end{split}
\end{equation}
\noindent where $h(\boldsymbol{\theta},\textbf{y}_u)=p\left(\textbf{O}|\boldsymbol{\theta},\textbf{y}_u\right)p(\textbf{y}_u \mid \boldsymbol{\theta})p\left(\boldsymbol{\theta}\right)$. Table~\ref{tab:model_yusummary} provides expressions for $h(\boldsymbol{\theta},\textbf{y}_u)$ for SEM under MAR and MNAR. The ELBO does not have a closed form solution in general. To maximise ELBO with respect to variational parameters, $\boldsymbol{\lambda}$, stochastic gradient ascent (SGA) methods are  used~\citep{nott2012regression,rezende2014stochastic,pmlr-v32-titsias14,NIPS2015_1373b284}. The  SGA method updates the initial value for $\boldsymbol{\lambda}$ (say $\boldsymbol{\lambda}^{0}$) according to the iterative scheme, 
\begin{equation}
    \label{eq:grad.ascent}    \boldsymbol{\lambda}^{(t+1)}=\boldsymbol{\lambda}^{(t)}+\mathbcal{a}_t\circ\widehat{\nabla_{\boldsymbol{\lambda}}\mathcal{L}(\boldsymbol{\lambda}^{(t)})},
\end{equation}

\noindent where 
$\widehat{\nabla_{\boldsymbol{\lambda}}\mathcal{L}(\lambda)}$ is an unbiased estimate of the gradient $\nabla_{\boldsymbol{\lambda}}\mathcal{L}(\lambda)$, $\mathbcal{a}_t$ ($t=0,1,\hdots$), is a sequence of vector-valued learning rates, and they are chosen to satisfy the Robbins-Monro conditions $\sum _t {\mathcal{a}_t}=\infty$ and $\sum _t ({\mathcal{a}_t})^2 \leq \infty$~\citep{robbins1951stochastic}, that ensure convergence of the sequence $\boldsymbol{\lambda}^{(t)}$ to a local optimum as $t \rightarrow \infty$, under regularity conditions~\citep{bottou2010large}. The symbol $\circ$ is the element-wise product of two vectors. The updating of Equation~\eqref{eq:grad.ascent} is done until a stopping criterion is satisfied. 

Adaptive learning rates are crucial for achieving rapid convergence of the algorithm. In this paper, we implement the ADADELTA algorithm proposed by~\citet{Zeiler2012ADADELTAAA} for calculating adaptive learning rates; see Section~\ref{sec:sup:ADADELTA} of the online supplement for the algorithm.
It is important to minimise the variance of the unbiased estimator of the gradient $\widehat{{\nabla_{\boldsymbol{\lambda}}\mathcal{L}(\boldsymbol{\lambda})}}$ in Equation~\eqref{eq:grad.ascent} since it influences both stability and convergence speed of the VB algorithm. In this study, we utilise the so-called reparameterisation trick~\citep{kingma2013auto,pmlr-v32-rezende14}, which is often much more efficient compared to alternative methods~\citep{xu2019variance}. 



\subsection{Joint Variational Bayes algorithm}
\label{sec:jvb}
In this section, we introduce the first VB algorithm, which we call the joint variational Bayes (JVB) algorithm, which approximates the joint posterior of $\theta$ and $\textbf{y}_u$ with a Gaussian variational approximation with a factor covariance structure~\citep{ong2018gaussian}. The variational distribution is parameterised as $q_{\boldsymbol{\lambda}}(\boldsymbol{\theta},\textbf{y}_u)\sim N((\boldsymbol{\theta},\textbf{y}_u); \boldsymbol{\mu},\textbf{B}\textbf{B}^\top+\textbf{D}^2)$, where $\boldsymbol{\mu}$ is the $(S+n_u) \times 1$ mean vector, $\textbf{B}$ is an $(S+n_u) \times p $ full rank matrix with $p << (S+n_u)$, and $\textbf{D}$ is an $(S+n_u) \times (S+n_u)$ diagonal matrix having diagonal elements $\textbf{d}=(d_1, \hdots, d_{S+n_u})$. We further impose the restriction that the upper triangular elements of $\textbf{B}$ are all zero.


The ELBO in Equation~\eqref{eq:ELBO} is an expectation with respect to  $q_{\boldsymbol{\lambda}}$,
\begin{equation}
    \label{eq:ELBO.expectation_wrt_q}
            \mathcal{L}(\boldsymbol{\lambda})=E_{q}\left [\text{log}~h(\boldsymbol{\theta},\textbf{y}_u)-\text{log}~  q_{\boldsymbol{\lambda}}(\boldsymbol{\theta},\textbf{y}_u) \right],
\end{equation}

\noindent where $E_{q}\left[\cdot\right]$ denotes the expectation with respect to $q_{\boldsymbol{\lambda}}$.
To apply the reparameterisation trick, we first need to generate samples from $q_{\boldsymbol{\lambda}}(\boldsymbol{\theta},\textbf{y}_u)$. This can be achieved by first drawing $\boldsymbol{\zeta}=(\boldsymbol{\eta}^\top,\boldsymbol{\epsilon}^\top)^\top$ (where $\boldsymbol{\eta}$ is $p$-dimensional and $\boldsymbol{\epsilon}$ is $(S+n_u)$-dimensional vectors) from a fixed density $f_{\boldsymbol{\zeta}}(\boldsymbol{\zeta})$ that does not depend on the variational parameters, and then calculating ($\boldsymbol{\theta},\textbf{y}_u)=u(\boldsymbol{\zeta,\boldsymbol{\lambda}})=\boldsymbol{\mu}+\textbf{B}\boldsymbol{\eta}+\textbf{d}\circ \boldsymbol{\epsilon}$. We let $\boldsymbol{\zeta}=(\boldsymbol{\eta}^\top,\boldsymbol{\epsilon}^\top)^\top\sim N(\textbf{0},\textbf{I}_{S+n_u+p})$, where $\textbf{0}$ is the zero mean vector of length $(S+n_u+p)$ and $\textbf{I}_{m+n_u+p}$ is the identity matrix of size $S+n_u+p$. i.e., the distribution $f_{\boldsymbol{\zeta}}\left(\cdot\right)$ is standard normal.
Then, the expectation in Equation~\eqref{eq:ELBO.expectation_wrt_q} is expressed with respect to the distribution $f_{\boldsymbol{\zeta}}$ as\begin{equation}
    \label{eq:ELBO.expectation_wrt_f}
    \begin{split}
                    \mathcal{L}(\boldsymbol{\lambda})&=E_{q}\left[\text{log}~h(\boldsymbol{\theta},\textbf{y}_u)-\text{log}~  q_{\boldsymbol{\lambda}}(\boldsymbol{\theta},\textbf{y}_u) \right]\\
                    &=E_{f_{\boldsymbol{\zeta}}}\left[\text{log}~h(u(\boldsymbol{\zeta,\boldsymbol{\lambda}}))-\text{log}~  q_{\boldsymbol{\lambda}}(u(\boldsymbol{\zeta,\boldsymbol{\lambda}})) \right],
    \end{split}
\end{equation}


\noindent and differentiating $\mathcal{L}(\boldsymbol{\lambda})$  under the integral sign, we obtain
\begin{equation}
    \label{eq:ELBO.grad.lamda}
    \begin{split}
                    \nabla_{\boldsymbol{\lambda}}\mathcal{L}(\boldsymbol{\lambda})&=E_{f_{\boldsymbol{\zeta}}}\left[\nabla_{\boldsymbol{\lambda}}~\text{log}~h(u(\boldsymbol{\zeta,\boldsymbol{\lambda}}))-\nabla_{\boldsymbol{\lambda}}~\text{log}~  q_{\boldsymbol{\lambda}}(u(\boldsymbol{\zeta,\boldsymbol{\lambda}})) \right],\\
                    &=E_{f_{\boldsymbol{\zeta}}}\left[\frac{du(\boldsymbol{\zeta,\boldsymbol{\lambda}})^\top}{d\boldsymbol{\lambda}}\{\nabla_{\boldsymbol{\theta},\textbf{y}_u}\text{log}~h(\boldsymbol{\theta},\textbf{y}_u)-\nabla_{\boldsymbol{\theta},\textbf{y}_u}\text{log}~  q_{\boldsymbol{\lambda}}(\boldsymbol{\theta},\textbf{y}_u) \}\right],\\
    \end{split}
\end{equation}

\noindent where $\frac{du(\boldsymbol{\zeta,\boldsymbol{\lambda}})}{d\boldsymbol{\lambda}}$ is the derivative of the transformation $u(\boldsymbol{\zeta,\boldsymbol{\lambda}})=\boldsymbol{\mu}+\textbf{B}\boldsymbol{\eta}+\textbf{d}\circ \boldsymbol{\epsilon}$ with respect to the variational parameters $\boldsymbol{\lambda}=(\boldsymbol{\mu}^\top
, \text{vech}(\textbf{B})^\top, \textbf{d}^\top)^\top$, where "vech" operator is the vectorisation of a matrix by stacking its columns from left to right. 
Algorithm~\ref{alg:VB1} gives the JVB algorithm. 
Analytical expressions for $\frac{du(\boldsymbol{\zeta,\boldsymbol{\lambda}})}{d\boldsymbol{\lambda}}$, $\nabla_{\boldsymbol{\theta},\textbf{y}_u}\text{log}~  q_{\boldsymbol{\lambda}}(\boldsymbol{\theta},\textbf{y}_u)$, $\nabla_{\boldsymbol{\theta},\textbf{y}_u}\text{log}~h(\boldsymbol{\theta},\textbf{y}_u)$, and the formulae for constructing an unbiased estimate $\widehat{{\nabla_{\boldsymbol{\lambda}}\mathcal{L}(\boldsymbol{\lambda})}}$ for $\nabla_{\boldsymbol{\lambda}}\mathcal{L}(\boldsymbol{\lambda})$ in step 4 of Algorithm~\ref{alg:VB1} are given in Section~\ref{sec:der.jvb} of the online supplement.


\begin{algorithm}
  \caption{JVB algorithm}
  \begin{algorithmic}[1]
  \label{alg:VB1}
   \STATE Initialize $\boldsymbol{\lambda}^{(0)}=(\boldsymbol{\mu}^{\top (0)},\textrm{vech}{(\textbf{B})}^{\top (0)},\textbf{d}^{\top (0)})^{\top}$ and set $t=0$ 
  \REPEAT 
      \STATE Generate $(\boldsymbol{\eta}^{(t)},\boldsymbol{\epsilon}^{(t)})\sim N(\textbf{0},\textbf{I}_{S+n_u+p})$
      \STATE Construct unbiased estimates $\widehat{{\nabla_{\boldsymbol{\mu}}\mathcal{L}(\boldsymbol{\lambda})}},\widehat{{\nabla_{\textrm{vech}(\textbf{B})}\mathcal{L}(\boldsymbol{\lambda})}}, \text{and}~ \widehat{{\nabla_{\textbf{d}}\mathcal{L}(\boldsymbol{\lambda})}}$ using Equations ~\eqref{eq:grad_wrt_mu},~\eqref{eq:grad_wrt_B} and ~\eqref{eq:grad_wrt_d} in Section~\ref{sec:der.jvb} of the online supplement at $\boldsymbol{\lambda}^{(t)}$.
      \STATE Set adaptive learning rates for the variational means, $\mathbcal{a}^{(t)}_{\boldsymbol{\mu}}$ and the variational parameters $\textrm{vech}(\textbf{B})$ and $\textbf{d}$, $\mathbcal{a}_{\textrm{vech}{(\textbf{B})}}^{(t)}$ and $\mathbcal{a}_{\textbf{d}}^{(t)}$, respectively, using ADADELTA described in Section~\ref{sec:sup:ADADELTA} of the online supplement.
      \STATE Set $\boldsymbol{\mu}^{(t+1)} = \boldsymbol{\mu}^{(t)} + \mathbcal{a}^{(t)}_{\boldsymbol{\mu}} \circ \widehat{\nabla_{\boldsymbol{\mu}}  \mathcal{L}(\boldsymbol{\lambda}^{(t)})}$.
      \STATE  Set $\textrm{vech}(\textbf{B})^{(t+1)} = \textrm{vech}(\textbf{B})^{(t)} + \mathbcal{a}_{\text{vech}(\textbf{B})}^{(t)} \circ \widehat{\nabla_{\textrm{vech}(\textbf{B}) } \mathcal{L}(\boldsymbol{\lambda}^{(t)})}$.
      \STATE Set $\textbf{d}^{(t+1)} = \textbf{d}^{(t)} + \mathbcal{a}_{\textbf{d}}^{(t)} \circ  \widehat{\nabla _{\textbf{d}}  \mathcal{L}(\boldsymbol{\lambda}^{(t)})}$.
      \STATE Set $\boldsymbol{\lambda}^{(t+1)} = (\boldsymbol{\mu}^{\top (t+1)}, \textrm{vech}(\textbf{B})^{\top (t+1)}, \textbf{d}^{\top (t+1)})$, and $t = t + 1$
      \UNTIL {some stopping rule is satisfied}
  \end{algorithmic}
\end{algorithm}




\subsection{Hybrid Variational Bayes algorithm}
\label{sec:augdvb}
In this section, we describe the second VB algorithm, which we call the hybrid variational Bayes (HVB) algorithm. 
The variational distribution  $q_{\boldsymbol{\lambda}}(\boldsymbol{\theta},\textbf{y}_u)$ is given by
\begin{equation}
    \label{eq:q}
    q_{\boldsymbol{\lambda}}(\boldsymbol{\theta},\textbf{y}_u)=p(\textbf{y}_u \mid \textbf{O},\boldsymbol{\theta})q_{\boldsymbol{\lambda}}^{0}(\boldsymbol{\theta}),
\end{equation}
\noindent where $p(\textbf{y}_u \mid \textbf{O},\boldsymbol{\theta})$ is the conditional distribution of missing data $\textbf{y}_u$ given observed data $\textbf{O}$ and the model parameters $\boldsymbol{\theta}$ and $q_{\boldsymbol{\lambda}}^{0}(\boldsymbol{\theta})$ is the Gaussian variational approximation with a factor covariance structure for approximating the posterior distribution of $\boldsymbol{\theta}$.

Given the variational approximation $q_{\boldsymbol{\lambda}}(\boldsymbol{\theta},\textbf{y}_u)$ in Equation~\eqref{eq:q}, the expectation in Equation~\eqref{eq:ELBO.expectation_wrt_q} is expressed as
\begin{equation}
    \label{eq:ELBO.expectation_wrt_qaug}
    \begin{split}
                    \mathcal{L}(\boldsymbol{\lambda})&=E_{q}\left(\text{log}~h(\boldsymbol{\theta},\textbf{y}_u)-\text{log}~  q_{\boldsymbol{\lambda}}(\boldsymbol{\theta},\textbf{y}_u) \right)\\
                    &=E_{q}\left(\text{log}~p(\textbf{O} \mid \textbf{y}_u,\boldsymbol{\theta})+\text{log}~p(\textbf{y}_u \mid \boldsymbol{\theta})+\text{log}~p(\boldsymbol{\theta})-\text{log}q_{\boldsymbol{\lambda}}^0(\boldsymbol{\theta})-\text{log}~p(\textbf{y}_u \mid \textbf{O},\boldsymbol{\theta})\right).\\   
    \end{split}
\end{equation}
\noindent Using the Bayes rule, we write $p(\textbf{y}_u \mid \textbf{O}, \boldsymbol{\theta})=p(\textbf{O}\mid \textbf{y}_u,\boldsymbol{\theta})p(\textbf{y}_u\mid \boldsymbol{\theta})/p(\textbf{O}\mid \boldsymbol{\theta})$. Substituting this into Equation~\eqref{eq:ELBO.expectation_wrt_qaug}, we obtain
\begin{equation}
    \label{eq:ELBO.expectation_wrt_qaug_Lo}
                    \mathcal{L}(\boldsymbol{\lambda})=E_{q}\left(\text{log}~p(\textbf{O}\mid \boldsymbol{\theta})+\text{log}~p(\boldsymbol{\theta})-\text{log}~  q_{\boldsymbol{\lambda}}^0(\boldsymbol{\theta}) \right)=  \mathcal{L}^0(\boldsymbol{\lambda}),\\
\end{equation}

\noindent where $\mathcal{L}^0(\boldsymbol{\lambda})$ is the ELBO resulting from approximating only the posterior distribution of model parameters; $p(\boldsymbol{\theta}\mid \textbf{O})$, directly via the variational distribution $q_{\boldsymbol{\lambda}}^0(\boldsymbol{\theta})$. 



We now describe the $q_{\boldsymbol{\lambda}}^0(\boldsymbol{\theta})$ in more detail. We assume that $q_{\boldsymbol{\lambda}}^0(\boldsymbol{\theta})\sim N(\boldsymbol{\theta}; \boldsymbol{\mu}_{\boldsymbol{\theta}},\textbf{B}_{\boldsymbol{\theta}}\textbf{B}_{\boldsymbol{\theta}}^\top+\textbf{D}_{\boldsymbol{\theta}}^2)$, where $\boldsymbol{\mu}_{\boldsymbol{\theta}}$ is a $S \times 1$ vector of variational means,  
$\textbf{B}_{\boldsymbol{\theta}}$ is an $S \times p$ matrix with the upper triangular elements are set to zero, and $\textbf{D}_{\boldsymbol{\theta}}$ is an $S \times S$ diagonal matrix having diagonal elements $\textbf{d}_{\boldsymbol{\theta}}=(d_{{\boldsymbol{\theta}},1}, \hdots, d_{{\boldsymbol{\theta}}, S})$. The vector of variational parameters is $\boldsymbol{\lambda}_{\boldsymbol{\theta}}=(\boldsymbol{\mu}_{\boldsymbol{\theta}}^\top,\text{vech}(\textbf{B}_{\boldsymbol{\theta}})^\top,\textbf{d}_{\boldsymbol{\theta}}^\top)^\top$.

To apply the reparameterisation trick, first we need to generate samples from $q_{\boldsymbol{\lambda}}^0(\boldsymbol{\theta})$. This can be achieved by first drawing ${\boldsymbol{\delta}^0}=({\boldsymbol{\eta}^0}^\top,{\boldsymbol{\epsilon}^0}^\top)^\top$ (where $\boldsymbol{\eta}^0$ is $p$-dimensional and $\boldsymbol{\epsilon}^0$ is $S$-dimensional vectors) from a density $f_{\boldsymbol{\delta}^0}(\boldsymbol{\delta}^0)$ that does not depend on the variational parameters, and then calculating $\boldsymbol{\theta}=t^0(\boldsymbol{\delta}^0,\boldsymbol{\lambda}_{\boldsymbol{\theta}})=\boldsymbol{\mu}_{\boldsymbol{\theta}}+\textbf{B}_{\boldsymbol{\theta}}\boldsymbol{\eta}^0+\textbf{d}_{\boldsymbol{\theta}}\circ \boldsymbol{\epsilon}^0$. We let ${\boldsymbol{\delta}^0}=({\boldsymbol{\eta}^0}^\top,{\boldsymbol{\epsilon}^0}^\top)^\top\sim N(\textbf{0},\textbf{I}_{S+p})$, where $\textbf{I}_{S+p}$ is the identity matrix of size $S+p$. i.e., the distribution $f_{\boldsymbol{\delta}^{0}}\left(\cdot\right)$ is standard normal.

Let $\boldsymbol{\delta}=({\boldsymbol{\delta}^0}^\top,\textbf{y}_u^\top)^\top$ with the product density $f_{\boldsymbol{\delta}}(\boldsymbol{\delta})=f_{\boldsymbol{\delta}^0}(\boldsymbol{\delta}^0)p(\textbf{y}_u \mid t^0(\boldsymbol{\delta}^0,\boldsymbol{\lambda}_{\boldsymbol{\theta}}),\textbf{O})$. There exists a vector-valued transformation $t$ from $\boldsymbol{\delta}$ to the parameter space and augmented missing value space given by $(\boldsymbol{\theta}^\top,\textbf{y}_u^\top)^\top=t(\boldsymbol{\delta},\boldsymbol{\lambda}_{\boldsymbol{\theta}})=(t^0(\boldsymbol{\delta}^0,\boldsymbol{\lambda}_{\boldsymbol{\theta}})^\top,\textbf{y}_u^\top)^\top=((\boldsymbol{\mu}_{\boldsymbol{\theta}}+\textbf{B}_{\boldsymbol{\theta}}\boldsymbol{\eta}^0+\textbf{d}_{\boldsymbol{\theta}}\circ \boldsymbol{\epsilon}^0)^\top,\textbf{y}_u^\top)^\top$. The reparameterisation gradient of the ELBO in Equation~\eqref{eq:ELBO.expectation_wrt_qaug_Lo} is obtained by differentiating under the integral sign as follows
\begin{equation}
    \label{eq:ELBO.grad.lamda.aug}
                    \nabla_{\boldsymbol{\lambda}}\mathcal{L}(\boldsymbol{\lambda})=E_{f_{\boldsymbol{\delta}}}\left[\frac{dt^0(\boldsymbol{\delta}^0,\boldsymbol{\lambda}_{\boldsymbol{\theta}})^\top}{d\boldsymbol{\lambda}_{\boldsymbol{\theta}}}(\nabla_{\boldsymbol{\theta}}\text{log}~h(\boldsymbol{\theta},\textbf{y}_u)-\nabla_{\boldsymbol{\theta}}\text{log}~  q_{\boldsymbol{\lambda}}^0(\boldsymbol{\theta}) )\right];
\end{equation}

\noindent 
where $\frac{dt^0(\boldsymbol{\delta}^0,\boldsymbol{\lambda}_{\boldsymbol{\theta}})}{d\boldsymbol{\lambda_{\boldsymbol{\theta}}}}$ is the derivative of the transformation $t^0(\boldsymbol{\delta}^0,\boldsymbol{\lambda}_{\boldsymbol{\theta}})=\boldsymbol{\mu}_{\boldsymbol{\theta}}+\textbf{B}_{\boldsymbol{\theta}}\boldsymbol{\eta}^0+\textbf{d}_{\boldsymbol{\theta}}\circ \boldsymbol{\epsilon}^0$ with respect to the variational parameters $\boldsymbol{\lambda}_{\boldsymbol{\theta}}=(\boldsymbol{\mu}_{\boldsymbol{\theta}}^\top,\text{vech}(\textbf{B}_{\boldsymbol{\theta}})^\top,\textbf{d}_{\boldsymbol{\theta}}^\top)^\top$. The proof is similar to \citet{loaiza2022fast} and can be found in Section~\ref{sec:der.hvb} of the online supplement. 

Algorithm \ref{alg:Augvb} gives the HVB algorithm. Analytical expressions for $\frac{dt^0(\boldsymbol{\delta}^0,\boldsymbol{\lambda}_{\boldsymbol{\theta}})}{d\boldsymbol{\lambda}_{\boldsymbol{\theta}}}$, $\nabla_{\boldsymbol{\theta}}\text{log}~  q_{\boldsymbol{\lambda}}^0(\boldsymbol{\theta})$,  $\nabla_{\boldsymbol{\theta}}\text{log}~h(\boldsymbol{\theta},\textbf{y}_u)$, and the formulae for constructing an unbiased estimate $\widehat{{\nabla_{\boldsymbol{\lambda}}\mathcal{L}(\boldsymbol{\lambda})}}$ for $\nabla_{\boldsymbol{\lambda}}\mathcal{L}(\boldsymbol{\lambda})$ in step 6 of Algorithm~\ref{alg:Augvb} using a single sample $\boldsymbol{\delta}=({\boldsymbol{\delta}^0}^\top,\textbf{y}_u^\top)^\top$ drawn from $f_{\boldsymbol{\delta}^0}$, and $p(\textbf{y}_u \mid \boldsymbol{\theta},\textbf{O})$ are detailed in Section~\ref{sec:der.hvb} of the online supplement.

\begin{algorithm}
  \caption{HVB algorithm}
  \begin{algorithmic}[1]
  \label{alg:Augvb}
   \STATE Initialize $\boldsymbol{\lambda}_{\boldsymbol{\theta}}^{(0)}=(\boldsymbol{\mu}_{\boldsymbol{\theta}}^{\top (0)},\textrm{vech}{(\textbf{B}_{\boldsymbol{\theta}})}^{\top (0)},\textbf{d}_{\boldsymbol{\theta}}^{\top (0)})$ and set $t=0$ 
  \REPEAT 
      \STATE Generate $({\boldsymbol{\eta}^0}^{(t)},{\boldsymbol{\epsilon}^0}^{(t)})\sim N(\textbf{0},\textbf{I}_{S+p})$
      \STATE Generate $\boldsymbol{\theta}^{(t)}\sim q_{{\boldsymbol{\lambda}}^{(t)}}^0(\boldsymbol{\theta})$ using its reparameterised representation.
      \STATE Generate $\textbf{y}_u^{(t)}~\sim p(\textbf{y}_u\mid \boldsymbol{\theta}^{(t)},\textbf{O})$
      \STATE Construct unbiased estimates $\widehat{{\nabla_{\boldsymbol{\mu}_{\boldsymbol{\theta}}}\mathcal{L}(\boldsymbol{\lambda})}},\widehat{{\nabla_{\textrm{vech}{(\textbf{B}_{\boldsymbol{\theta}})}}\mathcal{L}(\boldsymbol{\lambda})}}, \text{and}\widehat{{\nabla_{\textbf{d}_{\boldsymbol{\theta}}}\mathcal{L}(\boldsymbol{\lambda})}}$ using Equations ~\eqref{eq:grad_wrt_mu_vb2},~\eqref{eq:grad_wrt_B_vb2} and ~\eqref{eq:grad_wrt_d_vb2} in Section~\ref{sec:der.hvb} of the online supplement.
      \STATE Set adaptive learning rates for the variational means $\mathbcal{a}^{(t)}_{\boldsymbol{\mu}_{\boldsymbol{\theta}}}$ and the variational parameters $\textrm{vech}(\textbf{B}_{\boldsymbol{\theta}})$ and $\textbf{d}_{\boldsymbol{\theta}}$, $\mathbcal{a}_{\text{vech}(\textbf{B}_{\boldsymbol{\theta}})}$ and $\mathbcal{a}_{\textbf{d}_{\boldsymbol{\theta}}}^{(t)}$, using ADADELTA described in Section~\ref{sec:sup:ADADELTA} of the online supplement.
      \STATE Set $\boldsymbol{\mu}_{\boldsymbol{\theta}}^{(t+1)} = \boldsymbol{\mu}_{\boldsymbol{\theta}}^{(t)} + \mathbcal{a}^{(t)}_{\boldsymbol{\mu}_{\boldsymbol{\theta}}} \circ \widehat{\nabla_{\boldsymbol{\mu}_{\boldsymbol{\theta}}}  \mathcal{L}(\boldsymbol{\lambda}^{(t)})}$.
      \STATE  Set $\textrm{vech}(\textbf{B}_{\boldsymbol{\theta}})^{(t+1)} = \textrm{vech}(\textbf{B}_{\boldsymbol{\theta}})^{(t)} + \mathbcal{a}_{\text{vech}(\textbf{B}_{\boldsymbol{\theta}})}^{(t)} \circ \widehat{\nabla_{\textrm{vech}(\textbf{B}_{\boldsymbol{\theta}})} \mathcal{L}(\boldsymbol{\lambda}^{(t)})}$.
      \STATE Set $\textbf{d}_{\boldsymbol{\theta}}^{(t+1)} = \textbf{d}_{\boldsymbol{\theta}}^{(t)} + \mathbcal{a}_{\textbf{d}_{\boldsymbol{\theta}}}^{(t)} \circ  \widehat{\nabla _{\textbf{d}_{\boldsymbol{\theta}}}  \mathcal{L}(\boldsymbol{\lambda}^{(t)})}$.
      \STATE Set $\boldsymbol{\lambda}_{\boldsymbol{\theta}}^{(t+1)} = (\boldsymbol{\mu}_{\boldsymbol{\theta}}^{\top (t+1)}, \textrm{vech}(\textbf{B}_{\boldsymbol{\theta}})^{\top (t+1)}, \textbf{d}_{\boldsymbol{\theta}}^{\top (t+1)})^{\top}$, and $t = t + 1$
      \UNTIL {some stopping rule is satisfied}
  \end{algorithmic}
\end{algorithm}




\subsection{HVB under MAR\label{subsec:HVB-MAR}}
\label{sec:HVB.MAR}
Implementing step 5 of Algorithm~\ref{alg:Augvb} involves generating the missing values $\mathbf{y}_u$ from their conditional distribution $p(\mathbf{y}_u \mid \boldsymbol{\theta}^{(t)}, \mathbf{O})$, where $\boldsymbol{\theta}^{(t)}$ represents the parameters generated in step 4 of $t^{th}$ iteration of the algorithm. Under MAR, the conditional distribution $p(\textbf{y}_u \mid \boldsymbol{\theta}^{(t)},\textbf{O})=p(\textbf{y}_u \mid \boldsymbol{\phi}^{(t)}, \textbf{y}_o)$ is available in closed form, which follows a multivariate Gaussian distribution with the mean vector given by $\textbf{X}_u\boldsymbol{\beta}-\textbf{M}_{y,uu}^{-1}\textbf{M}_{y,uo}(\textbf{y}_o-\textbf{X}_o\boldsymbol{\beta})$ and the covariance matrix given by $\sigma^2_{\textbf{y}}\textbf{M}_{y,uu}^{-1}$, see~\citet{suesse2017computational}. 


As the total number of observations $n$ and the number of missing values $n_u$ increase, sampling directly from $p(\textbf{y}_u\mid \boldsymbol{\theta}^{(t)},\textbf{y}_o)$ becomes computationally expensive. We now discuss how to improve the efficiency of step 5 of the HVB algorithm given in Algorithm~\ref{alg:Augvb} under MAR. We start with partitioning the unobserved responses vector into $k$ blocks, such that $\textbf{y}_u=(\textbf{y}_{u_1}^\top,\dots,\textbf{y}_{u_k}^\top)^\top$. Then we implement a Gibbs step to update $\textbf{y}_u$ one block at a time by sampling from the full conditional distribution $p(\textbf{y}_{u_j}\mid \boldsymbol{\phi}^{(t)},\textbf{y}_o, \textbf{y}_u^{(-j)})$, for $j=1,..,k$, where $\textbf{y}_{u_j}$ is the updated block and $\textbf{y}_u^{(-j)}=(\textbf{y}_{u_{1}}^\top, \dots, \textbf{y}_{u_{j-1}}^\top,\textbf{y}_{u_{j+1}}^\top, \dots, \textbf{y}_{u_k}^\top)^\top$ is the remaining blocks. The complete response vector $\textbf{y}$ can now be written as $\textbf{y}=({\textbf{y}^\top_{s_j}}, \textbf{y}_{u_j}^{\top})^\top$, where ${\textbf{y}_{s_j}}=(\textbf{y}_o^{\top},{\textbf{y}_u^{(-j)}}^\top)^\top$. 
Based on this partitioning of $\textbf{y}$, the following partitioned matrices are defined:
\begin{equation}
\label{mat:portions_of_xwMB}
\textbf{X}=
\begin{pmatrix}
    \textbf{X}_{s_j}\\
   \textbf{X}_{u_j}
\end{pmatrix},
~\textbf{W}=
\begin{pmatrix}
    \textbf{W}_{{s_j}{s_j}} &  \textbf{W}_{{s_j}{u_j}}\\
    \textbf{W}_{{u_j}{s_j}} & \textbf{W}_{{u_j}{u_j}}
\end{pmatrix},
~\textbf{M}_{\textbf{y}}=
\begin{pmatrix}
    \textbf{M}_{y,{s_j}{s_j}}&  \textbf{M}_{y,{s_j}{u_j}}\\
    \textbf{M}_{y,{u_j}{s_j}} & \textbf{M}_{y,{u_j}{u_j}}
\end{pmatrix},
\end{equation}

\noindent where $\textbf{X}_{s_j}$ is the corresponding design matrix for the observed responses, and the unobserved responses that are not in the $j^{th}$ block (i.e. $\textbf{X}_{s_j}=(\textbf{X}_o^{\top},{\textbf{X}_u^{(-j)}}^\top)^\top$) and $\textbf{X}_{u_j}$ is the corresponding design matrix for $j^{th}$ block of unobserved responses. Similarly, $\textbf{W}_{{s_j}{s_j}}$, $\textbf{W}_{{s_j}{u_j}}$, $\textbf{W}_{{u_j}{s_j}}$, and $\textbf{W}_{{u_j}{u_j}}$ represent the sub-matrices of $\textbf{W}$, and  $\textbf{M}_{y,{s_j}{s_j}}$, $\textbf{M}_{y,{s_j}{u_j}}$, $\textbf{M}_{y,{u_j}{s_j}}$, and $\textbf{M}_{y,{u_j}{u_j}}$ are sub-matrices of $\textbf{M}_{\textbf{y}}$. 

Algorithm~\ref{alg:Gibbs} outlines the proposed Gibbs sampling steps. The full conditional distribution $p(\textbf{y}_{u_j} | \boldsymbol{\phi}^{(t)},\textbf{y}_o, \textbf{y}_u^{(-j)})=p(\textbf{y}_{u_j}\mid \boldsymbol{\phi}^{(t)},\textbf{y}_{s_j})$, follows a multivariate Gaussian distribution with the mean $\textbf{X}_{u_j}\boldsymbol{\beta}-\textbf{M}_{y,{u_j}{u_j}}^{-1}\textbf{M}_{y,{u_j}{s_j}}(\textbf{y}_{s_j}-\textbf{X}_{s_j}\boldsymbol{\beta})$ and the covariance matrix $\sigma^2_{\textbf{y}}\textbf{M}_{y,{u_j}{u_j}}^{-1}$. 
The HVB algorithm implemented using the Gibbs steps presented in Algorithm~\ref{alg:Gibbs}, which we call HVB-G in subsequent sections,  accelerates the generation of samples of missing values from their conditional distribution $p(\textbf{y}_u\mid \boldsymbol{\phi},\textbf{y}_o)$ when $n$ and $n_u$ are large. We replace step 5 of the HVB algorithm given in Algorithm~\ref{alg:Augvb} by the proposed Gibbs steps when $n_u$ is more than $1,000$ as directly sampling from 
$p(\textbf{y}_u\mid \boldsymbol{\phi},\textbf{y}_o)$ requires inverting a 
$1000 \times 1000$ covariance matrix. In the Gibbs sampling steps, we must specify the block size ($k^{*}$) and the number of Gibbs iterations ($N_1$). After some experimentation, we set $N_1=5$ with a block size of $k^*=500$, as these values consistently produce accurate inference results within a reasonable computational time.

\begin{algorithm}
  \caption{Gibbs steps within the $t^{th}$ iteration of the HVB algorithm under MAR when $n$ and $n_u$ are large}
  \begin{algorithmic}[1]
  \label{alg:Gibbs}
 \STATE Initialise missing values $\textbf{y}_{u,0}=(\textbf{y}_{u_1,0}^\top,\dots,\textbf{y}_{u_k,0}^\top)^\top~\sim p(\textbf{y}_u\mid \boldsymbol{\phi}^{(t)}, \textbf{y}_o)$
  \FOR{i=1, \dots, $N_1$} 
  \FOR{j=1, \dots, k}
      \STATE Sample $\textbf{y}_{u_j,i}$ from the conditional distribution $p(\textbf{y}_{u_j,i}\mid \boldsymbol{\phi}^{(t)},\textbf{y}_o,\textbf{y}_{u,i-1}^{(-j)})$, where $\textbf{y}_{u,i-1}^{(-j)}=(\textbf{y}^\top_{u_1,i}, \textbf{y}^\top_{u_2,i}, \hdots ,\textbf{y}^\top_{u_{j-1},i},\textbf{y}^\top_{u_{j+1},i-1},\hdots, ,\textbf{y}^\top_{u_{k},i-1})^\top$. 
  \ENDFOR
   \STATE $\textbf{y}_{u,i}=(\textbf{y}_{u_1,i}^{\top},\hdots, \textbf{y}_{u_k,i}^{\top})^\top$
  \ENDFOR
\STATE Output $\textbf{y}_u^{(t)}=\textbf{y}_{u,N_1}$
  \end{algorithmic}
\end{algorithm}

\subsection{HVB under MNAR\label{subsec:HVB-MNAR}}

Under MNAR, direct sampling from the conditional distribution $p(\mathbf{y}_u \mid \boldsymbol{\theta}^{(t)}, \mathbf{O}) = p(\mathbf{y}_u \mid \boldsymbol{\phi}^{(t)},\boldsymbol{\psi}^{(t)}, \mathbf{y}_o, \mathbf{m})$ is not feasible, as the conditional posterior distribution is not available in closed form. To sample from $p(\mathbf{y}_u \mid \boldsymbol{\phi}^{(t)},\boldsymbol{\psi}^{(t)}, \mathbf{y}_o, \mathbf{m})$, we employ the MCMC steps presented in Algorithm~\ref{alg:AugvbMCMCstep}, which employs $p(\textbf{y}_u\mid \boldsymbol{\phi}^{(t)},\textbf{y}_o)$ as the proposal. 

\begin{algorithm}
  \caption{MCMC steps within the $t^{th}$ iteration of the HVB algorithm under MNAR}
  \begin{algorithmic}[1]
  \label{alg:AugvbMCMCstep}
  \STATE Initialize missing values $\textbf{y}_{u,0}~\sim p(\textbf{y}_u\mid \boldsymbol{\phi}^{(t)}, \textbf{y}_o)$
  \FOR{i=1, \dots, $N_1$} 
      \STATE Sample $\textbf{y}_u^{*}$ from the proposal distribution $p(\textbf{y}_u^{*}\mid \boldsymbol{\phi}^{(t)},\textbf{y}_o)$, which follows a multivariate Gaussian with the mean vector given by $\textbf{X}_u\boldsymbol{\beta}-\textbf{M}_{y,uu}^{-1}\textbf{M}_{y,uo}(\textbf{y}_o-\textbf{X}_o\boldsymbol{\beta})$ and the covariance matrix given by $\sigma^2_{\textbf{y}}\textbf{M}_{y,uu}^{-1}$. 
      \STATE Sample $u$ from uniform distribution, $u~\sim \mathcal{U}(0,1)$
      \STATE Calculate $a=\text{min}\left (1,\frac{p(\textbf{m} \mid \textbf{y}^*,\boldsymbol{\psi}^{(t)})}{p(\textbf{m} \mid \textbf{y}_{{i-1}},\boldsymbol{\psi}^{(t)})}\right)$, where $\textbf{y}^*=(\textbf{y}_o^{\top},{\textbf{y}_u^{*}}^\top)$ and $\textbf{y}_{i-1}=(\textbf{y}^\top_o,{\textbf{y}^\top_{u,{i-1}}})$
      \IF{$a>u$}
        \STATE $\textbf{y}_{u,i}=\textbf{y}_u^{*}$
      \ELSE
        \STATE $\textbf{y}_{u,i}=\textbf{y}_{u,i-1}$
      \ENDIF
  \ENDFOR
\STATE Output $\textbf{y}^{(t)}_{u}=\textbf{y}_{u,N_1}$
  \end{algorithmic}
\end{algorithm}


The MCMC steps in Algorithm~\ref{alg:AugvbMCMCstep} generates samples from the conditional distribution $p(\textbf{y}_u\mid \boldsymbol{\phi}^{(t)},\boldsymbol{\psi}^{(t)},\textbf{y}_o,\textbf{m})$. However, as $n$ and $n_u$ increase, the HVB algorithm implemented using these MCMC steps does not estimate the parameters accurately because of the low acceptance percentage. After some experimentation, we found that achieving an acceptance percentage between $20\%$ and $30\%$ is necessary to balance between accurate posterior inference and computational cost. To improve the acceptance percentage, we partition $\textbf{y}_{u}$ into $k$ blocks as discussed in Section~\ref{sec:HVB.MAR}, and update one block at a time using proposals from $p(\textbf{y}_{u_j}\mid \boldsymbol{\phi}^{(t)},\textbf{y}_o, \textbf{y}_u^{(-j)})$. Algorithm~\ref{alg:AugvbMCMCstepB} outlines the MCMC steps for sampling the missing values one block at a time.

Updating all $k$ blocks for each of the $N_1$ MCMC iterations is computationally expensive, as for each block, the mean vector and covariance matrix of the conditional posterior distribution of $p(\textbf{y}_{u_j}\mid \boldsymbol{\phi},\textbf{y}_{s_j})$ must be calculated. This computational bottleneck can be overcome by updating only a randomly selected set of blocks at each iteration, so the computation of the conditional distribution is limited to the number of updated blocks, not the number of total blocks, $k$. We suggest that updating randomly selected $3$ blocks in each MCMC iteration is sufficient to obtain reliable inference with a smaller computational cost; see simulation results in Section~\ref{sec:SESM.estimates} for further details. The MCMC scheme for updating a randomly selected number of $k^{\prime}$ blocks can be obtained by modifying the MCMC scheme in Algorithm~\ref{alg:AugvbMCMCstepB}.




\begin{algorithm}
  \caption{MCMC steps within the $t^{th}$ iteration of the HVB algorithm under MNAR. The missing values are updated one block at a time}
  \begin{algorithmic}[1]
  \label{alg:AugvbMCMCstepB}
  \STATE Initialize missing values $\textbf{y}_{u,0}=(\textbf{y}^\top_{u_1,0},\dots,\textbf{y}^\top_{u_k,0})^\top~\sim p(\textbf{y}_u\mid \boldsymbol{\phi}^{(t)}, \textbf{y}_o)$
  \FOR{i=1, \dots, $N_1$} 
  \FOR{j=1, \dots, $k$}
      \STATE Sample $\textbf{y}_{u_j}^{*}$ from the proposal distribution $p(\textbf{y}_{u_j}^{*}\mid \boldsymbol{\phi}^{(t)},\textbf{y}_o,\textbf{y}_{u,{i-1}}^{(-j)})$, where $\textbf{y}_{u,i-1}^{(-j)}=(\textbf{y}^\top_{u_1,i}, \textbf{y}^\top_{u_2,i}, \hdots ,\textbf{y}^\top_{u_{j-1},i},\textbf{y}^\top_{u_{j+1},i-1},\hdots, ,\textbf{y}^\top_{u_{k},i-1})^\top$.  
      \STATE Sample $u$ from uniform distribution, $u~\sim \mathcal{U}(0,1)$
      \STATE Calculate $a=\text{min}\left (1,\frac{p(\textbf{m} \mid \textbf{y}^*,\boldsymbol{\psi}^{(t)})}{p(\textbf{m} \mid \textbf{y}_{i-1},\boldsymbol{\psi}^{(t)})}\right)$, where $\textbf{y}^*=(\textbf{y}_o^{\top},{\textbf{y}_u^{(-j)}}^\top,{\textbf{y}_{u_j}^{*}}^\top)^\top$ and $\textbf{y}_{i-1}=(\textbf{y}_o^{\top},{\textbf{y}_u^{(-j)}}^\top,{\textbf{y}_{u_j,{i-1}}}^\top)^\top$
      \IF{$a>u$}
        \STATE $\textbf{y}_{u_j,i}=\textbf{y}_{u_j}^{*}$
      \ELSE
        \STATE $\textbf{y}_{u_j,i}=\textbf{y}_{{u_j,i-1}}$
      \ENDIF
  \ENDFOR
   \STATE $\textbf{y}_{u,i}=(\textbf{y}_{u_1,i}^\top,\hdots ,\textbf{y}_{u_k,i}^\top)^\top$
  \ENDFOR
\STATE  Output $\textbf{y}^{(t)}_{u}=\textbf{y}_{u,N_1}$
  \end{algorithmic}
\end{algorithm}



For the remaining sections, the HVB algorithm implemented via the MCMC scheme in Algorithm~\ref{alg:AugvbMCMCstep} without blocking $\textbf{y}_u$ is called the HVB-No Block abbreviated as HVB-NoB. The HVB algorithm implemented through the MCMC scheme in Algorithm~\ref{alg:AugvbMCMCstepB} is called HVB-All Block abbreviated as HVB-AllB, and the HVB using the MCMC scheme that updates only randomly selected three blocks is referred to as HVB-3B. The criteria for setting tuning parameters of the proposed MCMC schemes, such as the number of MCMC steps $N_1$ and block size $k^{*} $, are discussed in Section~\ref{sec:SESM.estimates}.

\section{Simulation result}
\label{sec:SimulationStudy}


This section investigates the performance of the VB methods to estimate the SEM under MAR and MNAR mechanisms. 
We compare the posterior density estimates from the VB methods to those obtained from the Hamiltonian Monte Carlo method (HMC). All examples are implemented using the R programming language. The HMC is implemented using the RStan interface~\citep{stan}. We use the HMC method from~\citet{hoffman2014no}, known as the No U-Turn Sampler (NUTS). For details on the generic HMC algorithm, see Section ~\ref{sec:HMC} of the online supplement.

In all simulation studies, we simulate $n$ observations from a standard SEM according to Equation~\eqref{eq:SEM} with $10$ covariates. Each covariate for every observation is drawn from a standard normal distribution $N(0,1)$. The weight matrices are constructed based on a regular grid of size $\sqrt{n} \times \sqrt{n}$, where neighbours are defined using the Rook neighbourhood method (refer to \citet{wijayawardhana2024statistical} for details on constructing the weight matrix based on the Rook neighbourhood). We generate the true values of the $11$ fixed effects ($\boldsymbol{\beta}$'s) randomly from discrete uniform random numbers between $1$ and $5$. We set $\sigma^2_{\boldsymbol{y}}=1$ and $\rho=0.8$. The subsequent steps in the simulation process vary depending on the missing value mechanisms. 

For the MAR mechanism, after simulating $n$ observations from a standard SEM, we randomly select $n_o$ units to form the observed data set. In the MNAR mechanism, we generate missing responses using the logistic regression model regressed on a randomly chosen covariate from the 10 covariates and the response variable of SEM ($\textbf{y}$) as covariates, which is given by
\begin{equation}
 \label{eq:logistic.SEM}
    p(\textbf{m} \mid \textbf{y}, \textbf{X}^{*},\boldsymbol{\psi}) = \prod_{i=1}^{n} \frac{e^{\textbf{x}^{*}_{i}\boldsymbol{\psi}_{\textbf{x}}+{y}_i\psi_{y}}}{1 + e^{\textbf{x}^{*}_{i}\boldsymbol{\psi}_{\textbf{x}}+{y}_i\psi_{y}}},
\end{equation}


\noindent where the design matrix $\textbf{X}^{*}$ contains a column of ones and the selected covariate, and  $\bold{x}_{i}^{*}$ denotes the $i^{th}$ row vector of $\bold{X}^{*}$. The vector $\boldsymbol{\psi}_\textbf{x}=(\psi_0,\psi_{\textbf{x}^{*}})$ contains the coefficient of the intercept ($\psi_0$), and the coefficient of the selected covariate ($\psi_{\textbf{x}^{*}}$). The coefficient corresponding to $\textbf{y}$ is $\psi_{\textbf{y}}$.



All VB and HMC algorithms in this work utilise the same prior distributions, $p(\boldsymbol{\theta})$, for the parameters (under MAR  $p(\boldsymbol{\theta})=p(\boldsymbol{\phi})$, and under MNAR $p(\boldsymbol{\theta})=p(\boldsymbol{\phi},\boldsymbol{\psi})=p(\boldsymbol{\phi})p(\boldsymbol{\boldsymbol{\psi}})$). The priors for the parameters are given in Section~\ref{sec:h_gradients} of the online supplement. 
When implementing VB algorithms, the initial values are set as follows: 
Under MAR, the ordinary least squares (OLS) estimates are used for the initial values for the fixed effect parameters ($\boldsymbol{\beta}$'s) and the error variance ($\sigma^2_{\textbf{y}}$). Additionally, we assign a value of 0.01 to $\rho$, reflecting a very weak spatial dependence. Under MNAR, we have additional three parameters $\psi_0,\psi_{\textbf{x}^{*}}$ and $\psi_{\textbf{y}}$.
For all these coefficients, we set the starting value to $0.01$. The initial values for the missing data under MAR and MNAR are simulated from the conditional distribution $p(\textbf{y}_u\mid \boldsymbol{\phi}^{(0)}, \textbf{y}_o)$, where $\boldsymbol{\phi}^{(0)}$ is the vector containing the initial parameter values.
We use $p=4$ factors for HVB and JVB for the simulation study. The results do not improve when we increase the number of factors. We run the VB and HMC methods for $10,000$ iterations for all simulation studies.  Section~\ref{sec:sup:convanal} of the online supplement presents convergence plots for the VB algorithms and trace plots of posterior samples from HMC.


In Section~\ref{sec:SEM.MAR.estimates}, we examine the accuracy and computational cost of proposed VB methods for estimating SEM under the MAR mechanism. Section~\ref{sec:SESM.estimates} presents the simulation results for estimating SEM under the MNAR mechanism.



\subsection{Simulation study under MAR}
\label{sec:SEM.MAR.estimates}
This section discusses the simulation results for estimating SEM under the MAR mechanism. 
We investigate the accuracy of the proposed VB methods using two sample sizes: $n = 625$  and  $n = 10,000$. For  $n = 625$, we consider scenarios with $25\%$ and $75\%$ missing data. For $n = 10,000$, we only consider the scenario with $75\%$ missing data. This section discusses the results for the $75\%$ missing data scenario for  $n = 625$ and  $n = 10,000$. The results for  $n = 625$ with 25\% missing data are provided in Section~\ref{sec:sup.sim.MAR} of the online supplement.

\begin{figure}[H]
    \centering
    \includegraphics[width=0.8\textwidth, keepaspectratio]{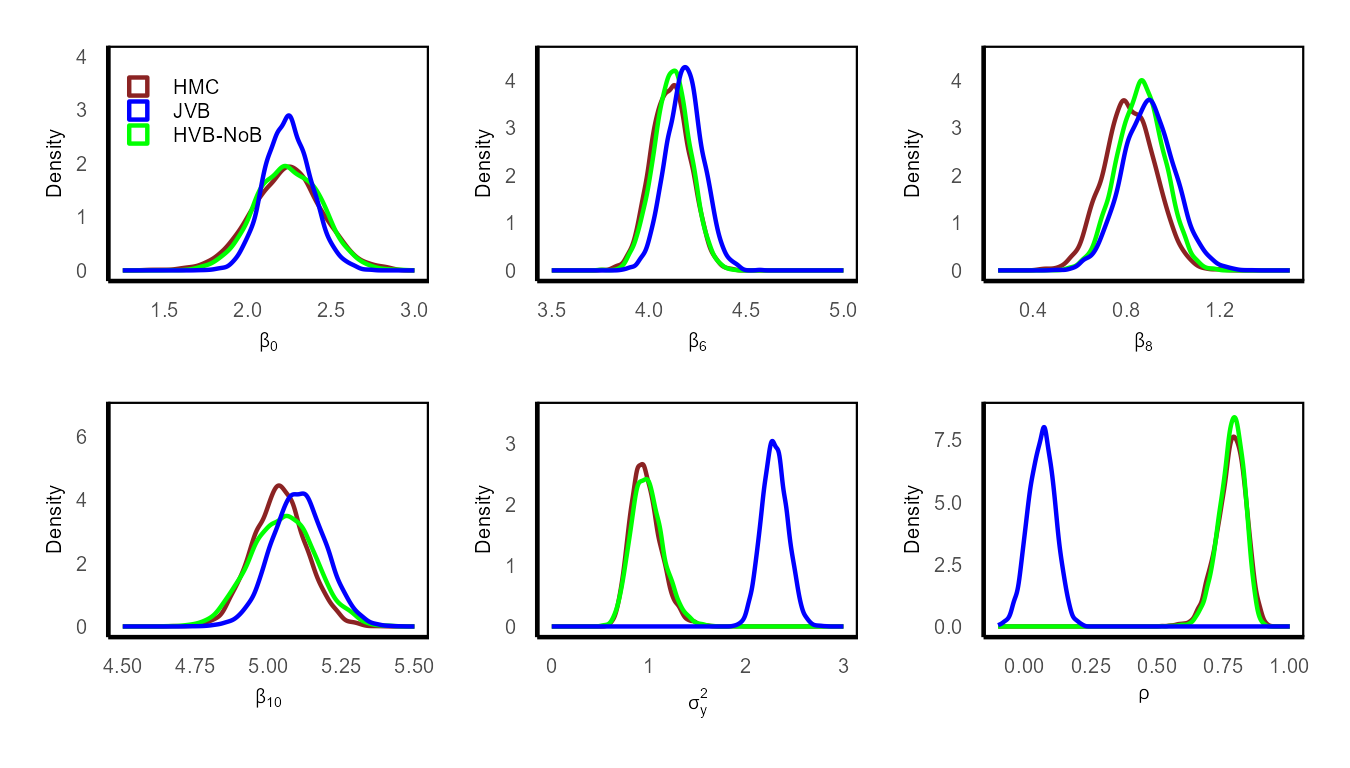}
    \caption{Posterior densities of SEM parameters under MAR for the simulated dataset with $n=625$ and $n_u=468$ ($75\%$ missing values) estimated using the HMC, JVB and HVB-NoB methods}
    \label{fig:kernal_SEM_MAR_625_miss_75}
\end{figure}

Figure~\ref{fig:kernal_SEM_MAR_625_miss_75} presents the posterior densities of SEM parameters estimated using the HMC, JVB, and HVB methods for the simulated dataset with $n=625$ and $75\%$ missing values ($n_u=468$). Since $n_u$ is small ($<1,000$), we use the standard HVB without the Gibbs steps, the HVB-NoB method, given in Algorithm~\ref{alg:Augvb}. For the fixed effects parameters ($\boldsymbol{\beta}$'s), the posterior densities from the JVB and HVB-NoB are nearly identical to those from HMC, with HVB-NoB being the closest to HMC. However, the posterior densities of $\sigma^2_{\textbf{y}}$ and $\rho$ from the JVB method exhibit significant deviations from those obtained by HMC, whereas the posterior densities from the HVB align well with those of HMC. 


\begin{figure}[H]
    \centering
    \begin{subfigure}{0.45\textwidth}
        \includegraphics[width=\linewidth]{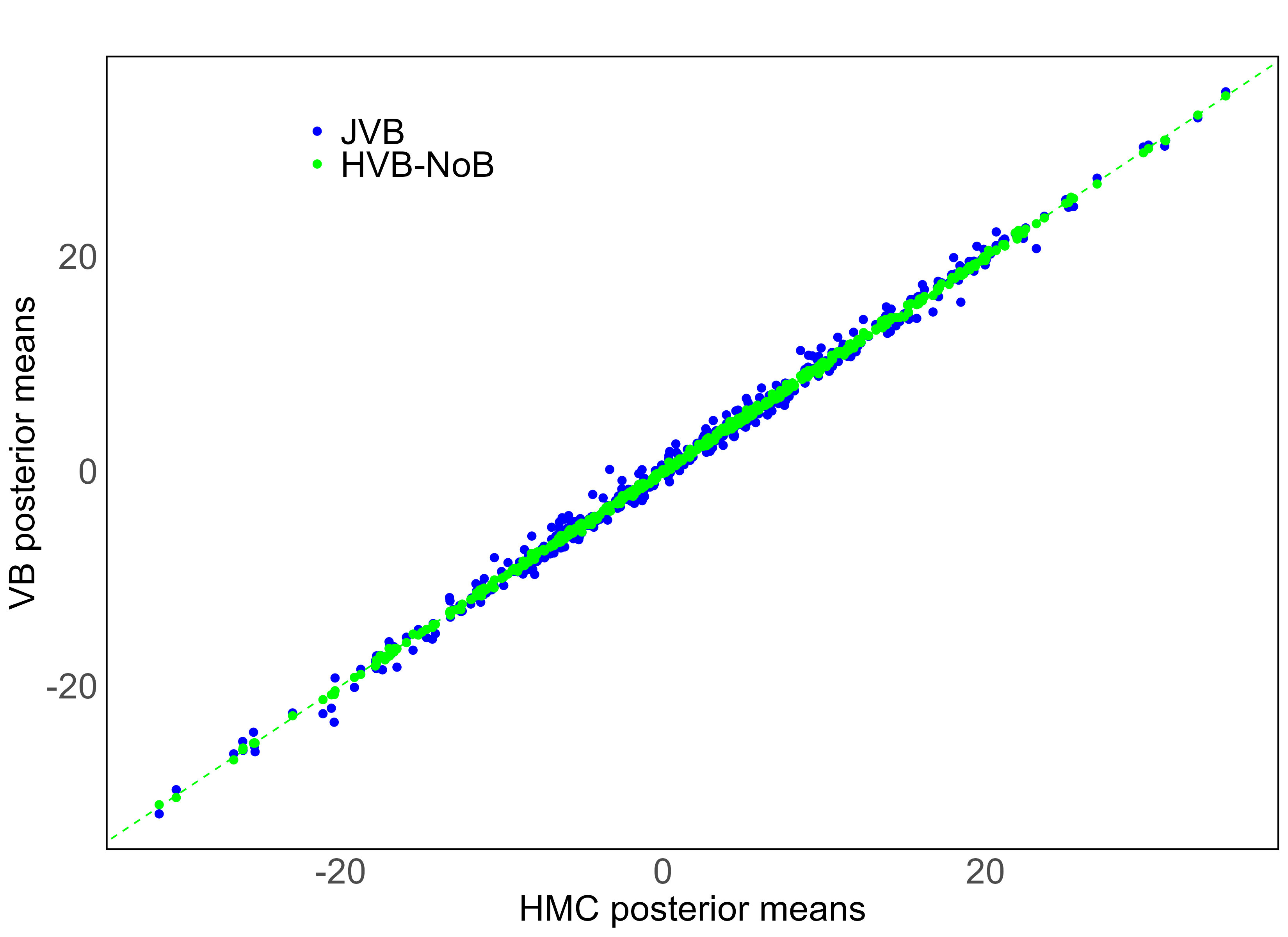}
        \caption{Posterior means}
        \label{fig:MAR_625_75p_vb_vs_HMC_mean}
    \end{subfigure}
    \hfill
    \begin{subfigure}{0.45\textwidth}
        \includegraphics[width=\linewidth]{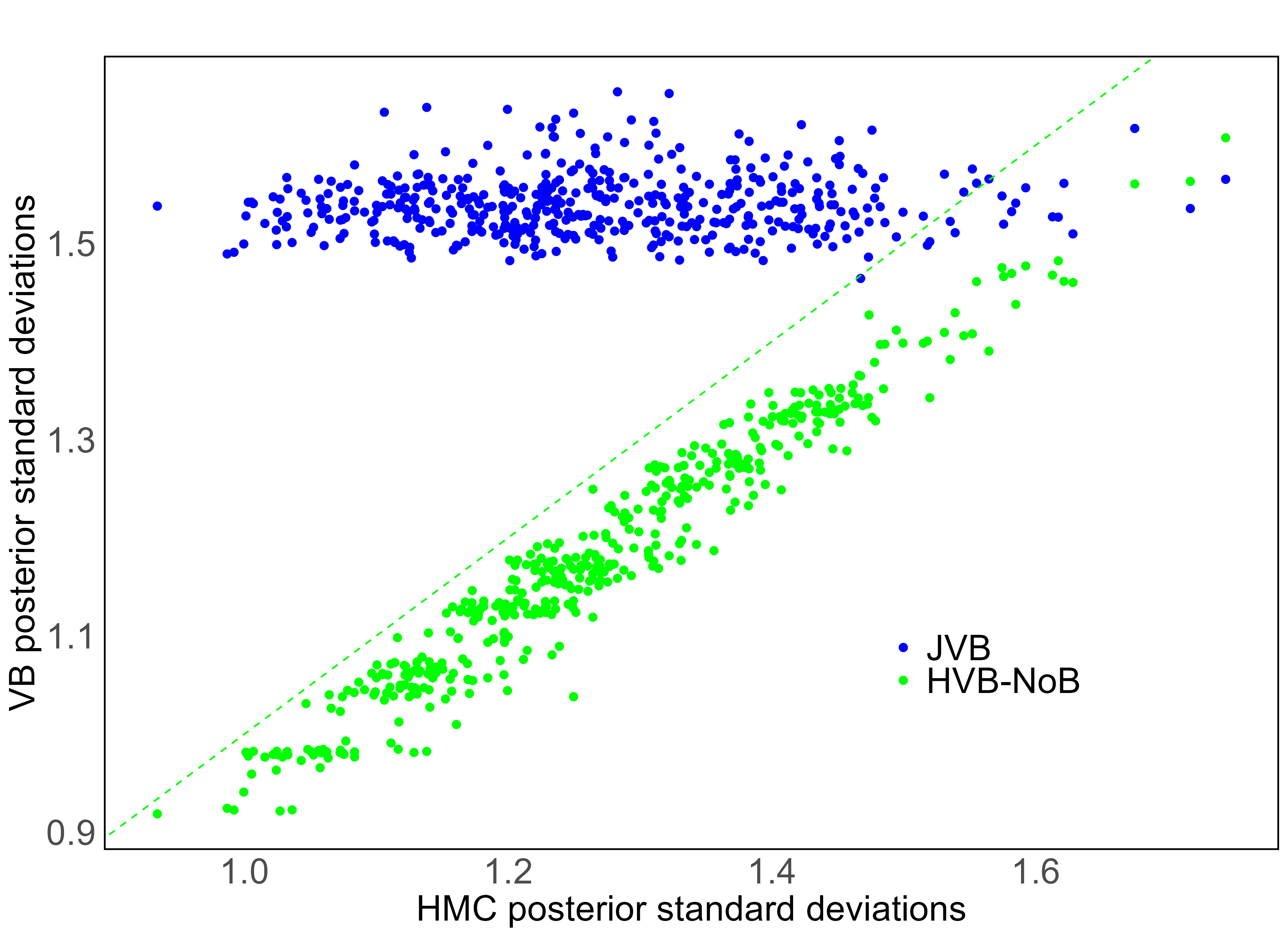}
        \caption{Posterior standard deviations}
        \label{fig:MAR_625_75p_vb_vs_HMC_sd}
    \end{subfigure}
    
    \caption{Comparison of the posterior means and standard deviations of the missing values obtained from JVB  and HVB-NoB with those of HMC under MAR for the simulated data with $n=625$ and $n_u=468$ ($75\%$ missing values)}
    \label{fig:MAR_625_75p_vb_vs_HMC_mean_sd}
\end{figure}

Figure~\ref{fig:MAR_625_75p_vb_vs_HMC_mean_sd} shows the comparison between the posterior means and standard deviations of the missing values obtained from the JVB, HVB-NoB, and HMC methods for the simulated dataset with $n=625$ and $n_u=468$. The posterior means obtained from the JVB and HVB-NoB methods are very close to those obtained from the HMC method, as shown in Figure \ref{fig:MAR_625_75p_vb_vs_HMC_mean}. However, the posterior standard deviations estimated from the JVB method are significantly different from those obtained using the HMC method, as shown in Figure~\ref{fig:MAR_625_75p_vb_vs_HMC_sd}. On the other hand, the posterior standard deviations estimated from the HVB-NoB method are very close to those of the HMC method. 


To investigate the accuracy of the VB methods with relatively large $n$ and missing values $n_u$, we simulated a dataset with $n=10,000$ and $n_u=7,500$ (i.e. missing percentage is $75\%$) under MAR. As indicated in Table~\ref{tab:MAR.times} in Section \ref{sec:sup.sim.MAR} of the online supplement, the average time taken per HMC iteration is notably high for high values of $n$ and $n_u$, making practical implementation of HMC infeasible. Further, when the number of units ($n_u$) is large (exceeding 1,000), utilising the standard HVB without the Gibbs steps becomes computationally intensive. We employ the HVB-G method with $N_1=5$ and $k^*=500$ and the JVB method. Table \ref{tbl:SEM_MAR_estimates_10000} compares the posterior means of the parameters obtained using the two VB methods with their true parameter values. Similar to the simulation results obtained from the simulated dataset with $n=625$, the HVB-G method accurately estimates the posterior means of SEM parameters, in particular for $\sigma^{2}_{y}$ and $\rho$ parameters, overcoming the inaccuracy of the JVB method. See Figures~\ref{fig:kernal_SEM_MAR_10000_miss_75} and \ref{fig:MAR_10000_75p_vbmean_vs_true} in Section~\ref{sec:sup:addit.sim.MAR} of the online supplement for a comparison of posterior densities of parameters and a comparison of estimated missing values from the two VB methods with the true missing values, respectively.



\begin{table}
    \centering
    \begin{tabular}{ccc} 
        \toprule
     & JVB & HVB-G \\
    \hline
    $\beta_0=1$ & \makecell{ 0.9349\\ (0.0199)} & \makecell{0.9387\\ (0.0611)}   \\
    $\beta_7=5$ & \makecell{ 4.9991\\ (0.0267)} & \makecell{ 4.9833\\ (0.0287)}  \\
        $\sigma^2_{\textbf{y}}=1$ & \makecell{ 2.1509 \\ (0.0443)} & \makecell{ 0.9998\\ (0.0385)}  \\
        $\rho=0.8$ & \makecell{ 0.0844\\ (0.0134)} & \makecell{  0.7971 \\ (0.0131)}  \\
    \bottomrule
    \end{tabular}
    \caption{Posterior means and standard deviations (inside brackets) of SEM parameters under MAR estimated using the JVB and HVB-G methods for the simulated dataset with $n=10,000$ and $n_u=7,500$ ($75\%$ missing values)}
    \label{tbl:SEM_MAR_estimates_10000}
\end{table}




Table~\ref{tab:MAR.times} in Section \ref{sec:sup.sim.MAR} of the online supplement displays the average computing cost per iteration (in seconds) for the VB and HMC methods for different $n$ and $n_u$ under the MAR mechanism. 
The HVB-G method is not implemented when $n_u$ is relatively small ($n_u<1,000$). The HMC method is computationally expensive when $n$ is large and is not implemented when $n > 5,000$. 
The HMC method is much more computationally expensive than the VB methods, regardless of the values of $n$ and $n_u$.
Although it can not accurately capture the posterior distributions of the parameters $\sigma^2_{\textbf{y}}$, $\rho$ and the posterior standard deviations of the missing values (see Figures \ref{fig:kernal_SEM_MAR_625_miss_75}, \ref{fig:MAR_625_75p_vb_vs_HMC_mean_sd} of the main paper, and Figure \ref{fig:kernal_SEM_MAR_10000_miss_75} in Section \ref{sec:sup:addit.sim.MAR} of the online supplement), the JVB method is generally the fastest among all the methods.
For smaller values of $n$ and $n_u$, the HVB-NoB algorithm is faster than the HVB-G method. The computational time of HVB-NoB increases rapidly as $n$ and $n_u$ increase, while HVB-G exhibits a lower computational cost than HVB-NoB, especially for higher missing value percentages.

\subsection{Simulation study under MNAR}
\label{sec:SESM.estimates}
This section discusses the simulation results for estimating SEM under the MNAR mechanism. 
When conducting simulations under MNAR, we set $\psi_{\textbf{x}^{*}}=0.5$ and $\psi_{\textbf{y}}=-0.1$ across all simulation studies. The parameter $\psi_0$ influences the percentages of missing values. We vary $\psi_0$ to obtain the desired missing value percentages. 
As discussed in Section~\ref{sec:methods}, when dealing with the MNAR mechanism, the parameters for the SEM and the missing data model in Equation~\eqref{eq:logistic.SEM} must be estimated to obtain accurate inference. The set of parameters to be estimated is $\boldsymbol{\theta}=(\boldsymbol{\phi}^\top,\boldsymbol{\psi}^\top)^\top=({\beta_0},\hdots,\beta_{10},\sigma^2_{\textbf{y}},\rho,\psi_0,\psi_{\textbf{x}^{*}},\psi_{\textbf{y}})^\top$. 


It is worth noting that properly selecting the tuning parameters for MCMC steps within the HVB-NoB, HVB-AllB, and HVB-3B algorithms is important for achieving accurate inference and rapid convergence, in particular for a large number of observations $n$ and a large number of missing values $n_u$. Our simulation studies showed that maintaining an acceptance percentage between $20\%$ and $30\%$ in the MCMC steps is necessary to balance between accurate inferences and computational cost. Adjusting the tuning parameters of the proposed MCMC schemes allows us to attain this desired acceptance percentage.
For the MCMC steps used in the HVB-NoB method presented in Algorithm~\ref{alg:AugvbMCMCstep}, there is one tuning parameter, which is the number of MCMC iterations, $N_1$. We set this to $N_1 = 10$ irrespective of the values $n$ and $n_u$. However, as $n_u$ increases, the acceptance percentage for this MCMC scheme drops rapidly. Increasing the value of $N_1$ does not improve the acceptance rate of the MCMC steps. The tuning parameters for the MCMC steps in Algorithm~\ref{alg:AugvbMCMCstepB} used in HVB-AllB and HVB-3B algorithms, include $N_1$, and the block size $k^*$. We fixed $N_1$ to $10$ irrespective of the values $n$ and $n_u$. If $n$ is small (say $n_u \leq 1,000$), we set the block size to $n_u \times 25\%$. This leads to the number of blocks being 4 or 5. When $n_u$ is large (say $n_u > 1,000$), we set the block size to $n_u \times 10\%$, resulting in 10 or 11 blocks.


We investigate the accuracy of the proposed VB methods for estimating SEM under MNAR with a small number of observations $n=625$ and a large number of observations $n=10,000$. For $n=625$, we consider $25\%$ and $75\%$ missing data percentages. This section discusses the results for the $75\%$ missing data percentage. The results for the $25\%$ missing data percentage are given in Section~\ref{sec:sup.sim.MNAR} of the online supplement. For the large number of observations $n=10,000$, we consider the $75\%$ missing data percentage.



\begin{figure}
    \centering
    \includegraphics[width=0.9\textwidth, keepaspectratio]{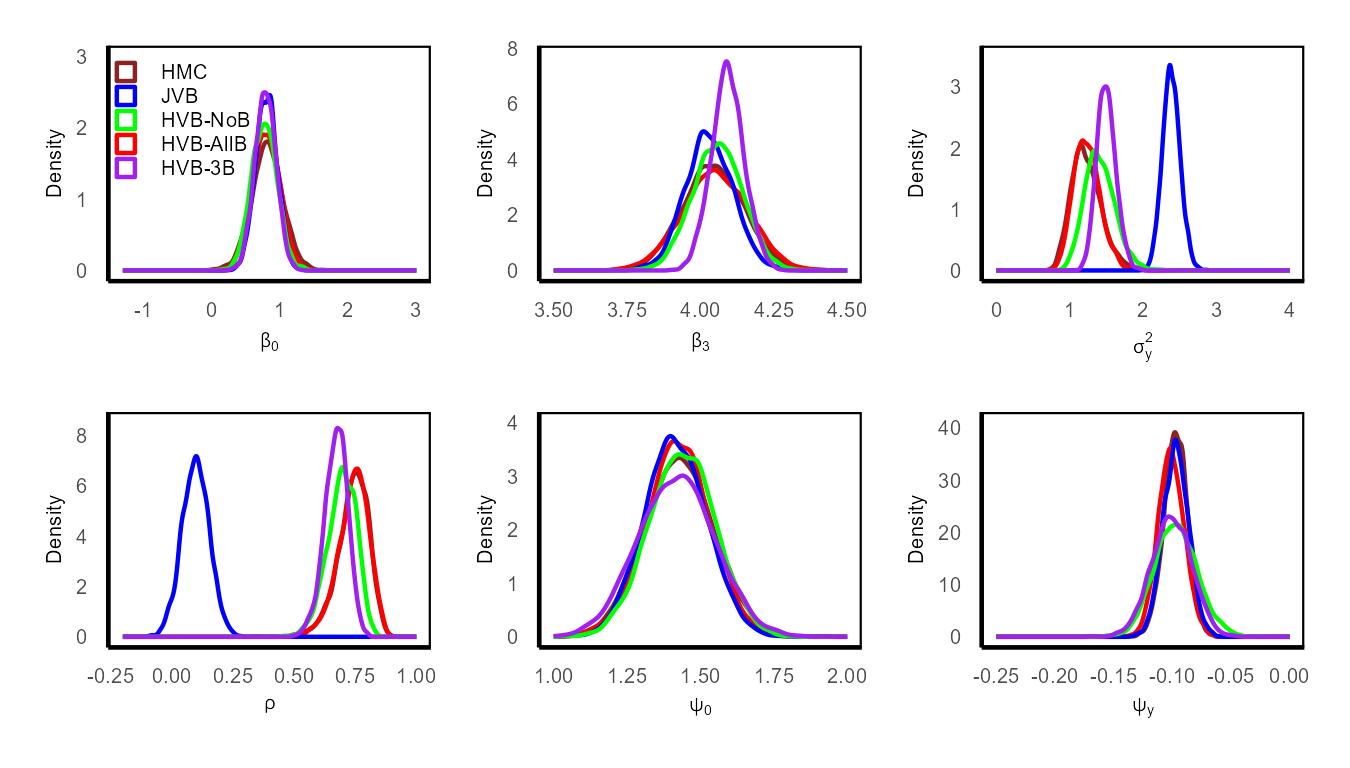}
    \caption{Posterior densities of SEM and missing data model parameters under MNAR for the simulated dataset with $n=625$ and $n_u=469$  (around $75\%$ missing) estimated using the HMC, JVB, HVB-NoB, HVB-AllB, and HVB-3B methods}
    \label{fig:kernal_SEM_MNAR_625_miss_75}
\end{figure}

Figure~\ref{fig:kernal_SEM_MNAR_625_miss_75} shows the posterior densities of SEM and missing model parameters estimated using different inference methods: HMC, JVB, HVB-NoB, HVB-AllB, and HVB-3B, for the simulated dataset with $n=625$ and around $75\%$ missing values ($n_n=469$). See Sections~\ref{sec:jvb}, \ref{sec:augdvb} to \ref{subsec:HVB-MNAR} for details on each VB method.  Similar to the inference under MAR, we observe significant differences in the posterior densities of $\sigma^2_{\textbf{y}}$ and $\rho$ from JVB compared to those from HMC, but the posterior densities from any variant of HVB methods closely resemble those from HMC. Among the three HVB variants, the posterior densities from HVB-AllB closely match those obtained from HMC for all parameters. However, despite its potential for more accurate inferences, HVB-AllB incurs a higher computing cost than HVB-NoB and HVB-3B, as detailed in Table~\ref{tab:SEM.MNAR.Times}.




\begin{figure}
    \centering
    \begin{subfigure}{0.45\textwidth}
        \includegraphics[width=\linewidth]{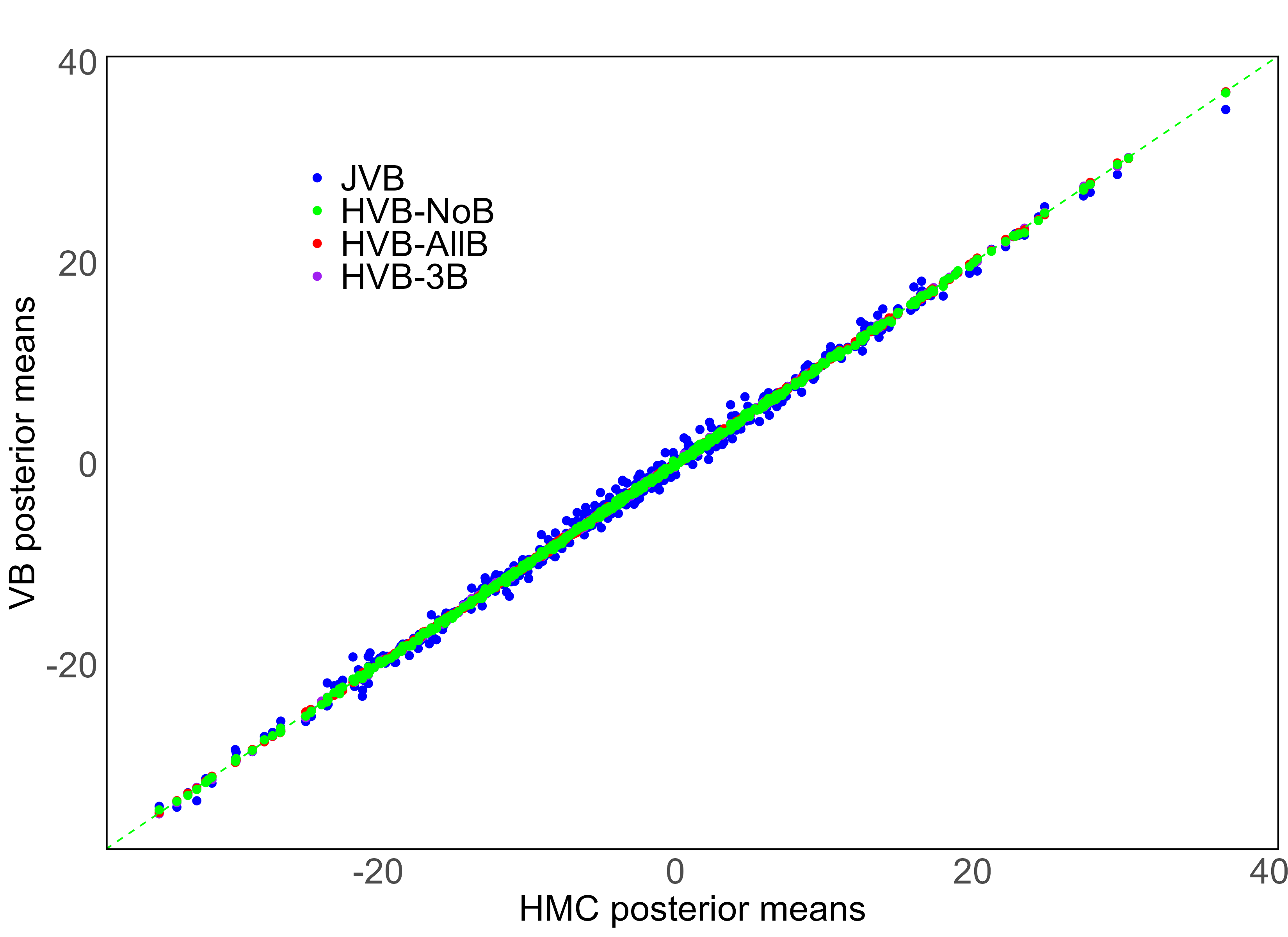}
        \caption{Posterior means}
        \label{fig:MNAR_625_75p_vb_vs_HMC_mean}
    \end{subfigure}
    \hfill
    \begin{subfigure}{0.45\textwidth}
        \includegraphics[width=\linewidth]{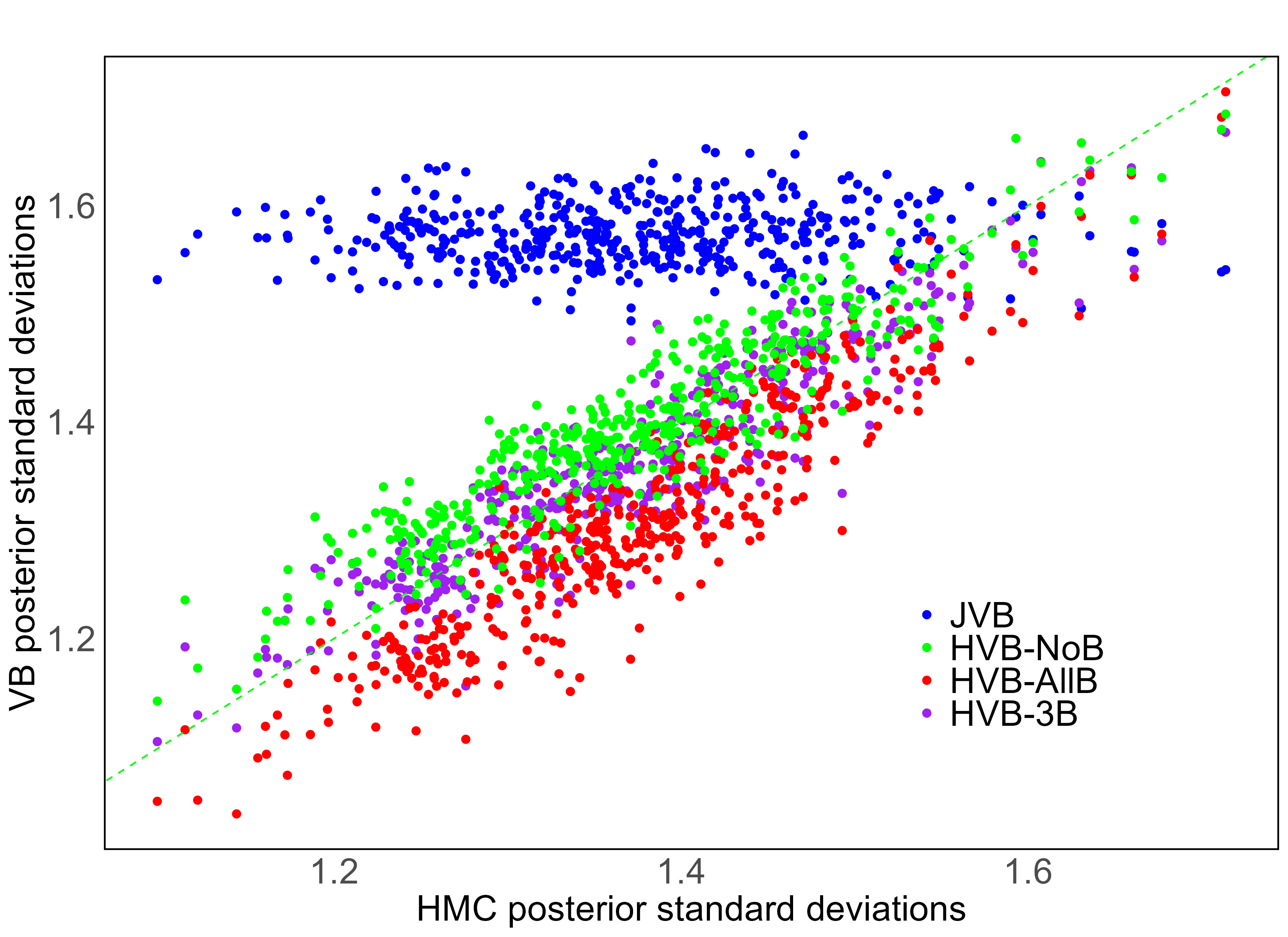}
        \caption{Posterior standard deviations}
         \label{fig:MNAR_625_75p_vb_vs_HMC_sd}
    \end{subfigure}
    
    \caption{Comparison of the posterior means and standard deviations of missing values from the JVB, HVB-NoB, HVB-AllB, and HVB-3B methods with those obtained using the HMC method under MNAR for the simulated dataset with $n=625$, and $n_u=469$ (around $75\%$ missing)}
    \label{fig:MNAR_625_75p_vb_vs_HMC_mean_sd}
\end{figure}

Figure~\ref{fig:MNAR_625_75p_vb_vs_HMC_mean_sd} compares the posterior means and standard deviations of missing values obtained from the JVB method and all three HVB methods with the HMC method for the simulated dataset with $n=625$ and $n_u=469$. The JVB posterior means and all the HVB posterior means are very close to those of HMC, as shown in Figure~\ref{fig:MNAR_625_75p_vb_vs_HMC_mean}. However, the posterior standard deviations estimated from the JVB method are significantly different from those obtained from the HMC method, as shown in Figure~\ref{fig:MNAR_625_75p_vb_vs_HMC_sd}, whereas the posterior standard deviations from all the HVB methods closely align with those of HMC. 


Similar to SEM under MAR, we also conducted a simulation study with $n=10,000$ with approximately $75\%$ of missing values under MNAR. As implementing HMC is infeasible for large $n$ (refer to Table~\ref{tab:SEM.MNAR.Times}), we implement the JVB, HVB-AllB, and HVB-3B methods on this dataset and compare the estimated posterior means of parameters obtained from VB methods with the true parameter values in Table~\ref{tbl:SEM_MNAR_estimates_10000}. The HVB-NoB algorithm is not implemented due to its high computational cost (as shown in Table~\ref{tab:SEM.MNAR.Times}). The posterior means obtained by HVB-AllB and HVB-3B algorithms accurately estimate the true parameter values, demonstrating superior accuracy compared to the JVB algorithm. See Figures~\ref{fig:kernal_SEM_MNAR_10000_miss_75} and \ref{fig:MNAR_10000_755p_vb_vs_HMC_mean} in Section~\ref{sec:sup:addit.sim.MNAR} of the online supplement for a comparison of posterior densities of model parameters and a comparison of estimated missing values from the three VB methods with the true missing values, respectively.

\begin{table}
    \centering
    \begin{tabular}{cccc} 
        \toprule
     True value& JVB & HVB-AllB & HVB-3B \\
    \hline
    $\beta_0=2$ & \makecell{ 2.0190\\ (0.0358)} & \makecell{1.9849\\ (0.0494)} & \makecell{1.9973 \\ (0.0547)}  \\
    $\beta_3=5$ & \makecell{ 5.0077\\ (0.0247)} & \makecell{5.0118 \\ (0.0214)} & \makecell{ 4.9898\\ (0.0081)} \\
        $\sigma^2_{\textbf{y}}=1$ & \makecell{  2.1769\\ (0.0441)} & \makecell{ 0.9682\\ (0.0286)} & \makecell{0.9209 \\ (0.0152)}  \\
        $\rho=0.8$ & \makecell{0.1443\\ (0.0129)} & \makecell{ 0.8128\\ (0.0100)} & \makecell{ 0.8190\\ (0.0066)} \\
    $\psi_0=1.5$ & \makecell{ 1.5062\\ (0.0255)} & \makecell{  1.5042\\ (0.0285)} & \makecell{ 1.5158\\ (0.0280)}  \\
    $\psi_k=0.5$ & \makecell{  0.5058\\ (0.0261)} & \makecell{ 0.4987\\ (0.0301)} & \makecell{ 0.5114\\ (0.0293)}  \\
    $\psi_{\textbf{y}}=-0.1$ & \makecell{-0.1003\\ (0.0057)} & \makecell{ -0.1098\\ (0.0090)} & \makecell{-0.1002 \\ (0.0083)}  \\
    \bottomrule
    \end{tabular}
    \caption{Posterior means and standard deviations (inside brackets) of SEM and missing data model parameters under MNAR estimated using the JVB and different HVB methods for the simulated dataset with $n=10,000$ and $n_u=7,542$ (around $75\%$ missing values)}
    \label{tbl:SEM_MNAR_estimates_10000}
\end{table}

\begin{table}
\centering
\small
\begin{tabular}{crrrrrrrrr}
\toprule
\makecell{$n$} & \multicolumn{3}{c}{\centering $n=625$} & \multicolumn{3}{c}{$n=1024$} & \multicolumn{3}{c}{$n=2500$} \\ \midrule
                           \makecell{missing \\ percentage}& \multicolumn{1}{c}{$25\%$}           & \multicolumn{1}{c}{$50\%$}  & \multicolumn{1}{c}{$75\%$}       & \multicolumn{1}{c}{$25\%$}        & \multicolumn{1}{c}{$50\%$}  & \multicolumn{1}{c}{$75\%$}   & \multicolumn{1}{c}{$25\%$}          & \multicolumn{1}{c}{$50\%$}  & \multicolumn{1}{c}{$75\%$} \\ \midrule
                           \makecell{HMC}               & 28.79  & 32.31 &  36.22                        &    42.79      & 98.59 & 157.52     &  1046.23       & 1100.89 &  1597.88 \\
\makecell{JVB}                &   0.03 & 0.04 &    0.04                       &     0.06    &  0.07 &  0.09    & 0.30       & 0.38 & 0.47 \\
\makecell{HVB-NoB}                & 0.05  &  0.07&     0.07                     &  0.07       & 0.12 &  0.20    &    0.36     & 0.88  & 1.91 \\
\makecell{HVB-AllB}                & 0.25  &  0.23  &   0.29                       &    0.31     & 0.36 &  0.43    &   0.66     & 1.38 &  1.60 \\
\makecell{HVB-3B}                &  0.18 & 0.19 &       0.20                    &     0.23    &  0.26 &   0.30    &      0.59   & 0.60 &  0.67  \\
 \midrule

\makecell{$n$} & \multicolumn{3}{c}{\centering $n=5,041$} & \multicolumn{3}{c}{$n=7,569$} & \multicolumn{3}{c}{$n=10,000$} \\ \midrule
                         \makecell{missing \\ percentage}   & \multicolumn{1}{c}{$25\%$}           & \multicolumn{1}{c}{$50\%$}  & \multicolumn{1}{c}{$75\%$}       & \multicolumn{1}{c}{$25\%$}         & \multicolumn{1}{c}{$50\%$}  & \multicolumn{1}{c}{$75\%$}   & \multicolumn{1}{c}{$25\%$}          & \multicolumn{1}{c}{$50\%$}  & \multicolumn{1}{c}{$75\%$} \\ \midrule
\makecell{HMC}                     &  - & \makecell{-} &     \makecell{-}                     & \makecell{-}        & \makecell{-} & \makecell{-}     & \makecell{-}        & \makecell{-} & \makecell{-} \\
\makecell{JVB}                & 1.34  &  1.59 &             2.02              &   3.52      & 4.16  &  5.29    &    9.23     & 9.33  & 13.35  \\
\makecell{HVB-NoB}                &  1.71 & 7.29 &    19.31                       &     4.90    &  26.62 &    90.57  &     11.543     & 71.82 & 228.29 \\
\makecell{HVB-AllB}                &  2.88 & 3.72 &    5.43                       &    5.87     & 8.57  &    14.54  &   12.13      & 17.05 & 26.29 \\
\makecell{HVB-3B}                & 1.74   & 1.98  &   2.50                       &   4.07       &  4.95 &    6.60  &    9.03     & 10.53  & 15.47 \\
\bottomrule
\end{tabular}
\caption{Average computing time (in seconds) of one iteration of the HMC, JVB, HVB-NoB, HVB-AllB, and HVB-3B methods for estimating SEM under MNAR for different $n$ and different missing value ($n_u$) percentages.
For HVB-AllB and HVB-3B, the tuning parameters, such as the block size and the number of MCMC iterations, are set according to the criteria described at the beginning of Section~\ref{sec:SESM.estimates}}
\label{tab:SEM.MNAR.Times}
\end{table}



Table~\ref{tab:SEM.MNAR.Times} presents the average computing time (in seconds) per iteration for the VB and HMC methods across different values of $n$ and $n_u$ under MNAR. Regardless of the values of $n$ and $n_u$, the HMC method is much more computationally expensive compared to the VB methods. Despite its limitations in accurately capturing the posterior distributions of $\sigma^2_{\textbf{y}}$ and $\rho$ (see Figure~\ref{fig:kernal_SEM_MNAR_625_miss_75} and Figure~\ref{fig:kernal_SEM_MNAR_10000_miss_75} in Section~\ref{sec:sup:addit.sim.MNAR} of the online supplement), and the posterior standard deviations of missing values (see Figure~\ref{fig:MNAR_625_75p_vb_vs_HMC_mean_sd}), the JVB method is faster than the any of HVB methods. For smaller values of $n$ and $n_u$, the HVB-NoB algorithm is significantly faster than its counterparts, HVB-AllB and HVB-3B. However, as $n$ and $n_u$ increase, the HVB-AllB and HVB-3B are faster than HVB-NoB, with HVB-3B exhibiting the lowest computing cost, as expected.

\section{Real example}
\label{sec:RealWorldAnalysis}
We utilise the proposed VB methods to analyse a dataset containing votes cast during the 1980 presidential election across 3,107 U.S. counties. This dataset is available in the R package spData ~\citep{spData}. The dataset includes county-level information on the following: the proportion of votes cast by the eligible population, the proportion of the eligible population with college degrees, the proportion of the eligible population that owns homes, and income per capita.~\citet{pace1997quick} applied the SDM to this dataset, choosing the logarithm of the proportion of votes cast as the dependent variable.
Furthermore, the dataset contains a pre-defined county-level weight matrix, with an average of 5-6 neighbours per unit (county). The weight matrix from the dataset is denser than those used in our simulation studies, which has an average of only 3-4 neighbours per unit. Therefore, implementing VB algorithms for this dataset requires more computing time than the simulation studies.

In our analysis, we treat the logarithm of the proportion of votes cast as the dependent variable. Additionally, we include the logarithms of the proportions of college degrees and homeownership, as well as income per capita, along with their interaction effects, as the set of covariates. Each covariate is standardized to have a mean of zero and a standard deviation of one.

\subsection{SEM under MAR \label{sec:realdataMAR}}

This section investigates the performance of VB methods to estimate SEM under MAR for the 1980 presidential election dataset. Given the full dataset, we randomly select $n_o$ units to form the observed dataset. The remaining $n_u$ units are treated as missing responses. We estimate SEM parameters and the missing values using the JVB and HVB-G algorithms.  Due to the moderately large number of observations  $n=3,107$, employing the HMC algorithm becomes computationally intensive, as detailed in Table~\ref{tab:MAR.times} in Section \ref{sec:sup.sim.MAR} of the online supplement. We compare the posterior mean estimates of the SEM parameters with those obtained from the marginal maximum likelihood (ML) method of \citet{suesse2018marginal}.


For both VB algorithms, we used the starting values as described in the simulation study in Section~\ref{sec:SimulationStudy} and ran the algorithms for $15,000$ iterations, at which point both algorithms were well-converged (see Section~\ref{sec:sup:con.real.MAR} of the online supplement for further details). For the HVB-G algorithm, we set the block size to $500$ and the number of Gibbs iterations 
$N_1$ to $10$.


Figure~\ref{fig:kernal_SEM_MAR_elec} in Section \ref{sec:sup:addit.real.MAR} of the online supplement presents the posterior densities of the SEM parameters estimated using the JVB and HVB-G methods with $75\%$ missing responses ($n_u=2,330$). The vertical lines indicate marginal ML estimates. The figure shows that the JVB method yields different posterior density estimates for the parameters $\sigma^2_{\textbf{y}}$ and $\rho$ compared to the HVB-G method. However, the posterior mean estimates of SEM parameters, obtained using the HVB-G method, closely align with the marginal ML estimates compared to those obtained from the JVB method. See Table~\ref{tbl:SEM_MAR_elec} for a summary of the estimation results.


Figure~\ref{fig:MAR_elec_vbmean_vs_true} in Section \ref{sec:sup:addit.real.MAR} of the online supplement compares the posterior means of missing values obtained from the JVB and HVB-G algorithms with the true missing values. It is evident that the posterior mean estimates of missing values from HVB-G are slightly closer to the true missing values than those from the JVB algorithm; see mean squared errors (MSEs) of
estimated missing values in Table~\ref{tbl:SEM_MAR_elec}. 

\begin{table}
    \centering
    \caption{Marginal ML estimates (with standard errors in brackets), and posterior means (with posterior standard deviations in brackets) of SEM parameters estimated by JVB and HVB-G algorithms for the 1980 presidential election dataset with $75\%$ ($n_u = 2,330$) missing values under MAR. The table also includes the MSE of missing values obtained from the JVB and HVB algorithms, along with computing time}
    \label{tbl:SEM_MAR_elec}
    \begin{tabular}{cccc} 
        \toprule
     & marginal ML& JVB & HVB-G \\
    \hline
    $intercept$ & \makecell{ -0.5723\\ (0.01087)}& \makecell{-0.5697\\ (0.0051)} & \makecell{-0.5696 \\ (0.0091)}   \\
        $\sigma^2_{\textbf{y}}$ & \makecell{0.0089 \\ (0.0006)}& \makecell{0.0253\\ (0.0007)} & \makecell{0.0093 \\ (0.0006)}   \\
        $\rho$  & \makecell{0.8306 \\ (0.0181)}& \makecell{0.1916 \\ (0.0282)} & \makecell{0.8346 \\ (0.0182)}\\
        MSE & - &44.8966 & 32.9627\\
        \makecell{Computing\\ time in seconds} & 277.26  & 10546.5 & 36939 \\
    \bottomrule
    \end{tabular}
\end{table}

Table~\ref{tbl:SEM_MAR_elec} presents the marginal ML estimates with their standard errors and the posterior means and standard deviations of the SEM parameters estimated by the two VB methods. The table also includes the computing time of each algorithm and the MSEs of estimated missing values from JVB and HVB. The table shows that the MSE of HVB-G is lower than that of JVB. See Table~\ref{tbl:SEM_MAR_elec_all} in Section~\ref{sec:sup:addit.real.MAR} of the online supplement for further details on the estimation results, including estimates for all the fixed effects.


\subsection{SEM under MNAR \label{sec:realSEM-MNAR}}

This section investigates the performance of VB methods to estimate SEM under MNAR for the 1980 presidential
election dataset. 
The logistic regression in Section \ref{sec:SimulationStudy} is used as the missing data model. We use the logarithms of the proportions of college degrees and the response variable $y$ of the SEM (logarithm of the proportion of votes cast) as the covariates in the missing data model. The missing data model has three parameters, denoted as $\boldsymbol{\psi}=(\psi_0,\psi_{\textbf{x}^{*}},\psi_{\textbf{y}})$. We set the values $\psi_0=1.4$, $\psi_{\textbf{x}^{*}}=0.5$, and $\psi_{\textbf{y}}=-0.1$, resulting in approximately $80\%$ of responses being missing ($n_u= 2,477$). 

The JVB, HVB-AllB, and HVB-3B methods are used to estimate the posterior densities of SEM and missing data model parameters. The starting values for all algorithms are chosen similarly to those in the simulation study. The tuning parameters for HVB-AllB and HVB-3B are selected as follows: the number of MCMC iterations is set to $N_1=20$, and since $n_u > 1,000$, the block size $k^*$ is set to $n_u \times 10\% \approx 247$. This led to a total of 11 blocks. All VB algorithms were run for $15,000$ iterations, at which point all algorithms had well converged. See Figure~\ref{fig:con_vb_MNAR_elec} in Section \ref{sec:sup:con.real.MNAR} of the online supplement for the convergence analysis.

Figure~\ref{fig:kernal_SEM_MNAR_elec} in Section \ref{sec:sup:addit.real.MNAR} of the online supplement compares the posterior densities of SEM and missing data model parameters obtained from the three VB methods. The figure shows that the posterior densities of the parameters, except for $\sigma^2_{\textbf{y}}$ and $\rho$, obtained from different methods, are almost identical. For $\sigma^2_{\textbf{y}}$ and $\rho$, the posterior densities obtained from HVB-AllB and HVB-3B are closer to each other compared to those from JVB.

Figure~\ref{fig:MNAR_elec_vb_mean_vs_yu} in Section \ref{sec:sup:addit.real.MNAR} of the online supplement compares the posterior means of missing values obtained from the JVB, HVB-AllB, and HVB-3B methods with the true missing values. The posterior means of the missing values obtained from the HVB-AllB method are slightly closer to the true missing values than those from the HVB-3B and JVB algorithms, as indicated by their greater concentration along the diagonal line. This is further supported by the lower MSE of estimated missing values from the HVB-AllB method in comparison to both the HVB-3B and JVB methods, as shown in Table~\ref{tbl:SEM_MNAR_elec}. The table also includes the posterior means and standard deviations of parameters obtained from the three VB algorithms, the true parameter values for the missing data model parameters, and the computing time for each method. Although the HVB-3B estimates of missing values have a slightly higher MSE than HVB-AllB, its computing time is nearly three times shorter. Therefore, HVB-3B is an computationally less expensive yet reasonably accurate alternative to the HVB-AllB method. Table~\ref{tbl:SEM_MNAR_elec_all} in Section~\ref{sec:sup:addit.real.MNAR} of the online supplement summarises estimates for all the fixed effects.



\begin{table}
    \centering
    \caption{Posterior means (with posterior standard deviations inside brackets) of SEM and missing data model parameters, and the true parameter values for the
missing data model parameters under MNAR obtained from the JVB, HVB-AllB, and HVB-3B methods for the 1980
presidential election datase with approximately $80\%$ missing values ($n_u=2,477$). The table also includes the MSE of missing values by JVB and HVB algorithms, along with computing time}
    \label{tbl:SEM_MNAR_elec}
    \begin{tabular}{ccccc} 
        \toprule
     & True value& JVB & HVB-AllB & HVB-3B \\
    \hline
    $intercept~(\beta_0)$ & \makecell{ NA}  & \makecell{-0.5815\\(0.0112)} & \makecell{-0.5773 \\(0.0163)} & \makecell{-0.5689\\(0.0078)} \\

        $\sigma^2_{\textbf{y}}$ & \makecell{ NA}  & \makecell{0.0242\\(0.0006)} & \makecell{0.0109 \\(0.0008)} & \makecell{0.0154\\(0.0006)} \\
        $\rho$ & \makecell{ NA}  & \makecell{0.1634\\(0.0259)} & \makecell{0.8004 \\(0.0171)} & \makecell{0.6990\\(0.0186)} \\
        $\psi_0$ & 1.4   & \makecell{1.6622\\(0.1977)} & \makecell{ 1.6504\\(0.1934)} & \makecell{1.5891\\(0.1455)} \\
          $\psi_{\textbf{x}^{*}}$ &  0.5   & \makecell{0.5227\\(0.0519)} & \makecell{0.5172 \\(0.0497)} & \makecell{0.5376\\(0.0454)} \\
            $\psi_{\textbf{y}}$ & -0.1  & \makecell{0.3540\\(0.3252)} & \makecell{ 0.3316\\(0.3269)} & \makecell{0.2286\\(0.2354)} \\
            MSE & NA & 46.9063 & 35.7157 & 37.6031\\
        \makecell{Computational\\ time in seconds} & \makecell{ NA}& 11620.5 & 68035.5 &23787\\
    \bottomrule
    \end{tabular}
\end{table}

\section{Conclusion\label{sec:Conclusion}}
Our article proposes VB methods for estimating SEM under missing at random (MAR) and missing not at random (MNAR) missing data mechanisms. The joint VB (JVB) and the class of hybrid VB (HVB) methods are proposed. The posterior densities estimated using the Hamiltonian Monte Carlo (HMC) method are considered as ground truth to assess the accuracy of the VB methods for a small to moderate number of observations $n$ and missing response values $n_u$. The HMC method for this model is infeasible when $n$ and $n_u$ are large. 

The empirical results show that: (1) All proposed VB methods are computationally less expensive compared to the HMC method; (2) All HVB methods produce posterior density estimates for all model parameters and missing response values that are similar to those obtained using the HMC method for estimating SEM under MAR and MNAR. However, as $n$ and $n_u$ increase, HVB-NoB produces inaccurate estimates due to the low acceptance percentage of the underlying MCMC steps; (3) The HVB-3B method generates slightly different posterior estimates compared to other HVB algorithms. This is expected because, for each MCMC step of the HVB-3B algorithm, updates are performed on only randomly selected 3 blocks; (4) HVB-3B is more scalable for large $n$ and $n_u$ compared to HVB-AllB while still providing nearly similar posterior density estimates; (5) The JVB method yields quite accurate posterior density estimates for fixed effect parameters and the posterior means of the missing response values. However, it provides inaccurate posterior density estimates for the parameters $\sigma^{2}_{y}$ and $\rho$, as well as for the posterior standard deviations of the missing response values under both MAR and MNAR; (6) Generally, all HVB algorithms tend to converge in fewer iterations compared to the JVB algorithm.



\section*{Statements and Declarations}
The authors declare no potential or apparent conflict of interest in this article. 

\begin{singlespace}
\bibliographystyle{apalike}
\bibliography{ref.bib}
\end{singlespace}

\pagebreak

\renewcommand{\thefigure}{S\arabic{figure}} 

\renewcommand{\theequation}{S\arabic{equation}} 

\renewcommand{\thetable}{S\arabic{table}}

\renewcommand{\thealgorithm}{S\arabic{algorithm}}

\renewcommand{\thesection} 
{S\arabic{section}}

\doublespacing
\section*{Online Supplement for Variational Bayes Inference for Spatial Error Models with Missing Data}

\setcounter{page}{1} 
\setcounter{section}{0} 

\setcounter{equation}{0} 
\setcounter{table}{0} 
\setcounter{figure}{0} 
\setcounter{algorithm}{0} 

We use the following notation in the online supplement. Eq.~(1), Table~1,
Figure~1, and Algorithm~1, etc, refer to the main paper, while Eq.~(S1),
Table~S1, Figure~S1,  and Algorithm~S1, etc, refer to the supplement.


\section{Derivation of VB algorithms}

\subsection{Derivation of the reparameterisation gradient for JVB algorithm}
\label{sec:der.jvb}

In the main paper, the reparameterisation gradient of $\mathcal{L}(\boldsymbol{\lambda})$ for JVB algorithm is given by

\begin{equation}
    \label{eq:ELBO.grad.lamda}
                    \nabla_{\boldsymbol{\lambda}}\mathcal{L}(\boldsymbol{\lambda})=E_{f_{\boldsymbol{\zeta}}}\left[\frac{du(\boldsymbol{\zeta,\boldsymbol{\lambda}})^\top}{d\boldsymbol{\lambda}}\{\nabla_{\boldsymbol{\theta},\textbf{y}_u}\text{log}~h(\boldsymbol{\theta},\textbf{y}_u)-\nabla_{\boldsymbol{\theta},\textbf{y}_u}\text{log}~  q_{\boldsymbol{\lambda}}(\boldsymbol{\theta},\textbf{y}_u) \}\right],\\
\end{equation}

\noindent where $\frac{du(\boldsymbol{\zeta,\boldsymbol{\lambda}})}{d\boldsymbol{\lambda}}$ is the derivative of the transformation $u(\boldsymbol{\zeta,\boldsymbol{\lambda}})=\boldsymbol{\mu}+\textbf{B}\boldsymbol{\eta}+\textbf{d}\circ \boldsymbol{\epsilon}$ with respect to the variational parameters $\boldsymbol{\lambda}=(\boldsymbol{\mu}^\top
, \text{vech}(\textbf{B})^\top, \textbf{d}^\top)^\top$, where "vech" operator is the vectorisation of a matrix by stacking its columns from left to right. We write that $u(\boldsymbol{\zeta,\boldsymbol{\lambda}})=\boldsymbol{\mu}+(\boldsymbol{\eta}^\top \otimes \textbf{I}_{S+n_u})\text{vech}(\textbf{B})+\textbf{d}\circ \boldsymbol{\epsilon}$, where $\otimes$ represents the  Kronecker product, and $\textbf{I}_{S+n_u}$ is the identity matrix of size $S+n_u$. It can be shown that  $\nabla_{\boldsymbol{\theta},\textbf{y}_u}\text{log}~  q_{\boldsymbol{\lambda}}(\boldsymbol{\theta},\textbf{y}_u)=-(\textbf{B}\textbf{B}^\top+\textbf{D}^2)^{-1}((\boldsymbol{\theta}^\top,\textbf{y}_u^\top)^\top-\boldsymbol{\mu})$, 
\begin{equation}
    \label{eq:du_by_dmu}
    \frac{du(\boldsymbol{\zeta},\boldsymbol{\lambda})}{d\boldsymbol{\mu}}=\textbf{I}_{S+n_u}\\~~~~~\text{and}~~~~~
    \frac{du(\boldsymbol{\zeta},\boldsymbol{\lambda})}{d\text{vech}(\textbf{B})}=\boldsymbol{\eta}^\top\otimes\textbf{I}_{S+n_u}.
\end{equation}
The derivatives of the lower bound with respect to variational parameters are: 
\begin{equation}
    \label{eq:grad_wrt_mu}
    \begin{split}
            \nabla_{\boldsymbol{\mu}}\mathcal{L}(\boldsymbol{\lambda})&=E_{f_{\boldsymbol{\zeta}}}[\nabla_{\boldsymbol{\theta},\textbf{y}_u}~\text{log}~h(\boldsymbol{\mu}+\textbf{B}\boldsymbol{\eta}+\textbf{d}\circ\boldsymbol{\epsilon})\\
            &+(\textbf{B}\textbf{B}^\top+\textbf{D}^2)^{-1}(\textbf{B}\boldsymbol{\eta}+\textbf{d}\circ\boldsymbol{\epsilon})],
    \end{split}
\end{equation}
\begin{equation}
    \label{eq:grad_wrt_B}
    \begin{split}
        \nabla_{\textbf{B}}\mathcal{L}(\boldsymbol{\lambda})&=E_{f_{\boldsymbol{\zeta}}}[\nabla_{\boldsymbol{\theta},\textbf{y}_u}\text{log}~h(\boldsymbol{\mu}+\textbf{B}\boldsymbol{\eta}+\textbf{d}\circ\boldsymbol{\epsilon})\boldsymbol{\eta}^\top\\
            &+(\textbf{B}\textbf{B}^\top+\textbf{D}^2)^{-1}(\textbf{B}\boldsymbol{\eta}+\textbf{d}\circ\boldsymbol{\epsilon})\boldsymbol{\eta}^\top],
    \end{split}
\end{equation}

and
\begin{equation}
    \label{eq:grad_wrt_d}
    \begin{split}
    \nabla_{\textbf{d}}\mathcal{L}(\boldsymbol{\lambda})&=E_{f_{\boldsymbol{\zeta}}}[\text{diag}(\nabla_{\boldsymbol{\theta},\textbf{y}_u}\text{log}~h(\boldsymbol{\mu}+\textbf{B}\boldsymbol{\eta}+\textbf{d}\circ\boldsymbol{\epsilon})\boldsymbol{\epsilon}^\top\\
            &+(\textbf{B}\textbf{B}^\top+\textbf{D}^2)^{-1}(\textbf{B}\boldsymbol{\eta}+\textbf{d}\circ\boldsymbol{\epsilon})\boldsymbol{\epsilon}^\top)],
    \end{split}
\end{equation}

\noindent where $\text{diag}(\cdot)$ is the vector of diagonal elements extracted from a square matrix. The analytical expressions for $\nabla_{\boldsymbol{\theta},\textbf{y}_u}\text{log}~h(\boldsymbol{\mu}+\textbf{B}\boldsymbol{\eta}+\textbf{d}\circ\boldsymbol{\epsilon})=\nabla_{\boldsymbol{\theta},\textbf{y}_u}\text{log}~h(\boldsymbol{\theta},\textbf{y}_u)$ in Equations~\eqref{eq:grad_wrt_mu}-\eqref{eq:grad_wrt_d} under MAR and MNAR mechanisms are provided in Section~\ref{sec:h_gradients}. 
The expectations in these gradients can be estimated using a single sample drawn from $f_{\boldsymbol{\zeta}}$, and they provide unbiased estimates $\widehat{{\nabla_{\boldsymbol{\lambda}}\mathcal{L}(\boldsymbol{\lambda})}} $ for $\nabla_{\boldsymbol{\lambda}}\mathcal{L}(\boldsymbol{\lambda})$. These estimates are utilised in the gradient calculation step (step 4) of Algorithm~\ref{alg:VB1} of the main paper. The adaptive learning rates (step sizes) utilised in Algorithm~\ref{alg:VB1} are determined through the ADADELTA algorithm ~\citep{Zeiler2012ADADELTAAA}, as detailed in Section~\ref{sec:sup:ADADELTA}.


Computing gradient estimates using Equations~\eqref{eq:grad_wrt_mu}, \eqref{eq:grad_wrt_B}, and \eqref{eq:grad_wrt_d} presents computational problems, in particular, when number of covariates and missing values is large. The inversion of $(S+n_u) \times (S+n_u)$ matrix, $(\textbf{B} \textbf{B}^T + \textbf{D}^2)$, is computationally expensive. Using the Woodbury formula~\citep[p. 427]{harville1997matrix}, this inversion can be reformulated as:


\begin{equation}
    \label{eq:woodbury}
    (\textbf{B} \textbf{B}^T + \textbf{D}^2)^{-1} = \textbf{D}^{-2} - \textbf{D}^{-2} \textbf{B} ( \textbf{I}_p + \textbf{B}^T D^{-2} \textbf{B} )^{-1} \textbf{B}^T \textbf{D}^{-2},
\end{equation}

\noindent where $\textbf{I}_p$ is the diagonal matrix of dimension $p \times p$.

On the right-hand side of Equation~\eqref{eq:woodbury}, the term $( \textbf{I}_p + \textbf{B}^T \textbf{D}^{-2} \textbf{B})$ is a square matrix of size $p \times p$ (where $p$ is much smaller than $S+n_u$), and $\textbf{D}$ is a diagonal matrix. Directly inverting $( \textbf{I} + \textbf{B}^T \textbf{D}^{-2} \textbf{B})$ has a computational complexity of $O(p^3)$. Consequently, computing $ (\textbf{B} \textbf{B}^T + \textbf{D}^2)^{-1}$ using this method also involves $O(p^3)$ complexity. Alternatively, without utilising the Woodbury formula, the complexity increases significantly to $O((S+n_u)^3)$.

\subsection{Derivation of the reparameterisation gradient for HVB algorithm}
\label{sec:der.hvb}

Since $\mathcal{L}(\boldsymbol{\lambda})=E_{q}\left(\text{log}~p(\textbf{O}\mid \boldsymbol{\theta})+\text{log}~p(\boldsymbol{\theta})-\text{log}~  q_{\boldsymbol{\lambda}}^0(\boldsymbol{\theta}) \right)=  \mathcal{L}^0(\boldsymbol{\lambda})$ as shown in Section~\ref{sec:augdvb} of the main paper, the reparameterisation gradient of $\mathcal{L}$ is the same as that of $\mathcal{L}^0$,
\begin{equation}
    \label{eq:ELBO.grad.lamda.aug_1}
                    \nabla_{\boldsymbol{\lambda}}\mathcal{L}(\boldsymbol{\lambda})=E_{f_{\boldsymbol{\delta}^0}}\left[\frac{dt^0(\boldsymbol{\delta}^0,\boldsymbol{\lambda}_{\boldsymbol{\theta}})^\top}{d\boldsymbol{\lambda_{\boldsymbol{\theta}}}}(\nabla_{\boldsymbol{\theta}}\text{log}~p(\boldsymbol{\theta})+\nabla_{\boldsymbol{\theta}}\text{log}~p(\textbf{O}\mid \boldsymbol{\theta})-\nabla_{\boldsymbol{\theta}}\text{log}~  q_{\boldsymbol{\lambda}}^0(\boldsymbol{\theta}) )\right],
\end{equation}

\noindent where, the random vector $\boldsymbol{\delta}^0$ has density $f_{\boldsymbol{\delta}^0}$, which follows a standard normal, and does not depend on $\boldsymbol{\lambda_{\boldsymbol{\theta}}}$, and $t^0$ is the one-to-one vector-valued
transformation from $\boldsymbol{\delta}^0=({\boldsymbol{\eta}^0}^\top,{\boldsymbol{\epsilon}^0}^\top)^\top$ to the parameter vector, such that $\boldsymbol{\theta}=t^0(\boldsymbol{\delta}^0,\boldsymbol{\lambda}_{\boldsymbol{\theta}})=\boldsymbol{\mu}_{\boldsymbol{\theta}}+\textbf{B}_{\boldsymbol{\theta}}\boldsymbol{\eta}^0+\textbf{d}_{\boldsymbol{\theta}}\circ \boldsymbol{\epsilon}^0$. 

The Fisher’s identity is given by 
\begin{equation}
    \label{eq:fishidentiy}
    \nabla_{\boldsymbol{\theta}}\text{log}~p(\textbf{O}\mid \boldsymbol{\theta})=\int  \nabla_{\boldsymbol{\theta}}\left[\text{log}~ (p(\textbf{O}\mid \textbf{y}_u, \boldsymbol{\theta})p(\textbf{y}_u \mid \boldsymbol{\theta}))\right] p(\textbf{y}_u \mid \textbf{O}, \boldsymbol{\theta})d\textbf{y}_u,
\end{equation}
\noindent see, for example,~\cite{poyiadjis2011particle}.

\noindent Substituting this expression into Equation~\eqref{eq:ELBO.grad.lamda.aug_1}, and writing $E_{f_{\boldsymbol{\delta}}}(\cdot)$ for expectation with respect to $f_{\boldsymbol{\delta}}(\boldsymbol{\delta})=f_{\boldsymbol{\delta}^0}(\boldsymbol{\delta}^0)p(\textbf{y}_u \mid \boldsymbol{\theta},\textbf{O})$ and because $h(\boldsymbol{\theta},\textbf{y}_u)=p(\textbf{O} \mid \textbf{y}_u,\boldsymbol{\theta})p(\textbf{y}_u\mid \boldsymbol{\theta})p(\boldsymbol{\theta})$, we get
\begin{equation}
    \label{eq:ELBO.grad.lamda.aug_2}
\begin{split}
                        \nabla_{\boldsymbol{\lambda}}\mathcal{L}(\boldsymbol{\lambda})&=E_{f_{\boldsymbol{\delta}}}\left [\frac{dt^0(\boldsymbol{\delta}^0,\boldsymbol{\lambda}_{\boldsymbol{\theta}})^\top}{d\boldsymbol{\lambda_{\boldsymbol{\theta}}}}(\nabla_{\boldsymbol{\theta}}\text{log}~p(\boldsymbol{\theta})+\nabla_{\boldsymbol{\theta}}\text{log}~p(\boldsymbol{\textbf{y}_u\mid \theta})+\nabla_{\boldsymbol{\theta}}\text{log}~p(\textbf{O} \mid \textbf{y}_u, \boldsymbol{\theta})-\nabla_{\boldsymbol{\theta}}\text{log}~  q_{\boldsymbol{\lambda}}^0(\boldsymbol{\theta}))\right]\\
                        &=E_{f_{\boldsymbol{\delta}}}\left[\frac{dt^0(\boldsymbol{\delta}^0,\boldsymbol{\lambda}_{\boldsymbol{\theta}})^\top}{d\boldsymbol{\lambda_{\boldsymbol{\theta}}}}(\nabla_{\boldsymbol{\theta}}\text{log}~h(\boldsymbol{\theta},\textbf{y}_u)-\nabla_{\boldsymbol{\theta}}\text{log}~  q_{\boldsymbol{\lambda}}^0(\boldsymbol{\theta}) )\right].\\                 
\end{split}
\end{equation}
The term $\frac{dt^0(\boldsymbol{\delta}^0,\boldsymbol{\lambda}_{\boldsymbol{\theta}})}{d\boldsymbol{\lambda_{\boldsymbol{\theta}}}}$ in Equation~\eqref{eq:ELBO.grad.lamda.aug_2} is the derivative of the transformation $t^0(\boldsymbol{\delta}^0,\boldsymbol{\lambda}_{\boldsymbol{\theta}})=\boldsymbol{\mu}_{\boldsymbol{\theta}}+\textbf{B}_{\boldsymbol{\theta}}\boldsymbol{\eta}^0+\textbf{d}_{\boldsymbol{\theta}}\circ \boldsymbol{\epsilon}^0$ with respect to the variational parameters $\boldsymbol{\lambda}_{\boldsymbol{\theta}}=(\boldsymbol{\mu}_{\boldsymbol{\theta}}^\top,\text{vech}(\textbf{B}_{\boldsymbol{\theta}})^\top,\textbf{d}_{\boldsymbol{\theta}}^\top)^\top$. We can express that $t^0(\boldsymbol{\delta}^0,\boldsymbol{\lambda}_{\boldsymbol{\theta}})=\boldsymbol{\mu}_{\boldsymbol{\theta}}+(\boldsymbol{\eta}^0 \otimes \textbf{I}_{S})\text{vech}(\textbf{B}_{\boldsymbol{\theta}})+\textbf{d}_{\boldsymbol{\theta}}\circ \boldsymbol{\epsilon}^0$, where $\textbf{I}_{S}$ is the identity matrix of size $S$, and it can be further shown that  $\nabla_{\boldsymbol{\theta}}\text{log}~  q^0_{\boldsymbol{\lambda}}(\boldsymbol{\theta})=-(\textbf{B}_{\boldsymbol{\theta}}\textbf{B}_{\boldsymbol{\theta}}^\top+\textbf{D}_{\boldsymbol{\theta}}^2)^{-1}(\boldsymbol{\theta}-\boldsymbol{\mu}_{\boldsymbol{\theta}})$, 

\begin{equation}
    \label{eq:du_by_dmu_aug}
    \frac{dt^0(\boldsymbol{\delta}^0,\boldsymbol{\lambda}_{\boldsymbol{\theta}})}{d\boldsymbol{\mu}_{\boldsymbol{\theta}}}=\textbf{I}_S\\~~~~~\text{and}~~~~~
    \frac{dt^0(\boldsymbol{\delta}^0,\boldsymbol{\lambda}_{\boldsymbol{\theta}})}{d\text{vech}(\textbf{B}_{\boldsymbol{\theta}})}={\boldsymbol{\eta}^0}^\top\otimes\textbf{I}_S.
\end{equation}

The derivatives of the lower bound with respect to variational parameters are:
    \begin{equation}
  \label{eq:grad_wrt_mu_vb2}
    \begin{split}
            \nabla_{\boldsymbol{\mu}_{\boldsymbol{\theta}}}\mathcal{L}(\boldsymbol{\lambda})&=E_{f_{\boldsymbol{\delta}}}(\nabla_{\boldsymbol{\theta}}~\text{log}~h(\boldsymbol{\mu}_{\boldsymbol{\theta}}+\textbf{B}_{\boldsymbol{\theta}}\boldsymbol{\eta}^0+\textbf{d}_{\boldsymbol{\theta}}\circ\boldsymbol{\epsilon}^0,\textbf{y}_u)\\
            &+(\textbf{B}_{\boldsymbol{\theta}}\textbf{B}_{\boldsymbol{\theta}}^\top+\textbf{D}_{\boldsymbol{\theta}}^2)^{-1}(\textbf{B}_{\boldsymbol{\theta}}\boldsymbol{\eta}^0+\textbf{d}_{\boldsymbol{\theta}}\circ\boldsymbol{\epsilon}^0)),
    \end{split}
\end{equation}
\begin{equation}
 \label{eq:grad_wrt_B_vb2}
    \begin{split}
            \nabla_{\textbf{B}_{\boldsymbol{\theta}}}\mathcal{L}(\boldsymbol{\lambda})&=E_{f_{\boldsymbol{\delta}}}(\nabla_{\boldsymbol{\theta}}~\text{log}~h(\boldsymbol{\mu}_{\boldsymbol{\theta}}+\textbf{B}_{\boldsymbol{\theta}}\boldsymbol{\eta}^0+\textbf{d}_{\boldsymbol{\theta}}\circ\boldsymbol{\epsilon}^0,\textbf{y}_u){\boldsymbol{\eta}^0}^\top\\
            &+(\textbf{B}_{\boldsymbol{\theta}}\textbf{B}_{\boldsymbol{\theta}}^\top+\textbf{D}_{\boldsymbol{\theta}}^2)^{-1}(\textbf{B}_{\boldsymbol{\theta}}\boldsymbol{\eta}^0+\textbf{d}_{\boldsymbol{\theta}}\circ\boldsymbol{\epsilon}^0){\boldsymbol{\eta}^0}^\top),
    \end{split}
\end{equation}
\begin{equation}
 \label{eq:grad_wrt_d_vb2}
    \begin{split}
    \nabla_{\textbf{d}_{\boldsymbol{\theta}}}\mathcal{L}(\boldsymbol{\lambda})&=E_{f_{\boldsymbol{\delta}}}(\text{diag}(\nabla_{\boldsymbol{\theta}}\text{log}~h(\boldsymbol{\mu}_{\boldsymbol{\theta}}+\textbf{B}_{\boldsymbol{\theta}}\boldsymbol{\eta}+\textbf{d}_{\boldsymbol{\theta}}\circ\boldsymbol{\epsilon}^0,\textbf{y}_u){\boldsymbol{\epsilon}^0}^\top\\
            &+(\textbf{B}_{\boldsymbol{\theta}}\textbf{B}_{\boldsymbol{\theta}}^\top+\textbf{D}^2)^{-1}(\textbf{B}_{\boldsymbol{\theta}}\boldsymbol{\eta}^0+\textbf{d}_{\boldsymbol{\theta}}\circ\boldsymbol{\epsilon}^0){\boldsymbol{\epsilon}^0}^\top)).
    \end{split}
\end{equation}

The analytical expressions for  $\nabla_{\boldsymbol{\theta}}\text{log}~h(\boldsymbol{\mu}_{\boldsymbol{\theta}}+\textbf{B}_{\boldsymbol{\theta}}\boldsymbol{\eta}+\textbf{d}_{\boldsymbol{\theta}}\circ\boldsymbol{\epsilon}^0,\textbf{y}_u)=\nabla_{\boldsymbol{\theta}}\text{log}~h(\boldsymbol{\theta},\textbf{y}_u)$ in Equations~\eqref{eq:grad_wrt_mu_vb2}-\eqref{eq:grad_wrt_d_vb2} under both missing data mechanisms are similar to that of for the JVB method, and can be found in Section~\ref{sec:h_gradients}. 
The expectations in these gradients can be estimated using a single sample $\boldsymbol{\delta}=({\boldsymbol{\delta}^0}^\top,\textbf{y}_u^\top)^\top$ drawn from $f_{\boldsymbol{\delta}^0}$, and $p(\textbf{y}_u \mid \boldsymbol{\theta},\textbf{O})=p(\textbf{y}_u \mid t^0(\boldsymbol{\delta}^0,\boldsymbol{\lambda}_{\boldsymbol{\theta}}),\textbf{O})$.

\subsection{Calculate adaptive learning rates using ADADELTA}
\label{sec:sup:ADADELTA}
The adaptive learning rates (step sizes) for the VB algorithms of the main paper are calculated using the ADADELTA algorithm~\citep{Zeiler2012ADADELTAAA}. The ADADELTA algorithm is now briefly described. Different step sizes are used for each element in variational parameters $\boldsymbol{\lambda}$. 


The update for the $i^{th}$ element of $\boldsymbol{\lambda}$ is
\begin{equation}
    \label{eq:ADADELTA.updating}
\boldsymbol{\lambda}^{(t+1)}_i = \boldsymbol{\lambda}^{(t)}_i + \Delta \boldsymbol{\lambda}^{(t)}_i,
\end{equation}

\noindent where, the step size $\Delta \boldsymbol{\lambda}^{(t)}_i$ is $a_i^{(t)} g_{\boldsymbol{\lambda}i}^{(t)}$. The term $g_{\boldsymbol{\lambda}i}^{(t)}$ denotes the $i^{th}$ component of $\widehat{\nabla_{\boldsymbol{\lambda}}\mathcal{L}(\boldsymbol{\lambda}^{(t)})}$ and $a_i^{(t)}$ is defined as:
\begin{equation}
    \label{eq:a_i_t}
    \mathcal{a}_i^{(t)}=\sqrt{\frac{E\left(\boldsymbol{\Delta}_{\boldsymbol{\lambda}_i}^2\right)^{(t-1)}+\alpha}{E\left(g^2_{\boldsymbol{\lambda}_i}\right)^{(t)}+\alpha}},
\end{equation}

\noindent where $\alpha$ is a small positive constant, $E\left(\boldsymbol{\Delta}_{\boldsymbol{\lambda_ i}}^2\right)^{(t)}$ and $E\left(g^2_{\boldsymbol{\lambda}_i}\right)^{(t)}$  are decayed moving average estimates of  ${\boldsymbol{\Delta}_{\boldsymbol{\lambda_i}}^{(t)}}^2$ and ${g_{\boldsymbol{\lambda}_i}^{(t)}}^2$, defined by
\begin{equation}
    \label{eq:E_Delta2}
    E\left(\boldsymbol{\Delta}_{\boldsymbol{\lambda }_i}^2\right)^{(t)}=\upsilon E\left(\boldsymbol{\Delta}_{\boldsymbol{\lambda }_i}^2\right)^{(t-1)} +(1-\upsilon) {\Delta \boldsymbol{\lambda}^{(t)}_i}^2,
\end{equation}
\noindent and
\begin{equation}
    \label{eq:E_g2}
    E\left(g^2_{\boldsymbol{\lambda}_i}\right)^{(t)}=\upsilon E\left(g^2_{\boldsymbol{\lambda}_i}\right)^{(t-1)} +(1-\upsilon) {g_{\boldsymbol{\lambda}_i}^{(t)}}^2,
\end{equation}

\noindent where the variable $\upsilon$ is a decay constant. We use the default tuning parameter choices $\alpha = 10^{-6}$ and $\upsilon = 0.95$, and initialize $ E\left(\boldsymbol{\Delta}_{\boldsymbol{\lambda }_i}^2\right)^{(0)}= E\left(g^2_{\boldsymbol{\lambda}_i}\right)^{(0)} = 0$.

\section{Prior distributions of model parameters and gradients \texorpdfstring{$\nabla_{\boldsymbol{\theta},\textbf{y}_u}\text{log}~h(\boldsymbol{\theta},\textbf{y}_u)$}{}}
\label{sec:h_gradients}

To map the parameters $\sigma^2_{\textbf{y}}$ and $\rho$ into the real line, we use the following transformations.
\begin{equation}
    \begin{split}
         \gamma & = \text{log}~\sigma^2_{\textbf{y}}\\
        \sigma^2_{\textbf{y}}&=e^{\gamma},
    \end{split}
\end{equation}
\noindent and
\begin{equation}
    \begin{split}
       \lambda &= \text{log}(1+\rho)-\text{log}(1-\rho)\\
        \rho&=\frac{e^{\lambda}-1}{e^{\lambda}+1}.
    \end{split}
\end{equation}
The prior distributions of SEM and missing data model parameters are given in Table~\ref{tab:priors}.

\begin{table}[H]
    \centering
    \begin{tabular}{ccc}
    \toprule
        \makecell{Parameter} & Prior distribution &  Hyperparameters \\
        \hline
         $\boldsymbol{\beta}$& $N(\textbf{0},\sigma^2_{\boldsymbol{\beta}}\textbf{I})$ & $\sigma^2_{\boldsymbol{\beta}}=10,000$ \\
         $\gamma$& $N(0,\sigma^2_{\gamma})$ & $\sigma^2_{\gamma}=10,000$ \\
         $\lambda$& $N(0,\sigma^2_{\lambda})$  & $\sigma^2_{\lambda}=10,000$\\
         $\boldsymbol{\psi}$& $N(\textbf{0},\sigma^2_{\boldsymbol{\psi}}\textbf{I})$ &  $\sigma^2_{\boldsymbol{\psi}}=10,000$\\
         \bottomrule
    \end{tabular}
    \caption{Prior distributions of parameters}
    \label{tab:priors}
\end{table}

\subsection{Derivation of the gradient \texorpdfstring{$\nabla_{\boldsymbol{\theta},\textbf{y}_u}\text{log}~h(\boldsymbol{\theta},\textbf{y}_u)$}{} under MAR}
\label{sec:grad_SEM_MAR}

Under MAR, the vector of parameters $\boldsymbol{\theta}$ contains the fixed effects $\boldsymbol{\beta}$, the variance $\sigma^2_{\textbf{y}}$ and the spatial dependence parameter $\rho$. Further, we also know that $\text{log}~h(\boldsymbol{\theta},\textbf{y}_u)=\text{log}~p(\textbf{y} \mid \boldsymbol{\phi})+ \text{log}~p(\boldsymbol{\phi})$; see Table~\ref{tab:model_yusummary} of the main paper. Note that, for $\sigma^2_{\textbf{y}}$ and $\rho$, we utilise transformed parameters, $ \gamma = \log \sigma^2_{\textbf{y}}$ and $\lambda = \log(1+\rho)-\log(1-\rho)$. This leads to 
\begin{equation} 
\label{eq:h_theta_yu_MAR}
\text{log}~h(\boldsymbol{\theta},\textbf{y}_u) \propto -\frac{n}{2}\gamma+\frac{1}{2}\textrm{log}|\textbf{M}_{\textbf{y}}|-\frac{e^{-\gamma}}{2}\textbf{r}^\top\textbf{M}_{\textbf{y}}\textbf{r}-\frac{\boldsymbol{\beta}^\top\boldsymbol{\beta}}{2 \sigma^2_{\boldsymbol{\beta}}}-\frac{\gamma^2}{2\sigma^2_{\gamma}}-\frac{\lambda^2}{2\sigma^2_{\lambda}},
\end{equation} 

\noindent where $\sigma^2_{\boldsymbol{\beta}}$, $\sigma^2_{\gamma}$, and $\sigma^2_{\lambda}$ are each set to $10,000$, as detailed in Table~\ref{tab:priors}.

The derivative of $\text{log}~h(\boldsymbol{\theta},\textbf{y}_u)$ in Equation~\eqref{eq:h_theta_yu_MAR} with respect to $\boldsymbol{\beta}$  is
\begin{equation}
    \label{eq:SEM.MAR.grad.beta}
     \frac{\partial \text{log} h(\boldsymbol{\theta},\textbf{y}_u)}{\partial \boldsymbol{\beta}}=e^{-\gamma}(\textbf{y}-\textbf{X}\boldsymbol{\beta})^\top\textbf{M}_{\boldsymbol{\textbf{y}}}\textbf{X}-\frac{\boldsymbol{\beta}^\top}{\sigma^2_{\boldsymbol{\beta}}},
\end{equation}

\noindent the derivative of $\text{log}~h(\boldsymbol{\theta},\textbf{y}_u)$ with respect to $\gamma$ is
\begin{equation}
    \label{eq:SEM.MAR.grad.gamma}
     \frac{\partial \text{log} h(\boldsymbol{\theta},\textbf{y}_u)}{\partial \gamma}=-\frac{n}{2}+\frac{e^{-\gamma}}{2}(\textbf{y}-\textbf{X}\boldsymbol{\beta})^\top\textbf{M}_{\boldsymbol{\textbf{y}}}(\textbf{y}-\textbf{X}\boldsymbol{\beta})-\frac{\gamma}{\sigma^2_{\gamma}},
\end{equation}

\noindent the derivative of $\text{log}~h(\boldsymbol{\theta},\textbf{y}_u)$ with respect to $\lambda$ is
\begin{equation}
    \label{eq:SEM.MAR.grad.lambda}
     \frac{\partial \text{log} h(\boldsymbol{\theta},\textbf{y}_u)}{\partial \lambda}=\frac{\partial \text{log}~|\textbf{M}_{\boldsymbol{\textbf{y}}}|}{2 \partial \lambda}-\frac{e^{-\gamma}}{2}(\textbf{y}-\textbf{X}\boldsymbol{\beta})^\top\left(\frac{\partial \textbf{M}_{\boldsymbol{\textbf{y}}}}{\partial \lambda}\right)(\textbf{y}-\textbf{X}\boldsymbol{\beta})-\frac{\lambda}{\sigma^2_{\lambda}},
\end{equation}
\noindent where 
\begin{equation*}
    \begin{split}
            \frac{\partial \textbf{M}_{\boldsymbol{\textbf{y}}}}{\partial \lambda}&=\frac{\partial \textbf{M}_{\boldsymbol{\textbf{y}}}}{\partial \rho}\times \frac{\partial \rho}{\partial \lambda},\\
                   \frac{\partial \textbf{M}_{\boldsymbol{\textbf{y}}}}{\partial \rho} &=-(\textbf{W}^\top+\textbf{W})+2\rho\textbf{W}^\top\textbf{W},\\
            \frac{\partial \rho}{\partial \lambda} &=\frac{2 e^{\lambda}}{(1+e^{\lambda})^2},\\
            \frac{\partial \text{log}~|\textbf{M}_{\boldsymbol{\textbf{y}}}|}{ \partial \lambda}&=\text{tr}\left\{\textbf{M}_{\boldsymbol{\textbf{y}}}^{-1}\left(\frac{\partial \textbf{M}_{\boldsymbol{\textbf{y}}}}{\partial \lambda}\right)\right\}.
    \end{split}
\end{equation*}

Additionally, for the JVB algorithm, we require the derivative of $\text{log}~h(\boldsymbol{\theta},\textbf{y}_u)$ with respect to  $\textbf{y}_u$, and it can be calculated by first calculating the derivative with respect to complete vector $\textbf{y}$, $ \frac{\partial \text{log} h(\boldsymbol{\theta},\textbf{y}_u)}{\partial \textbf{y}}$ 
using

\begin{equation}
    \label{eq:SEM.MAR.grad.y}
     \frac{\partial \text{log} h(\boldsymbol{\theta},\textbf{y}_u)}{\partial \textbf{y}}=-e^{-\gamma}(\textbf{y}-\textbf{X}\boldsymbol{\beta})^\top\textbf{M}_{\boldsymbol{\textbf{y}}},
\end{equation}

\noindent and then we extract the sub-vector, which corresponds to the missing values, $\textbf{y}_u$.

\subsection{Derivation of the gradient \texorpdfstring{$\nabla_{\boldsymbol{\theta},\textbf{y}_u}\text{log}~h(\boldsymbol{\theta},\textbf{y}_u)$}{} under MNAR}
\label{sec:grad_SEM_MNAR}

Under MNAR, the vector of parameters $\boldsymbol{\theta}$ contains the fixed effects of the SEM $\boldsymbol{\beta}$, the variance $\sigma^2_{\textbf{y}}$, the spatial dependence parameter $\rho$ and the fixed effects of the missing data model $\boldsymbol{\psi}$. We also know that $\text{log}~h(\boldsymbol{\theta},\textbf{y}_u)=\text{log}~p(\textbf{y} \mid \boldsymbol{\phi})+ \text{log}~p(\textbf{m}\mid \textbf{y},\boldsymbol{\psi})+ \text{log}~p(\boldsymbol{\phi})+\text{log}~p(\boldsymbol{\psi})$; see Table~\ref{tab:model_yusummary} of the main paper. Note that, for $\sigma^2_{\textbf{y}}$ and $\rho$, we utilise transformed parameters $\gamma$ and $\lambda$, where $ \gamma = \log \sigma^2_{\textbf{y}}$ and $\lambda = \log(1+\rho)-\log(1-\rho)$. This leads to 
\begin{equation} 
\label{eq:h_theta_yu_MNAR}
\begin{split}
    \text{log}~h(\boldsymbol{\theta},\textbf{y}_u) &\propto -\frac{n}{2}\gamma+\frac{1}{2}\textrm{log}|\textbf{M}_{\textbf{y}}|-\frac{e^{-\gamma}}{2}\textbf{r}^\top\textbf{M}_{\textbf{y}}\textbf{r}+\sum_{i=1}^{n}m_i(\textbf{x}^*_i\boldsymbol{\psi}_{\textbf{x}}+{y}_i\boldsymbol{\psi}_{y})\\
    &-\text{log}(1+e^{(\textbf{x}^*_i\boldsymbol{\psi}_{\textbf{x}}+{y}_i\boldsymbol{\psi}_{y})})\\
    &-\frac{\boldsymbol{\beta}^\top\boldsymbol{\beta}}{2 \sigma^2_{\boldsymbol{\beta}}}-\frac{\gamma^2}{2\sigma^2_{\gamma}}-\frac{\lambda^2}{2\sigma^2_{\lambda}}-\frac{\boldsymbol{\psi}^\top\boldsymbol{\psi}}{2 \sigma^2_{\boldsymbol{\psi}}},\\
\end{split}
\end{equation} 
\noindent where $\sigma^2_{\boldsymbol{\beta}}$, $\sigma^2_{\gamma}$, $\sigma^2_{\lambda}$ and $\sigma^2_{\boldsymbol{\psi}}$ are each set to $10,000$, as detailed in Table~\ref{tab:priors}.
The derivatives of $\text{log}~h(\boldsymbol{\theta},\textbf{y}_u)$ in Equation~\eqref{eq:h_theta_yu_MNAR} with respect to $\boldsymbol{\beta}$, $\gamma$ and $\lambda$ are similar to that of under MAR given in Equations~\eqref{eq:SEM.MAR.grad.beta}~\eqref{eq:SEM.MAR.grad.gamma}, and ~\eqref{eq:SEM.MAR.grad.lambda}.
The derivative of $\text{log}~h(\boldsymbol{\theta},\textbf{y}_u)$ with respect to $\boldsymbol{\psi}$ is
\begin{equation}
    \label{eq:SEM.MNAR.grad.psi}
     \frac{\partial \text{log} h(\boldsymbol{\theta},\textbf{y}_u)}{\partial \boldsymbol{\psi}}=\sum_{i=1}^n\left(m_i-\frac{e^{\textbf{x}^*_i\boldsymbol{\psi}_{\textbf{x}}+{y}_i\boldsymbol{\psi}_{y}}}{1+e^{\textbf{x}^*_i\boldsymbol{\psi}_{\textbf{x}}+{y}_i\boldsymbol{\psi}_{y}}}\right)\textbf{z}_i-\frac{\boldsymbol{\psi}^\top}{\sigma^2_{\boldsymbol{\psi}}},
\end{equation}

\noindent where $\textbf{z}_i=({\textbf{x}^*_i}^\top,y_i)$ is the vector containing the $i^{th}$ row vector of matrix $\textbf{X}^*$, and the $i^{th}$ element of the vector $\textbf{y}$, see Section~\ref{sec:missmach} of the main paper.

Similar to MAR case, for the JVB algorithm, we require the derivative of $\text{log}~h(\boldsymbol{\theta},\textbf{y}_u)$ with respect to  $\textbf{y}_u$, and it can be calculated in two steps. First, calculate the derivatives of $\text{log}~p(\textbf{y} \mid \boldsymbol{\phi})$ and $\text{log}~p(\textbf{m}\mid \textbf{y},\boldsymbol{\psi})$ with respect to $\textbf{y}_u$ separately and sum them up.

We first focus on the derivative of $\text{log}~p(\textbf{m}\mid \textbf{y},\boldsymbol{\psi})$ with respect to $\textbf{y}_u$. The derivative with respect to the $i^{th}$ missing value $y_{u_i}$ is
\begin{equation}
    \label{eq:SEM.MNAR.grad.yui}
       \frac{\partial \text{log}~p(\textbf{m}\mid \textbf{y},\boldsymbol{\psi})}{\partial y_{u_i}}=\left (1-\frac{e^{\textbf{z}_i\boldsymbol{\psi}}}{1+e^{\textbf{z}_i\boldsymbol{\psi}}}\right)\psi_\textbf{y}.
\end{equation}

Now, by stacking partial derivatives with respect to the  individual missing values we obtain $ \frac{\partial \text{log}~p(\textbf{m}\mid \textbf{y},\boldsymbol{\psi})}{\partial \textbf{y}_{u}}$ as
\begin{equation}
 \label{eq:SEM.MNAR.grad.yu}
 \frac{\partial \text{log}~p(\textbf{m}\mid \textbf{y},\boldsymbol{\psi})}{\partial \textbf{y}_{u}}=\begin{bmatrix}
   \frac{\partial \text{log}~p(\textbf{m}\mid \textbf{y},\boldsymbol{\psi})}{\partial y_{u_1}}\\
    \frac{\partial \text{log}~p(\textbf{m}\mid \textbf{y},\boldsymbol{\psi})}{\partial y_{u_2}} \\
    \vdots \\
    \frac{\partial \text{log}~p(\textbf{m}\mid \textbf{y},\boldsymbol{\psi})}{\partial y_{u_{n_u}}} \\
\end{bmatrix}.
\end{equation}


For the derivative of $\text{log}~p(\textbf{y} \mid \boldsymbol{\phi})$ with respect to $\textbf{y}_u$, we first calculate the derivative with respect to complete vector $\textbf{y}$,
\begin{equation}
    \label{eq:SEM.MAR.grad.py_by_y}
     \frac{\partial\text{log}~p(\textbf{y} \mid \boldsymbol{\phi})}{\partial \textbf{y}}=-e^{-\gamma}(\textbf{y}-\textbf{X}\boldsymbol{\beta})^\top\textbf{M}_{\boldsymbol{\textbf{y}}},
\end{equation}
\noindent and then we extract the sub-vector, which corresponds to the missing values $\textbf{y}_u$.

Finally, the gradient of $\text{log}h(\boldsymbol{\theta,\textbf{y}_u})$ with respect to $\textbf{y}_u$ is   
\begin{equation}
    \label{eq:SEM.MNAR.grad.yu_final}
     \frac{\partial \text{log} h(\boldsymbol{\theta,\textbf{y}_u})}{\partial \textbf{y}_u}= \frac{\partial \text{log}~p(\textbf{m}\mid \textbf{y},\boldsymbol{\psi})}{\partial \textbf{y}_{u}}+ \frac{\partial\text{log}~p(\textbf{y} \mid \boldsymbol{\phi})}{\partial \textbf{y}_u}.
\end{equation}

\section{Hamiltonian Monte Carlo}
\label{sec:HMC}
We compare the performance of the proposed VB methods with the Hamiltonian Monte Carlo (HMC) method, which was initially introduced by ~\citet{duane1987hybrid}, and was primarily developed for calculations within the field of lattice quantum chromodynamics.~\citet{neal1995bayesian} introduced the HMC methods into applied statistics in the field of Bayesian
neural networks. With
the rise of high-performance software implementations such as Stan~\citep{carpenter2017stan}, the HMC method has now become a pervasive tool across many scientific, medical,
and industrial applications. 

HMC is a method for generating random samples from a desired probability distribution. This approach proves especially useful when obtaining samples directly from the target distribution poses difficulties~\citep{mcelreath2018statistical}. It achieves this by mimicking the dynamics of a system using Hamiltonian dynamics and a numerical integrator, such as the leapfrog integrator.

The main difference between conventional MCMC sampling methods and HMC lies in their proposal mechanisms and exploration strategies. MCMC methods typically make small changes to the current values, which can be inefficient in high-dimensional spaces with complex distributions, such as posterior distributions with many parameters and a large number of missing values. HMC, on the other hand, uses the gradient of the log posterior to simulate the trajectory of parameters and missing values governed by Hamilton equations. This approach enables more efficient exploration of the parameter space and missing value space, particularly in high dimensions, leading to faster convergence and improved sampling efficiency.

Consider the problem of sampling from the joint posterior distribution of the parameter vector $\boldsymbol{\theta}$ and the missing values vector $\textbf{y}_u$ of SEM with missing data. For simplicity of the illustration, let $\boldsymbol{\chi}=(\boldsymbol{\theta}^\top,\textbf{y}_u^\top)^\top$. Let $\textbf{s}$ be an auxiliary parameter vector  with the same
dimensions as  $\boldsymbol{\chi}$. The Hamiltonian's equation, $\mathcal{H}(\boldsymbol{\chi},\textbf{s})$ is a function that combines the potential energy; $\mathcal{U}(\boldsymbol{\chi})$ and the kinetic energy; $\mathcal{K}(\textbf{s})$  of a system through
\begin{equation}
    \label{eq:Ham.H}
    \mathcal{H}(\boldsymbol{\chi}, \textbf{s}) = \mathcal{U}(\boldsymbol{\chi}) + \mathcal{K}(\textbf{s}),
\end{equation}
\noindent with
\begin{equation}
    \label{eq:U.K}
    \mathcal{U}(\boldsymbol{\chi}) = -\text{log}(h(\boldsymbol{\chi})),
\\ ~~~~~
    \mathcal{K}(\textbf{s}) = \frac{1}{2}\textbf{s}^\top\textbf{R}^{-1}\textbf{s},
\end{equation}

\noindent where $h(\boldsymbol{\chi})=h(\boldsymbol{\theta},\textbf{y}_u)$ is given in Table~\ref{tab:model_yusummary} for MAR and MNAR mechanisms, and $\textbf{R}$ is a positive definite mass matrix, usually chosen as the identity matrix.


In the HMC algorithm, the numerical integration of the Hamiltonian equations is performed using the leapfrog integrator. This involves updating the momentum variable $\textbf{s}$ and the position of $\boldsymbol{\chi}$ over a series of $L$ iterations. The three steps of the leapfrog algorithm are: (1) half-step momentum update: 
\begin{equation}
   \boldsymbol{s} = \boldsymbol{s} + \frac{\epsilon}{2} \nabla_{\boldsymbol{\chi}} \mathcal{U}(\boldsymbol{\chi}),
\end{equation}
\noindent (2) full-step Position Update: 
\begin{equation}
    \boldsymbol{\chi} = \boldsymbol{\chi} + \epsilon \mathbf{R}^{-1} \boldsymbol{s},
\end{equation}
\noindent and, (3) half-step momentum update:
\begin{equation}
   \boldsymbol{s} = \boldsymbol{s} + \frac{\epsilon}{2} \nabla_{\boldsymbol{\chi}} \mathcal{U}(\boldsymbol{\chi}),
\end{equation}

\noindent where $\nabla_{\boldsymbol{\chi}} \mathcal{U}(\boldsymbol{\chi})$ is the gradient of the negative log-posterior with respect to $\boldsymbol{\chi}$, see Equation~\eqref{eq:U.K}. The Leapfrog algorithm is shown in Algorithm~\ref{alg:hmc_leapfrog}. The  HMC algorithm for sampling from the joint posterior $\boldsymbol{\theta}$ and $\textbf{y}_u$ is described in Algorithm~\ref{alg:hmc}.

\begin{algorithm}
\caption{The Hamiltonian Monte Carlo (HMC) Algorithm for sampling from joint posterior of $\boldsymbol{\theta}$ and $\textbf{y}_u$ of SEM with missing values}
\label{alg:hmc}
\begin{algorithmic}[1]
\STATE Set number of samples $N$, number of leapfrog steps $L$, step size $\epsilon$
\STATE Initialize starting position $\boldsymbol{\chi}^{(1)}=(\boldsymbol{\theta}^{(1) \top },\textbf{y}_{u}^{(1) \top})^\top$
\FOR{$i = 1$ to $N$}
    \STATE Sample momentum $\textbf{s}^{(i)} \sim N(\textbf{0},\textbf{R})$
        \STATE Compute initial Hamiltonian $H(\boldsymbol{\chi}^{(i)}, \textbf{s}^{(i)}) = \mathcal{U}(\boldsymbol{\chi}^{(i)}) + \frac{1}{2}\textbf{s}^{(i)\top}\textbf{R}^{-1}\textbf{s}^{(i)}$
            \STATE Set $\boldsymbol{\chi}^{*}\leftarrow \boldsymbol{\chi}^{(i)}$ and $\textbf{s}^*\leftarrow \textbf{s}^{(i)}$
        \FOR{$j=1$ to $L$}
    \STATE $ \text{Set}(\boldsymbol{\chi}^{*},\textbf{s}^{*}) \leftarrow \text{Leapfrog}(\boldsymbol{\chi}^{*},\textbf{s}^{*},\epsilon) $. The Leapfrog function is given in Algorithm~\ref{alg:hmc_leapfrog}.
    \ENDFOR
    \STATE Compute final Hamiltonian $H^*(\boldsymbol{\chi}^{*}, \textbf{s}^{*}) = \mathcal{U}(\boldsymbol{\chi}^{*}) + \frac{1}{2}\textbf{s}^{*\top}\textbf{R}^{-1}\textbf{s}^{*}$
    \STATE Accept sample with probability $\min(1, \exp(H^{*} (\boldsymbol{\chi}^{*}, \textbf{s}^{*}) - H(\boldsymbol{\chi}^{(i)}, \textbf{s}^{(i)})))$
    \IF{Accepted}
        \STATE Set $\boldsymbol{\chi}^{(i+1)} \leftarrow \boldsymbol{\chi}^{*}$
    \ELSE
        \STATE Set $\boldsymbol{\chi}^{(i+1)} \leftarrow \boldsymbol{\chi}^{(i)}$
    \ENDIF
\ENDFOR
\end{algorithmic}
\end{algorithm}

\begin{algorithm}
\caption{The Leapfrog Algorithm}
\label{alg:hmc_leapfrog}
\begin{algorithmic}[1]
    \STATE function Leapfrog($\boldsymbol{\chi}$,$\textbf{s}$,$\epsilon$)
        \STATE Set: $ \boldsymbol{s}^* = \boldsymbol{s} + \frac{\epsilon}{2} \nabla_{\boldsymbol{\chi}} \mathcal{U}(\boldsymbol{\chi})$
        \STATE Set: $    \boldsymbol{\chi}^* = \boldsymbol{\chi} + \epsilon \mathbf{R}^{-1} \boldsymbol{s}^*$
        \STATE Set: $ \boldsymbol{s}^* = \boldsymbol{s}^* + \frac{\epsilon}{2} \nabla_{\boldsymbol{\chi}} \mathcal{U}(\boldsymbol{\chi}^*)$
        \STATE return $\boldsymbol{\chi}^*, \boldsymbol{s}^*$
\end{algorithmic}
\end{algorithm}

We employ the widely-used R package RSTAN~\citep{Stan_Development_Team2020-sz} to perform HMC sampling as described in Algorithm~\ref{alg:hmc} for sampling from the joint posterior of parameters and missing values of the models under consideration. 
In particular, we use the No U-Turn Sampler (NUTS) \citep{hoffman2014no}, which adaptively selects the number of leapfrogs $L$ and the step size $\epsilon$. 
We apply the same prior distributions for the model parameters as in the two VB algorithms.


\section{Simulation study with small missing value percentages}
\label{sec:sup.sim}

In Section~\ref{sec:SimulationStudy} of the main paper, we present the results of simulation studies conducted with a large percentage of missing data (approximately $75\%$). The results presented here relate to a scenario using a small percentage of missing values, specifically $n=625$ with $25\%$ missing data.

\subsection{Simulation study under MAR}
\label{sec:sup.sim.MAR}

\begin{figure}[H]
    \centering
    \includegraphics[width=0.9\textwidth, keepaspectratio]{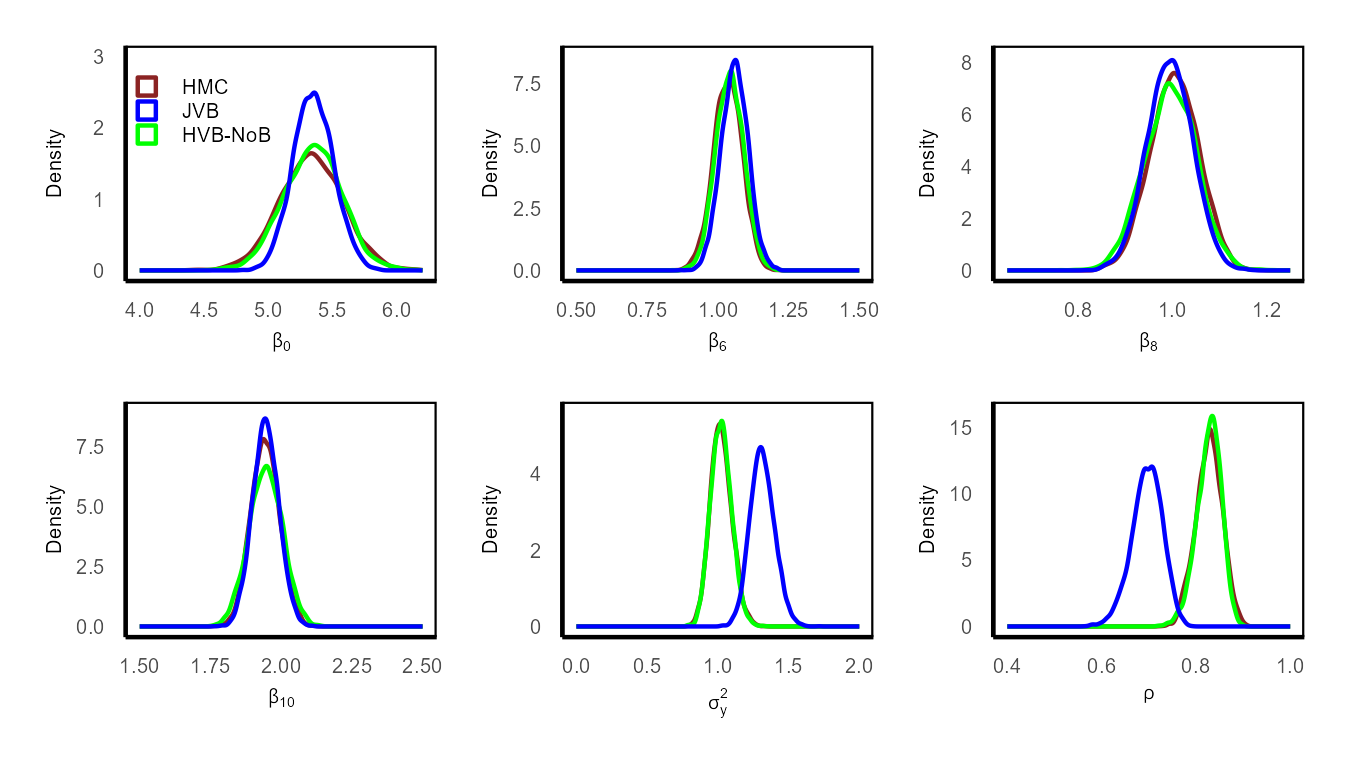}
    \caption{Posterior densities of SEM parameters under MAR for the simulated dataset with $n=625$ and $25\%$ missing values ($n_u=156$) estimated using the HMC, JVB and HVB-NoB methods}
    \label{fig:kernal_SEM_MAR_625_miss_25}
\end{figure}

\begin{figure}[H]
    \centering
    \begin{subfigure}{0.45\textwidth}
        \includegraphics[width=\linewidth]{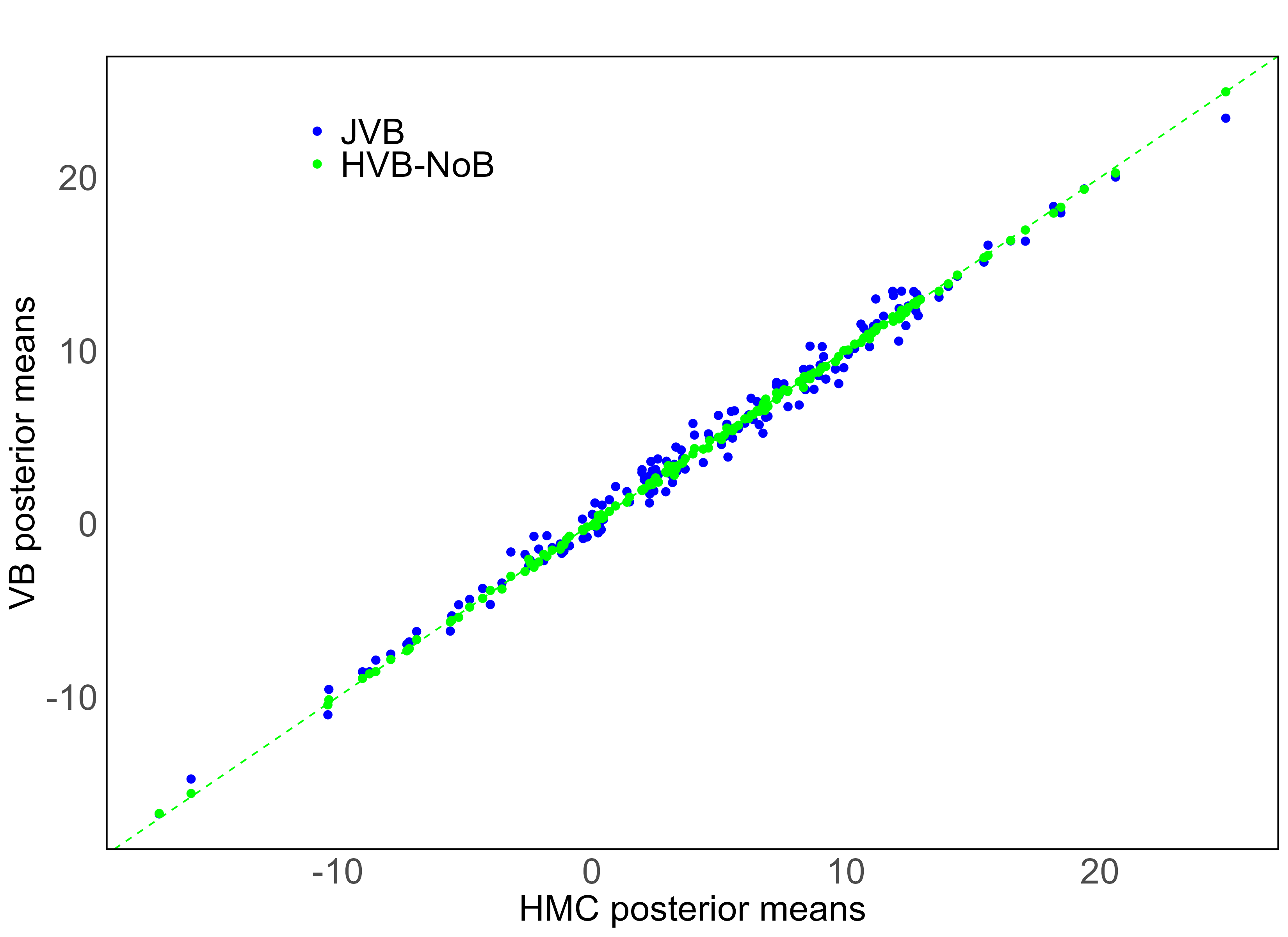}
        \caption{Posterior means}
    \end{subfigure}
    \hfill
    \begin{subfigure}{0.45\textwidth}
        \includegraphics[width=\linewidth]{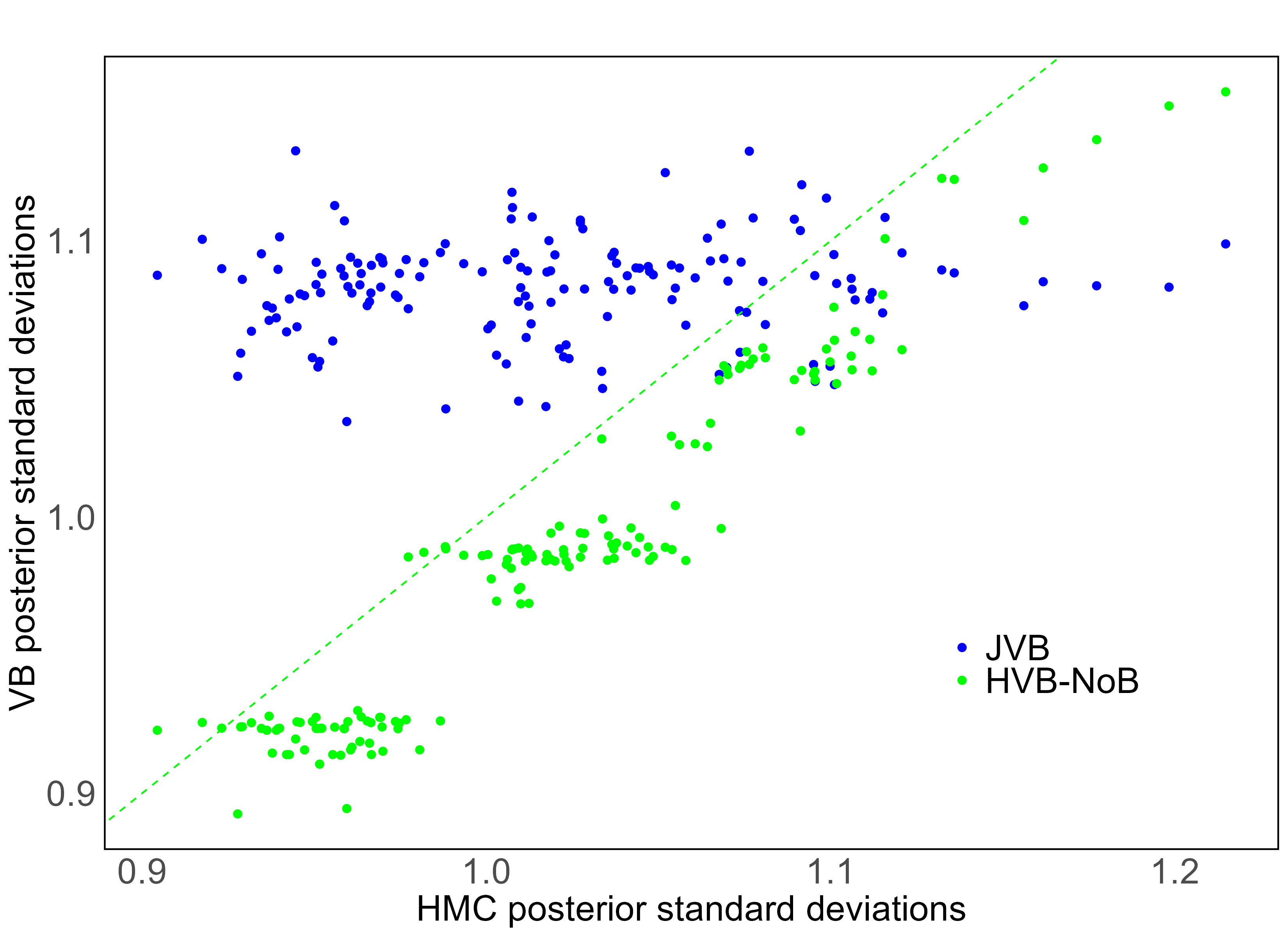}
        \caption{Posterior standard deviations}
    \end{subfigure}
    
    \caption{
    Comparison of the posterior means and standard deviation of  missing
values obtained using the JVB and HVB-NoB methods to those obtained using the HMC method for SEM under MAR with
$n = 625$ and $25\%$ missing values ($n_u = 156$)}
    \label{fig:MAR_625_25p_vb_vs_HMC_mean_sd}
\end{figure}

Under MAR, with $n=625$ and $25\%$ missing values ($n_u=156$), the posterior distributions of SEM parameters obtained from the HVB-NoB algorithm are closer to the posterior distributions obtained from the HMC method than those from the JVB algorithm, as shown in Figure~\ref{fig:kernal_SEM_MAR_625_miss_25}. The posterior means of missing values obtained using the HMC, JVB, and HVB-NoB methods are similar. However, the posterior standard deviations of missing values obtained from the JVB method are different from those obtained using the HMC method; see Figure~\ref{fig:MAR_625_25p_vb_vs_HMC_mean_sd} for further details. These findings are very similar to the simulation results conducted with $n=625$ and $75\%$ missing values under MAR, presented in Section~\ref{sec:SEM.MAR.estimates} of the main paper.

\begin{table}
\centering
\small

\begin{tabular}{rrrrrrrrrr}
\toprule
\makecell{$n$} & \multicolumn{3}{c}{\centering $n=625$} & \multicolumn{3}{c}{$n=1024$} & \multicolumn{3}{c}{$n=2500$} \\ \midrule
                           \makecell{missing \\ percentage}& \multicolumn{1}{c}{$25\%$}           & \multicolumn{1}{c}{$50\%$}  & \multicolumn{1}{c}{$75\%$}       & \multicolumn{1}{c}{$25\%$}        & \multicolumn{1}{c}{$50\%$}  & \multicolumn{1}{c}{$75\%$}   & \multicolumn{1}{c}{$25\%$}          & \multicolumn{1}{c}{$50\%$}  & \multicolumn{1}{c}{$75\%$} \\ \midrule
                           \makecell{HMC}                 & 9.74  & 12.19 &  11.39                         &  39.64     & 34.03 &   51.53   &     533.38    & 593.77  &  637.38 \\
\makecell{JVB}               &  0.04 & 0.48  &  0.06                       &  0.07      & 0.08 &   0.10   & 0.38   &   0.38   &  0.46\\
\makecell{HVB-\\NoB}    & 0.04   &  0.05  &                0.06        &   0.06    & 0.10 &  0.17   & 0.35        & 0.79  & 1.84 \\
\makecell{HVB-\\G} & \makecell{-}  & \makecell{-} & \makecell{-} & \makecell{-} & 0.24 & 0.27 & 0.47 & 0.74&  1.00\\
 \midrule

\makecell{$n$} & \multicolumn{3}{c}{\centering $n=5,041$} & \multicolumn{3}{c}{$n=7,569$} & \multicolumn{3}{c}{$n=10,000$} \\ \midrule
                         \makecell{missing \\ percentage}   & \multicolumn{1}{c}{$25\%$}           & \multicolumn{1}{c}{$50\%$}  & \multicolumn{1}{c}{$75\%$}       & \multicolumn{1}{c}{$25\%$}         & \multicolumn{1}{c}{$50\%$}  & \multicolumn{1}{c}{$75\%$}   & \multicolumn{1}{c}{$25\%$}          & \multicolumn{1}{c}{$50\%$}  & \multicolumn{1}{c}{$75\%$} \\ \midrule
\makecell{HMC}                   & \makecell{-}  & \makecell{-} & \makecell{-}                         &  \makecell{-}       & \makecell{-} &  \makecell{-}    &  \makecell{-}       & \makecell{-} &  \makecell{-}\\
\makecell{JVB}     & 1.285  & 1.610 &        2.066                  &    3.519     & 4.18 &  5.09    &   7.95      &  7.93 & 10.46  \\
 \makecell{HVB-\\ NoB}    &  1.63 & 6.49 & 16.39                          &     4.49    & 26.61 & 72.13     & 10.44        & 69.92 &  201.14\\
 \makecell{HVB-\\G} & 1.69&2.66 & 3.20 &4.04 & 5.77 & 6.38& 9.06& 10.88& 12.59\\
\bottomrule
\end{tabular}

\caption{
Average computing time (in seconds) of one iteration of the HMC, JVB, and HVB methods for estimating the SEM under MAR for different $n$ and different missing value ($n_u$) percentages. For HVB-G, the tuning parameters, such as the block size ($k^*$) and the number of draws ($N_1$) of the Gibbs iteration scheme, are set to $N_1=5$ and $k^*=500$, respectively. The HVB-G is not implemented for $n_u<1000$. The HMC is not implemented for $n>5000$}
\label{tab:MAR.times}
\end{table}

Table~\ref{tab:MAR.times} displays the average computing time per iteration (in seconds) for the VB and HMC methods for different $n$ and $n_u$ under the MAR mechanism. 
The HVB-G method is not implemented when $n_u$ is relatively small ($n_u<1,000$). The HMC method is computationally expensive when $n$ is large and is not implemented when $n > 5,000$. 
The HMC method is much more computationally expensive than the VB methods, regardless of the values of $n$ and $n_u$.
Although it cannot accurately capture the posterior distributions of the parameters $\sigma^2_{\textbf{y}}$, $\rho$ and the posterior standard deviation of the missing values (see Figures \ref{fig:kernal_SEM_MAR_625_miss_75} and \ref{fig:MAR_625_75p_vb_vs_HMC_mean_sd} of the main paper, and Figures~\ref{fig:kernal_SEM_MAR_625_miss_25} and \ref{fig:kernal_SEM_MAR_10000_miss_75} of the online supplement), the JVB method is generally the fastest among all the methods.
For smaller values of $n$ and $n_u$, the HVB-NoB algorithm is faster than the HVB-G method. The computing time of HVB-NoB increases rapidly as $n$ and $n_u$ increase, while HVB-G exhibits lower computing time than HVB-NoB, especially for high missing value percentages.

\subsection{Simulation study under MNAR}
\label{sec:sup.sim.MNAR}

\begin{figure}[H]
    \centering
    \includegraphics[width=0.9\textwidth, keepaspectratio]{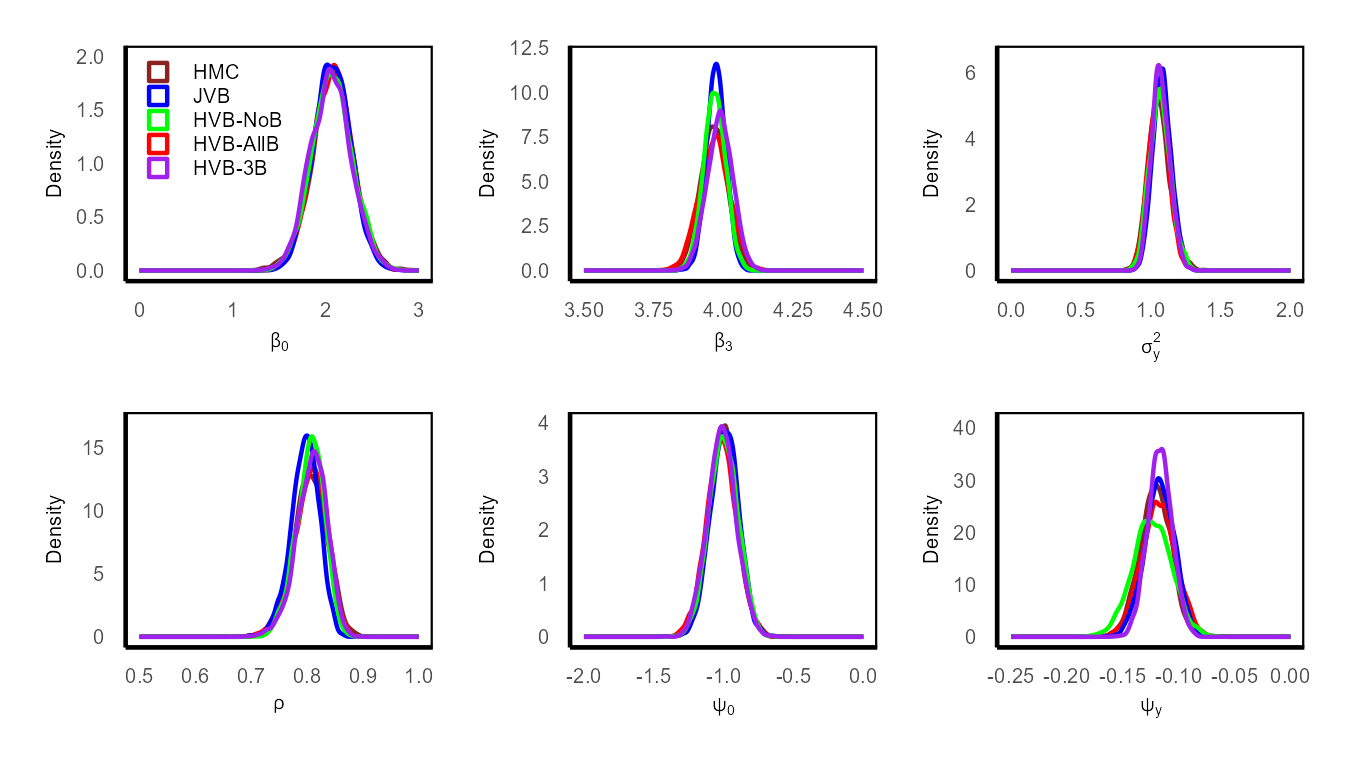}
    \caption{Posterior densities of SEM and missing value model parameters obtained using the HMC, JVB, HVB-NoB, HVB-AllB, and HVB-3B methods, under MNAR
with $n=625$  and $n_u=170$ (around $25\%$ missing) }
    \label{fig:kernal_SEM_MNAR_625_miss_25}
\end{figure}

\begin{figure}[H]
    \centering
    \begin{subfigure}{0.45\textwidth}
        \includegraphics[width=\linewidth]{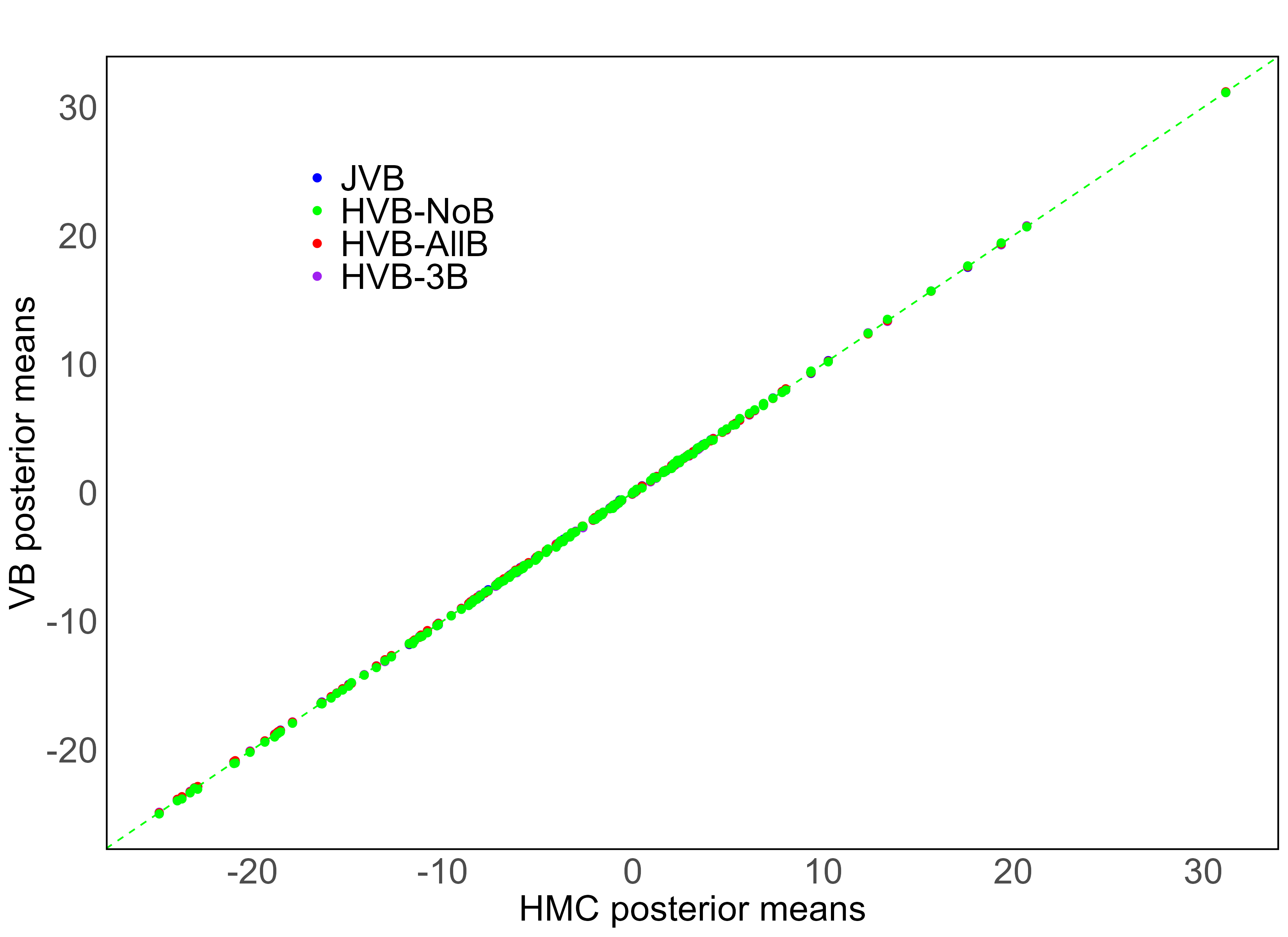}
        \caption{Posterior means}
    \end{subfigure}
    \hfill
    \begin{subfigure}{0.45\textwidth}
        \includegraphics[width=\linewidth]{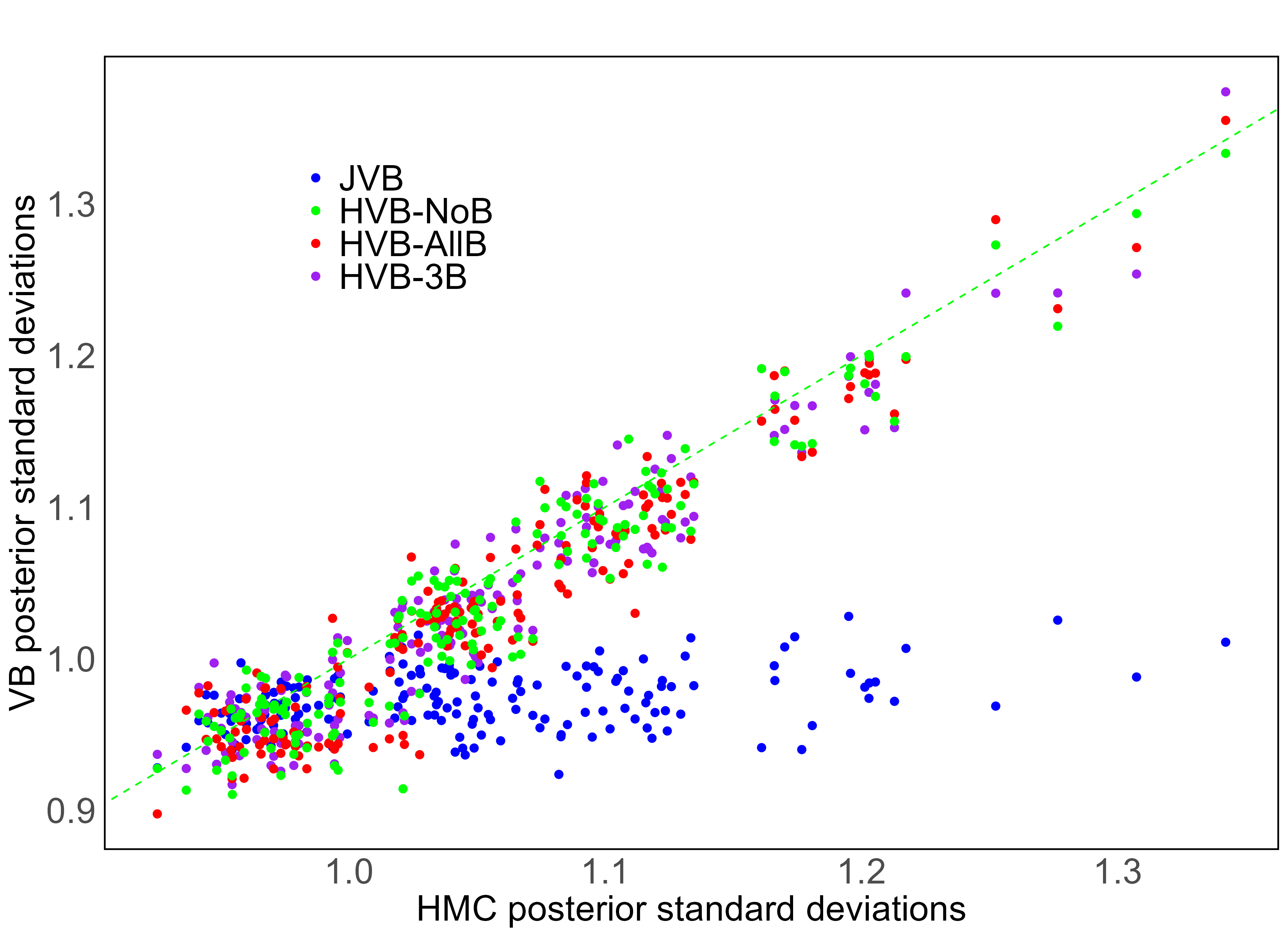}
        \caption{Posterior standard deviations}
    \end{subfigure}
    \caption{
    Comparison of the posterior means and standard deviations of missing values from JVB and HVBs compared to that of HMC under MNAR
with $n=625$ and around $75\%$ missing 
($n_u=170$)}
\label{fig:MNAR_625_25p_vb_vs_HMC_mean_sd}
\end{figure}

Under MNAR, with $n=625$ and $25\%$ missing values percentage, the posterior distributions of the SEM and missing value model parameters obtained from the JVB algorithm and all three HVB algorithms are close to those obtained from the HMC method as shown in Figure~\ref{fig:kernal_SEM_MNAR_625_miss_25}. While the posterior means of missing values are nearly identical across HMC, JVB, and three HVB methods, the posterior standard deviations of missing values obtained using the JVB method are slightly different from those obtained from HMC, as shown in Figure~\ref{fig:MNAR_625_25p_vb_vs_HMC_mean_sd}.

\section{Additional figures from simulation study section of the main paper}

This section provides additional figures related to the simulation study presented in Section~\ref{sec:SimulationStudy} of the main paper.

\subsection{Simulation study under MAR}
\label{sec:sup:addit.sim.MAR}
\begin{figure}[H]
    \centering
    \includegraphics[width=0.9\textwidth, keepaspectratio]{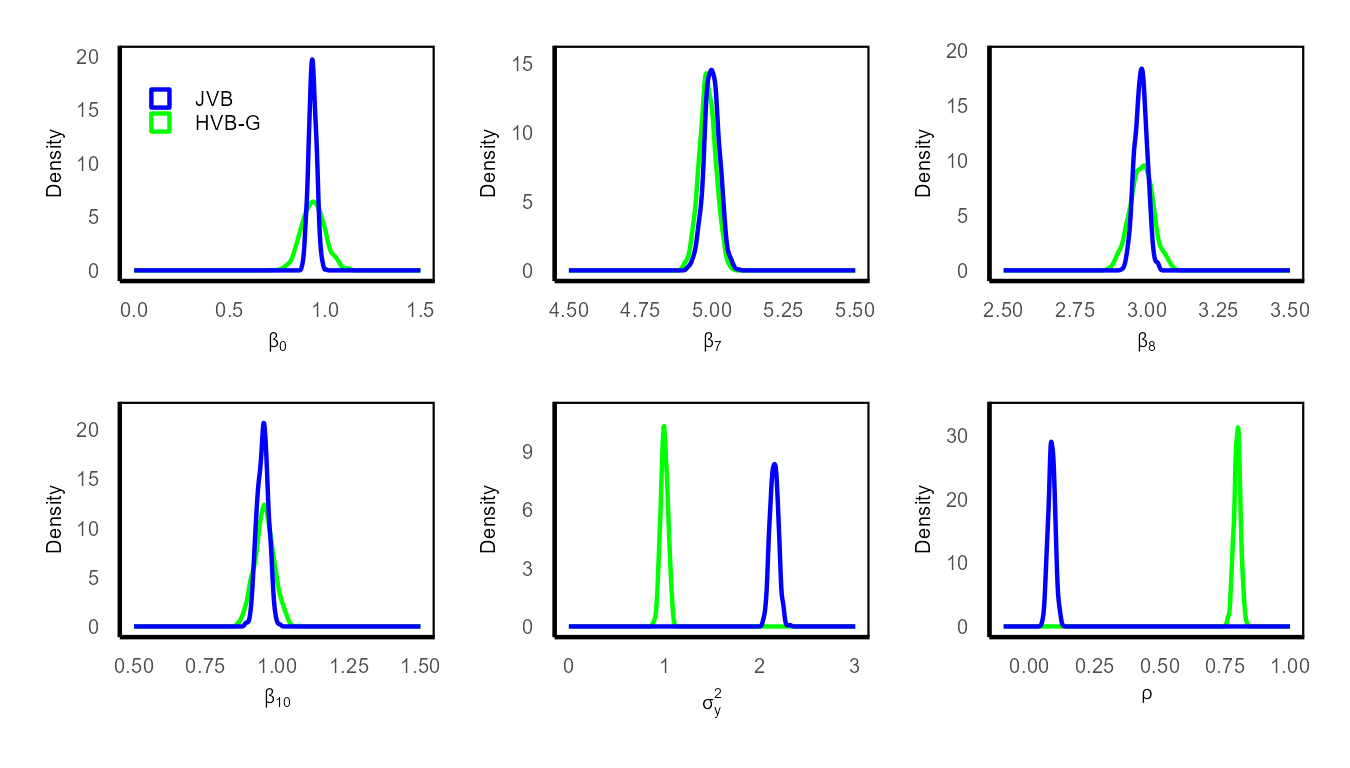}
    \caption{
    Posterior densities of SEM parameters under MAR for the simulated data with $n = 10,000$ and $75\%$
    missing values ($n_u=7,500$) estimated using the JVB and HVB-G methods}
    \label{fig:kernal_SEM_MAR_10000_miss_75}
\end{figure}

\begin{figure}[H]
    \centering
        \includegraphics[width=0.5\textwidth, keepaspectratio]{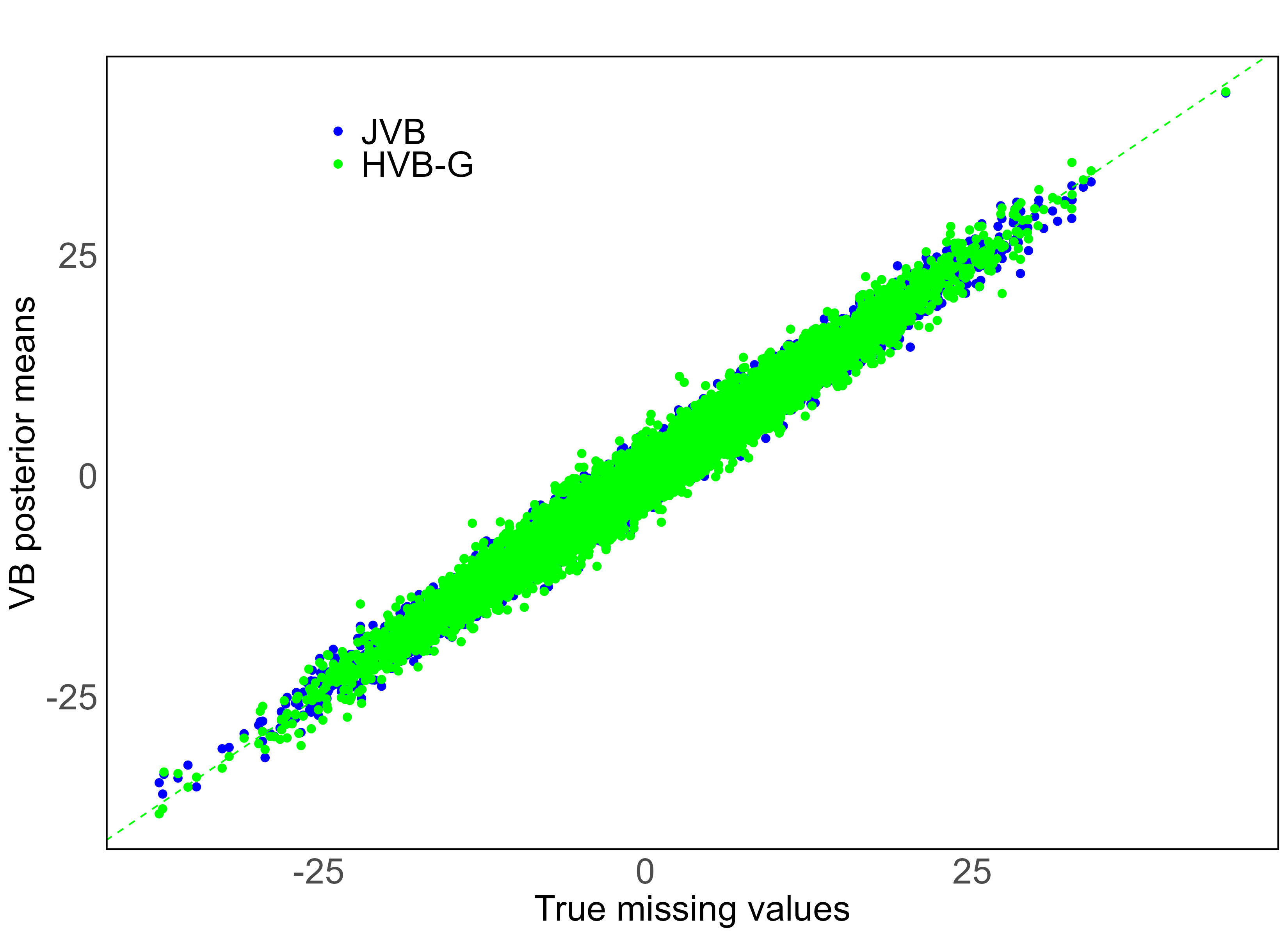}
    \caption{
    Comparison of the posterior means of missing values estimated using the JVB  and HVB-G methods with the true missing values under MAR for the simulated dataset with $n=10,000$ and  $75\%$ missing values ($n_u=7,500$)}
\label{fig:MAR_10000_75p_vbmean_vs_true}
\end{figure}

In Figure~\ref{fig:kernal_SEM_MAR_10000_miss_75}, we compare the posterior densities of SEM parameters obtained using the JVB and HVB-G algorithms for 
$n=10,000$ and $n_u=7,500$ under MAR. The posterior distributions of 
$\rho$ and $\sigma^2_{\textbf{y}}$ obtained from JVB are different from their true values. In contrast, the posterior means from HVB-G align with the true values for all parameters. Figure~\ref{fig:MAR_10000_75p_vbmean_vs_true} compares the posterior means of the missing values estimated by the two VB methods with the true missing values. The posterior means of the missing values obtained from both methods are close to the true values.
\subsection{Simulation study under MNAR}
\label{sec:sup:addit.sim.MNAR}

\begin{figure}[H]
    \centering
    \includegraphics[width=0.9\textwidth, keepaspectratio]{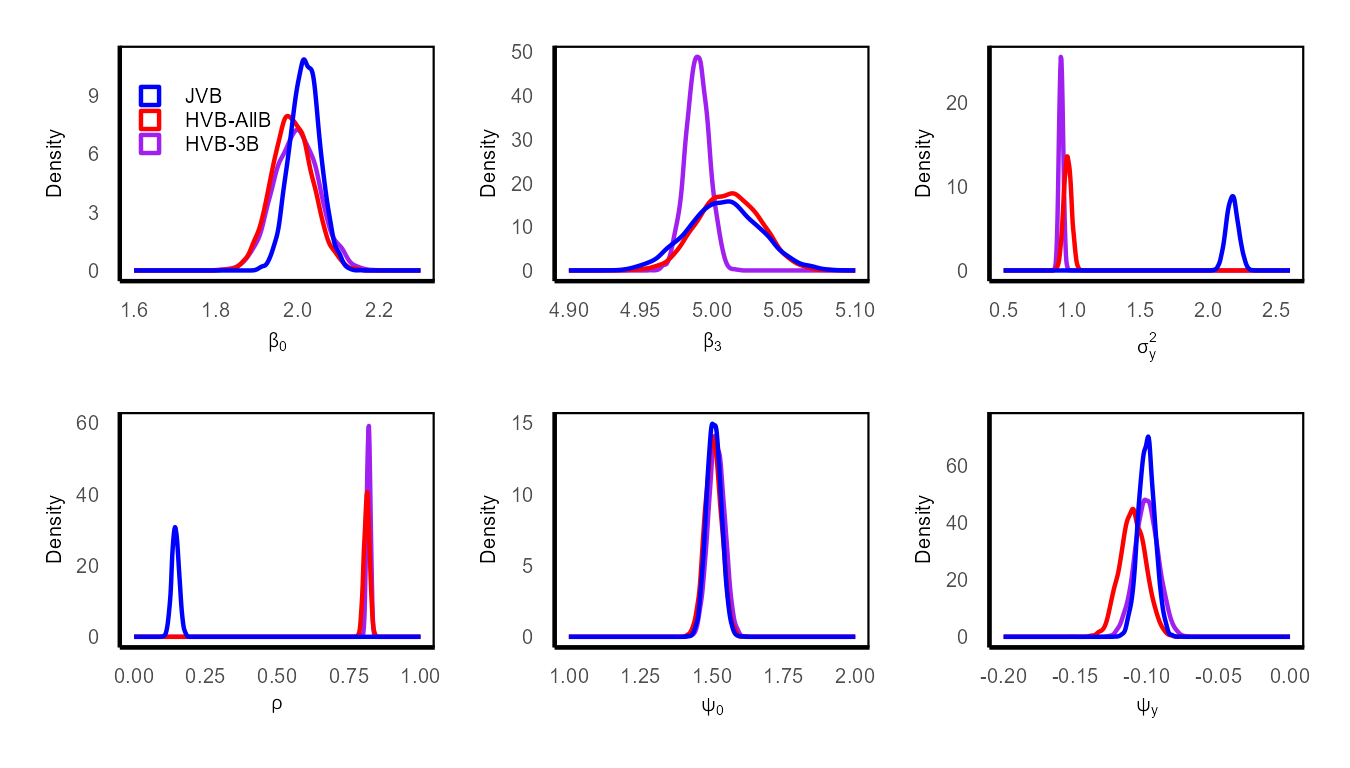}
    \caption{Posterior densities of SEM and missing value model parameters under MNAR with $n=10,000$ and $n_u=7,542$  (around $75\%$ missing) using the JVB and different HVB methods}
    \label{fig:kernal_SEM_MNAR_10000_miss_75}
\end{figure}

In Figure~\ref{fig:kernal_SEM_MNAR_10000_miss_75}, we compare the posterior densities of SEM and missing data model parameters obtained using the JVB, HVB-AllB, and HVB-NoB algorithms for $n=10,000$ and $n_u = 7,542$ under MNAR. The posterior distributions obtained from the HVB-AllB and HVB-3B algorithms are almost identical, except for a slight difference in the posterior distribution of 
$\beta_{0}$. The posterior distributions of $\rho$ and $\sigma^2_{\textbf{y}}$ obtained from JVB are significantly different from their true values. In contrast, the posterior means of all parameters from HVB-AllB and HVB-3B align with the true values.

\begin{figure}[H]
    \centering
        \includegraphics[width=0.5\textwidth, keepaspectratio]{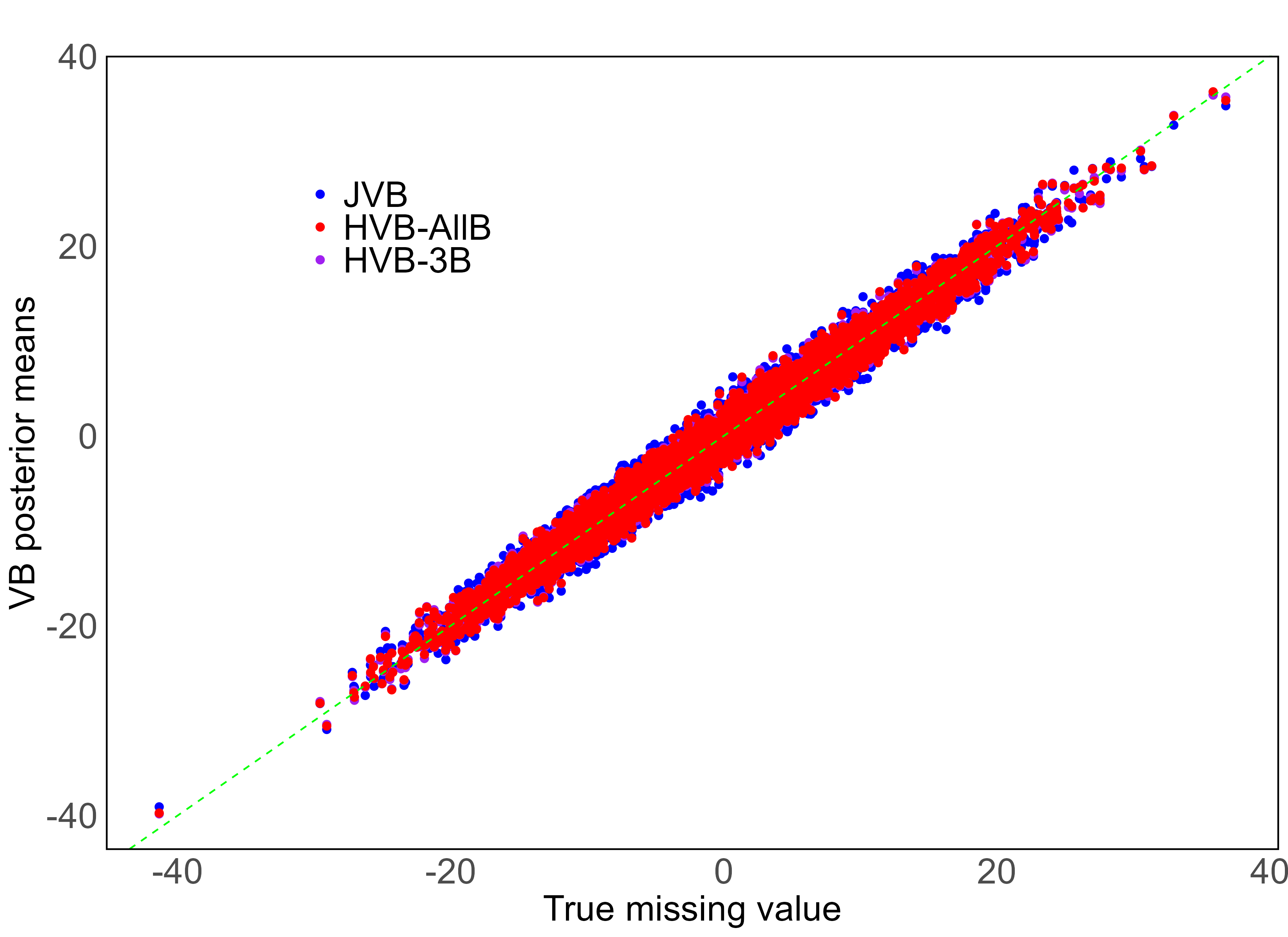}
   
    \caption{
    Comparison of the posterior means of the missing values obtained from JVB, HVB-AllB, and HVB-3B with the true missing values under MNAR with $n=7,542$, and $n_u=10,000$ (around $75\%$ missing)}
\label{fig:MNAR_10000_755p_vb_vs_HMC_mean}
\end{figure}

Figure~\ref{fig:MNAR_10000_755p_vb_vs_HMC_mean} compares the posterior means of the missing values estimated by the three VB methods with the true missing values. The posterior means of the missing values obtained from all methods are close to the true values. However, the posterior means of the missing values from the HVB-AllB and HVB-3B methods are closer to the true values, as they are more concentrated along the diagonal line compared to JVB.

\section{Additional figures and tables from real data section of the main paper}

This section provides additional figures and tables related to the real data application presented in Section~\ref{sec:RealWorldAnalysis} of the main paper.
\label{sec:add:tbl:real}
\subsection{SEM under MAR}   
\label{sec:sup:addit.real.MAR}

\begin{figure}
    \centering
    \includegraphics[width=0.9\textwidth, keepaspectratio]{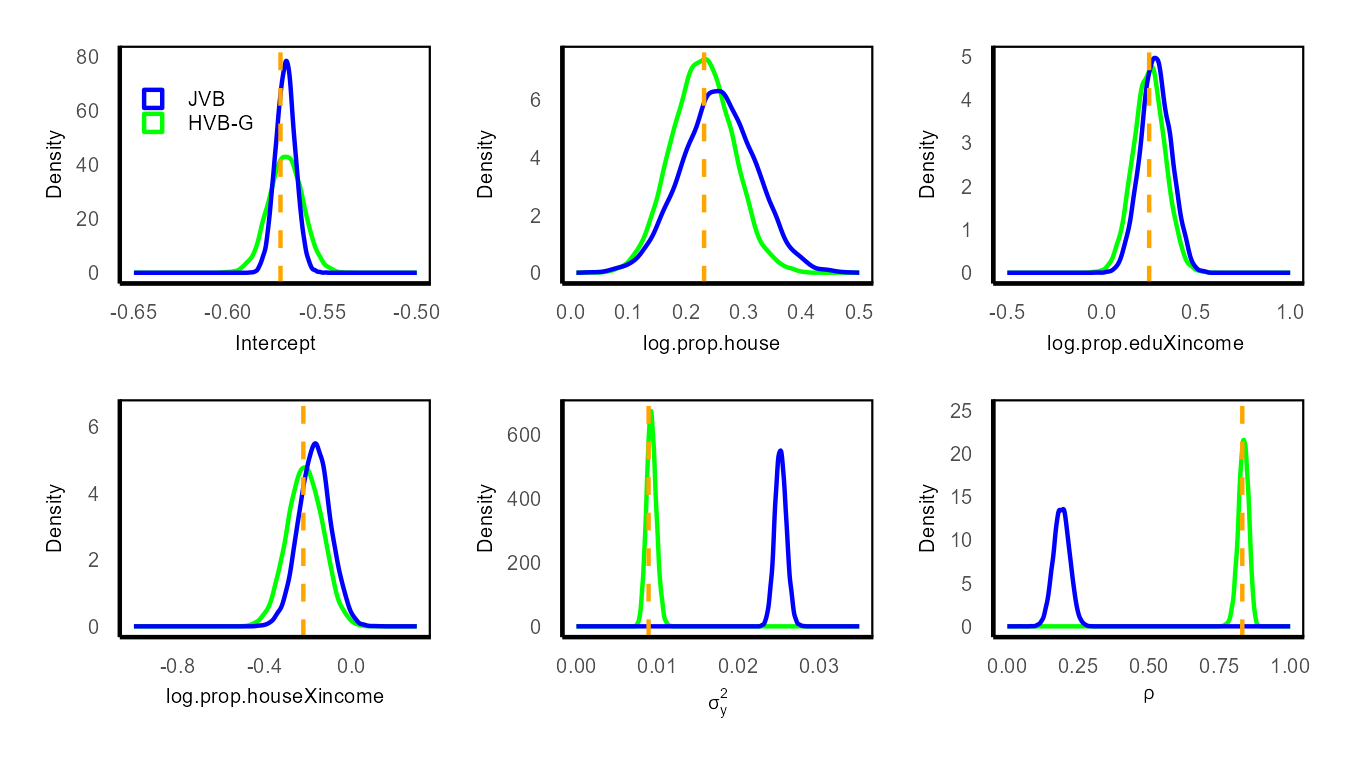}
    \caption{Posterior densities of SEM parameters for 1980 presidential election dataset under MAR using JVB and HVB-G
with $75\%$ missing values ($n_u=2,330$). The vertical lines indicate marginal ML estimates}
    \label{fig:kernal_SEM_MAR_elec}
\end{figure}

Figure~\ref{fig:kernal_SEM_MAR_elec} presents the posterior densities of SEM parameters estimated using the JVB and HVB-G methods with $75\%$ of missing responses ($n_u=2,330$). The vertical lines indicate the marginal ML estimates. The figure shows that the JVB method yields different posterior density estimates for the parameters $\sigma^2_{\textbf{y}}$ and $\rho$ compared to the HVB-G method. The posterior mean estimates obtained using the HVB-G algorithm are closer to the marginal ML estimates than those from the JVB method. 

\begin{figure}
    \centering
        \includegraphics[width=0.5\textwidth, keepaspectratio]{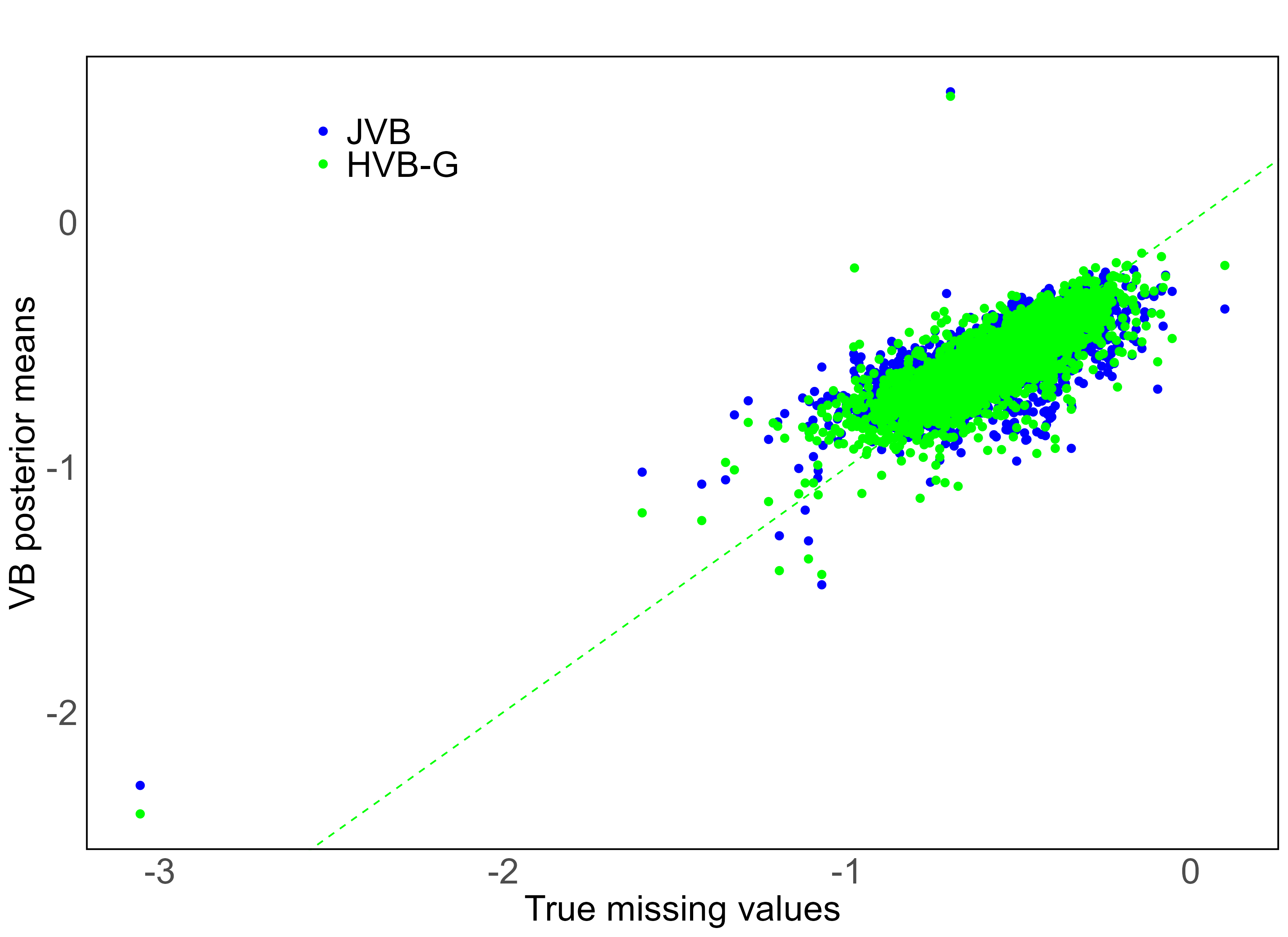}

    \caption{
    Comparison of the posterior means of missing values obtained from the JVB  and HVB-G algorithms with the true missing response values for the 1980 presidential election dataset under MAR with $75\%$ missing values ($n_u=2,330$)}
\label{fig:MAR_elec_vbmean_vs_true}
\end{figure}

Figure~\ref{fig:MAR_elec_vbmean_vs_true} compares the posterior means of missing values obtained from the JVB and HVB-G algorithms with the true missing values. The figure shows that the estimates of missing values from HVB-G are slightly closer to the true missing values than those obtained from the JVB algorithm, see also the MSE values in Table~\ref{tbl:SEM_MAR_elec} of the main paper.

Table~\ref{tbl:SEM_MAR_elec_all} presents the ML estimates with their standard errors and the posterior means with their standard deviations obtained from the JVB and HVB-G methods for SEM parameters for the 1980 presidential election dataset with 75\% missing values under MAR. The posterior means and standard deviations obtained from HVB-G align more closely with the estimates from marginal ML than those from JVB.

\begin{table}
    \centering
    \caption{Marginal ML estimates (with standard errors in brackets), and posterior means (with posterior standard deviations in brackets) of SEM parameters estimated by the JVB and HVB-G algorithms for the 1980 presidential election dataset with $75\%$ ($n_u = 2,330$) missing values under MAR}
    \label{tbl:SEM_MAR_elec_all}
    \begin{tabular}{cccc} 
        \toprule
     & marginal ML& JVB & HVB-G \\
    \hline
    $intercept~(\beta_0)$ & \makecell{ -0.5723\\ (0.01087)}& \makecell{-0.5697\\ (0.0051)} & \makecell{-0.5696 \\ (0.0091)}   \\
             $log.prop.edu~(\beta_1)$ & \makecell{-0.1721\\(0.1246)}  & \makecell{-0.0826\\( 0.0815)} & \makecell{-0.1816\\(0.0964)} \\
                $log.prop.house~(\beta_2)$  & \makecell{0.2312\\(0.0657)}  & \makecell{0.2563\\(  0.0637)} & \makecell{ 0.2280\\( 0.0527)} \\
                        $income~(\beta_3)$   & \makecell{ -0.1233\\(0.0847)}  & \makecell{-0.0946\\(0.0525)}& \makecell{-0.1234\\(0.0669)} \\
                        
        $log.prop.edu\times log.prop.house~(\beta_4)$ & \makecell{-0.1424\\(0.1353)}  & \makecell{ -0.0780\\(0.0804)}  & \makecell{-0.1444\\(0.1059)} \\
                $log.prop.edu \times income~(\beta_5)$ &  \makecell{ 0.2521\\(0.0970)}  & \makecell{0.2870\\(0.0797)}& \makecell{0.2517\\( 0.0843)} \\
                        $log.prop.house \times income~(\beta_6)$   & \makecell{ -0.2213\\(0.1066)}  & \makecell{-0.1690\\(0.0733)} & \makecell{-0.2137\\(0.0818)} \\
                              \makecell{  $log.prop.edu \times log.prop.house$\\ $\times income~(\beta_7)$} & \makecell{0.2076\\(0.1209)}  & \makecell{0.2582\\(0.0989)} & \makecell{0.2203\\(0.1073)} \\
        $\sigma^2_{\textbf{y}}$ & \makecell{0.0089 \\ (0.0006)}& \makecell{0.0253\\ (0.0007)} & \makecell{0.0093 \\ (0.0006)}   \\
        $\rho$  & \makecell{0.8306 \\ (0.0181)}& \makecell{0.1916 \\ (0.0282)} & \makecell{0.8346 \\ (0.0182)}\\
    \bottomrule
    \end{tabular}
\end{table}

\subsection{SEM under MNAR}
\label{sec:sup:addit.real.MNAR}

\begin{figure}[H]
    \centering
    \includegraphics[width=0.9\textwidth, keepaspectratio]{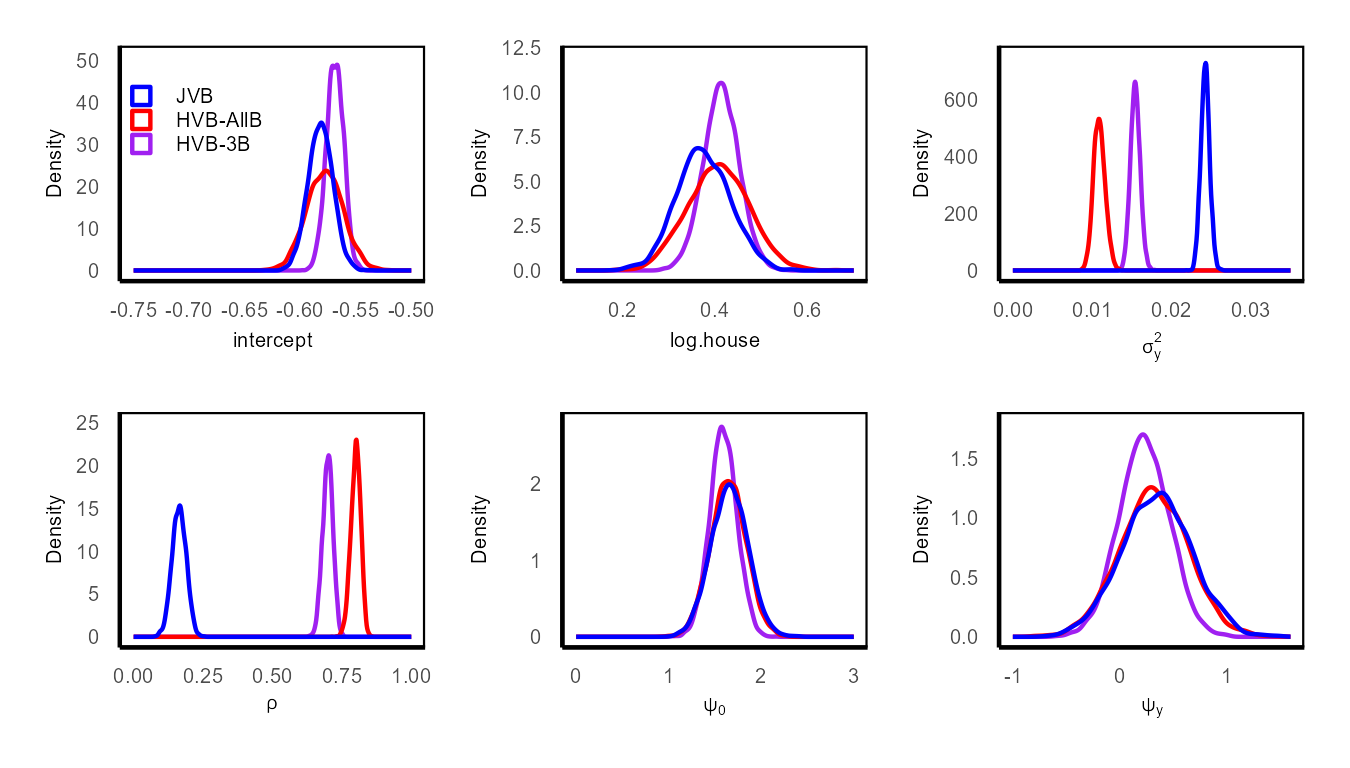}
    \caption{Posterior densities of SEM and missing value model parameters for the 1980 presidential
election dataset under MNAR obtained using the JVB, HVB-AllB, and HVB-3B methods 
with approximately $80\%$ missing values ($n_u=2,477$)}
    \label{fig:kernal_SEM_MNAR_elec}
\end{figure}

Figure~\ref{fig:kernal_SEM_MNAR_elec} compares the posterior densities of SEM and missing data model parameters obtained from the JVB, HVB-AllB, and HVB-3B methods for the 1980 presidential election dataset with around $80\%$ missing values
under MNAR. The figure shows that, except for $\sigma^2_{\textbf{y}}$ and $\rho$, the posterior densities of SEM and missing data model parameters obtained from different algorithms are almost identical. For $\sigma^2_{\textbf{y}}$ and $\rho$, the posterior densities obtained from HVB-AllB and HVB-3B differ from those obtained using JVB.

\begin{figure}
    \centering
        \includegraphics[width=0.5\textwidth, keepaspectratio]{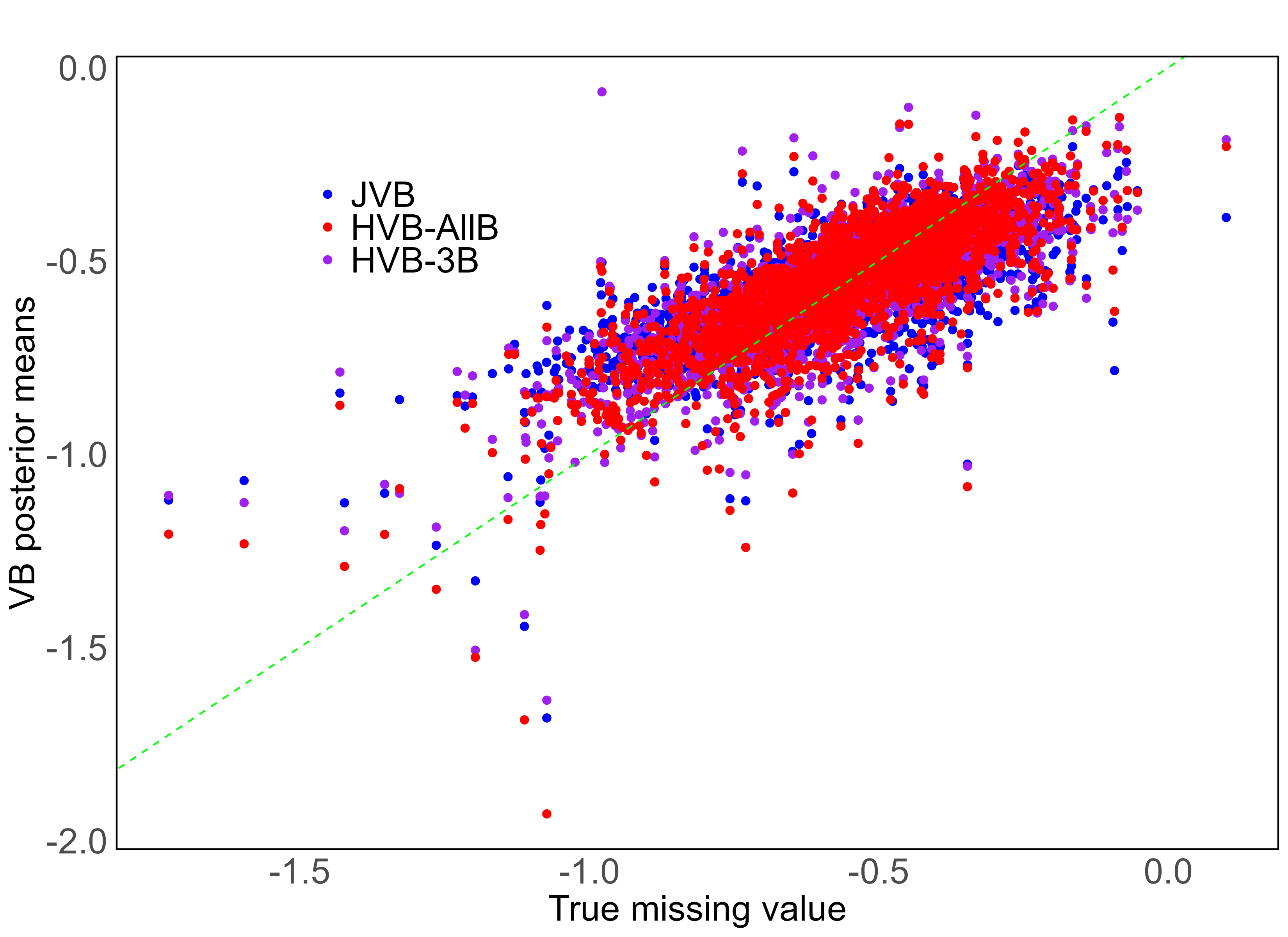}

    \caption{
    Comparison of the posterior means of the missing values obtained from JVB, HVB-AllB, and HVB-3B with the true missing values for the 1980 presidential election dataset under MNAR with $80\%$ missing values ($n_u=2,477$)}
\label{fig:MNAR_elec_vb_mean_vs_yu}
\end{figure}

Figure~\ref{fig:MNAR_elec_vb_mean_vs_yu} compares the posterior means of missing values obtained from the JVB, HVB-AllB, and HVB-3B methods with the true missing values. The estimates from the HVB-AllB method are slightly closer to the true missing values than those from the HVB-3B and JVB methods.

Table~\ref{tbl:SEM_MNAR_elec_all} presents the posterior means and standard deviations of SEM and missing data model parameters obtained from the JVB, HVB-AllB, and HVB-3B algorithms, and the true values for missing data model parameters. {The posterior means and standard deviations from the HVB-AllB and HVB-3B algorithms are very close}. However, the estimates from the JVB algorithm differ, particularly for $\rho$ and $\sigma^2_{\textbf{Y}}$. 


\begin{table}
    \centering
    \caption{Posterior means (with posterior standard deviations inside brackets) of SEM and missing data model parameters, and the true parameter values for the
missing data model parameters under MNAR obtained from the JVB, HVB-AllB, and HVB-3B methods for the 1980
presidential election datase with approximately $80\%$ missing values ($n_u=2477$)}
    \label{tbl:SEM_MNAR_elec_all}
    \begin{tabular}{ccccc} 
        \toprule
     & True value& JVB & HVB-AllB & HVB-3B \\
    \hline
    $intercept~(\beta_0)$ & \makecell{ NA}  & \makecell{-0.5815\\(0.0112)} & \makecell{-0.5773 \\(0.0163)} & \makecell{-0.5689\\(0.0078)} \\
        $log.prop.edu~(\beta_1)$ & \makecell{ NA}  & \makecell{ 0.1345\\(0.0851)} & \makecell{0.1078 \\(0.0975)} & \makecell{ 0.0913\\(0.0539)} \\
                $log.prop.house~(\beta_2)$ & \makecell{ NA}  & \makecell{0.3732\\(0.0583)} & \makecell{0.4094 \\(0.0665)} & \makecell{ 0.4139\\(0.0398)} \\
                        $income~(\beta_3)$ & \makecell{ NA}  & \makecell{ -0.2617\\(0.0629)} & \makecell{  -0.2941 \\( 0.0728)} & \makecell{-0.2684\\(0.0460)} \\
                        
        $log.prop.edu\times log.prop.house~(\beta_4)$ & \makecell{ NA}  & \makecell{0.2177\\(0.0905)} & \makecell{ 0.2362\\(0.0953)} & \makecell{0.2108\\( 0.0738)} \\
                $log.prop.edu \times income~(\beta_5)$ & \makecell{ NA}  & \makecell{ 0.0582\\(0.0499)} & \makecell{0.0312 \\(0.0487)} & \makecell{0.0736\\(0.0395)} \\
                        $log.prop.house \times income~(\beta_6)$ & \makecell{ NA}  & \makecell{-0.4237\\(0.0918)} & \makecell{-0.5056 \\(0.0956)} & \makecell{ -0.4701\\(0.0556)} \\
                              \makecell{  $log.prop.edu \times log.prop.house$\\ $\times income~(\beta_7)$} & \makecell{ NA}  & \makecell{   -0.0607\\(0.0605)} & \makecell{-0.1038 \\(0.0659)} & \makecell{-0.0848\\(0.0457)} \\
        $\sigma^2_{\textbf{y}}$ & \makecell{ NA}  & \makecell{0.0242\\(0.0006)} & \makecell{0.0109 \\(0.0008)} & \makecell{0.0154\\(0.0006)} \\
        $\rho$ & \makecell{ NA}  & \makecell{0.1634\\(0.0259)} & \makecell{0.8004 \\(0.0171)} & \makecell{0.6990\\(0.0186)} \\
        $\psi_0$ & 1.4   & \makecell{1.6622\\(0.1977)} & \makecell{ 1.6504\\(0.1934)} & \makecell{1.5891\\(0.1455)} \\
          $\psi_{\textbf{x}^{*}}$ &  0.5   & \makecell{0.5227\\(0.0519)} & \makecell{0.5172 \\(0.0497)} & \makecell{0.5376\\(0.0454)} \\
            $\psi_{\textbf{y}}$ & -0.1  & \makecell{0.3540\\(0.3252)} & \makecell{ 0.3316\\(0.3269)} & \makecell{0.2286\\(0.2354)} \\
    \bottomrule
    \end{tabular}
\end{table}

\section{Convergence analysis of the VB methods}
\label{sec:sup:convanal}


To evaluate the convergence of the proposed JVB method, we plot the lower bound over iterations. For the HVB algorithms, we analyse trajectories of variational means of the parameters across iterations for the simulation study and the real application.


\subsection{Convergence Analysis for the Simulation Studies}
This subsection provides convergence analysis plots for the simulation studies presented in Section~\ref{sec:SimulationStudy} of the main paper, as well as in Section~\ref{sec:sup.sim} of the online supplement. Generally, HVB algorithms converge more rapidly compared to JVB algorithms.

\subsubsection{Convergence analysis for the simulation studies for the SEM under MAR}

\begin{figure}[H]
    \centering
    \includegraphics[width=0.9\textwidth, keepaspectratio]{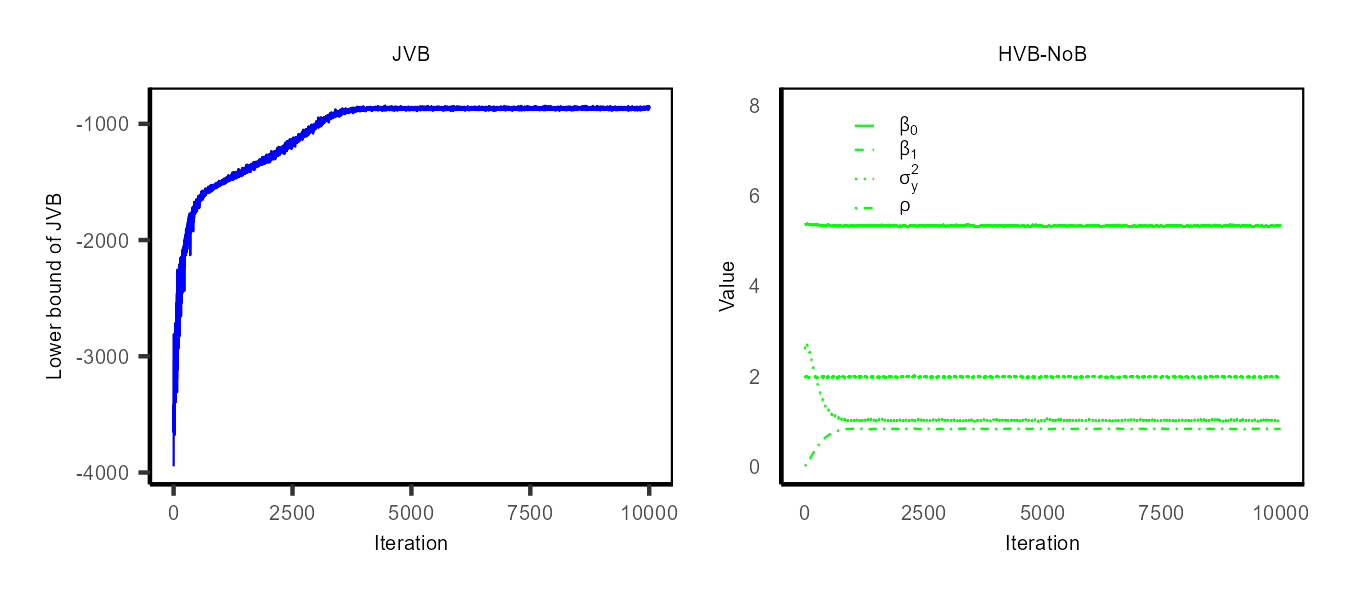}
    \caption{The lower bound for the JVB algorithm (left figure), and the trajectories of variational means of SEM parameters for the HVB-NoB algorithm over iterations (right figure), under MAR for the simulated dataset with $n=625$ and $n_u$= 156 (i.e. missing percentage is 25\%)}
    \label{fig:con_vb_MAR_625_25}
\end{figure}

Figures~\ref{fig:con_vb_MAR_625_25}, \ref{fig:con_vb_MAR_625_75}, and~\ref{fig:con_vb_MAR_10000_75} illustrate lower bounds for the JVB algorithm (the left figure), and the trajectories of variational means of SEM parameters for the HVB algorithm (the right figure) across VB iterations, under MAR, for different values of $n$ and $n_u$. All VB algorithms achieve convergence well before the final iteration. The HVB algorithms consistently reach convergence in fewer iterations than the JVB algorithm.

\begin{figure}[H]
    \centering
    \includegraphics[width=0.9\linewidth]{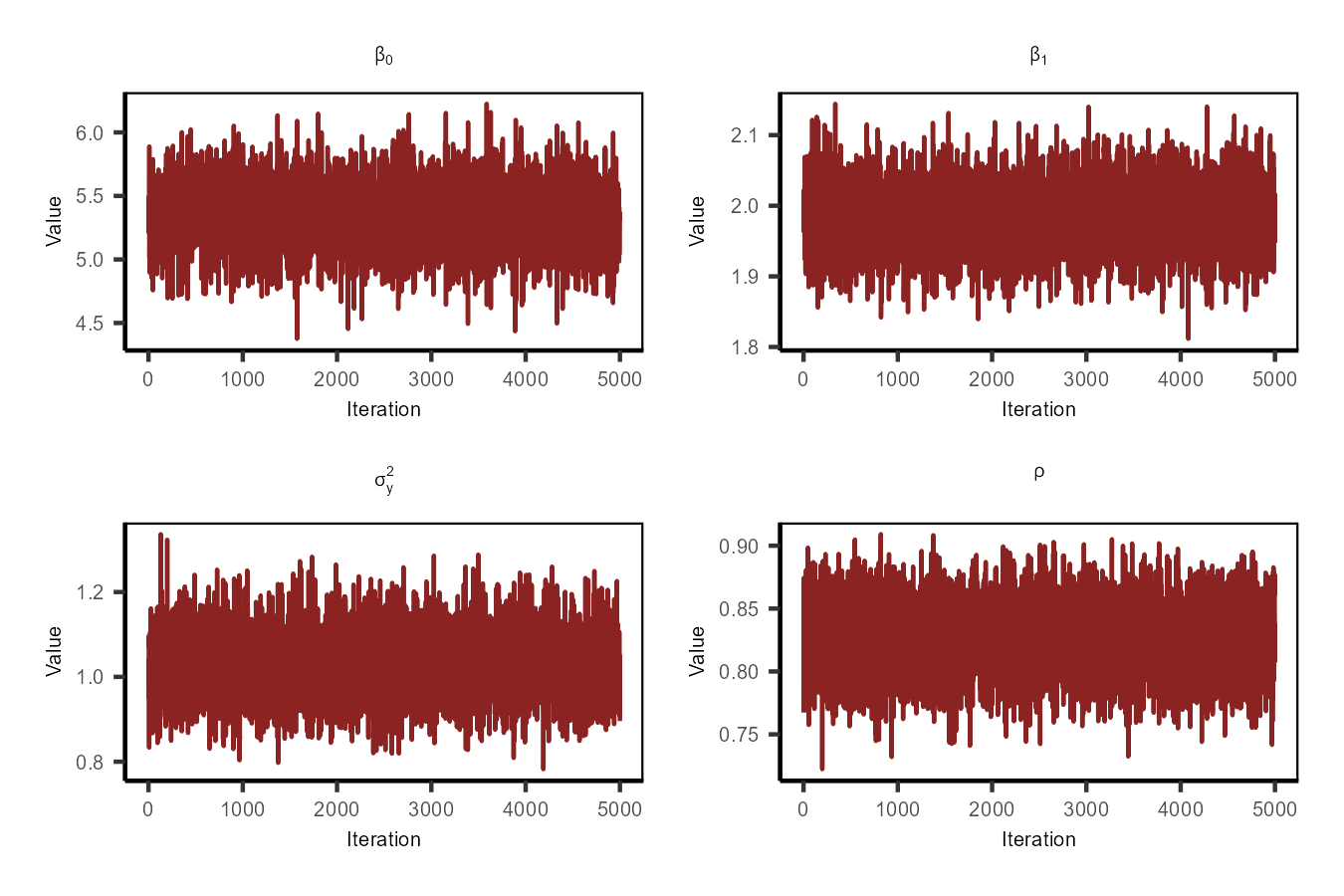}
    \caption{Trace plots of SEM posterior samples of SEM parameters under MAR from the HMC method after excluding burn-in iterations for the simulated dataset with $n=625$ and $n_u=156$ (i.e., the missing percentage is 25\%)}
    \label{fig:HMC_MAR_25_625_TP}
\end{figure}

\begin{figure}[H]
    \centering
    \includegraphics[width=0.9\textwidth, keepaspectratio]{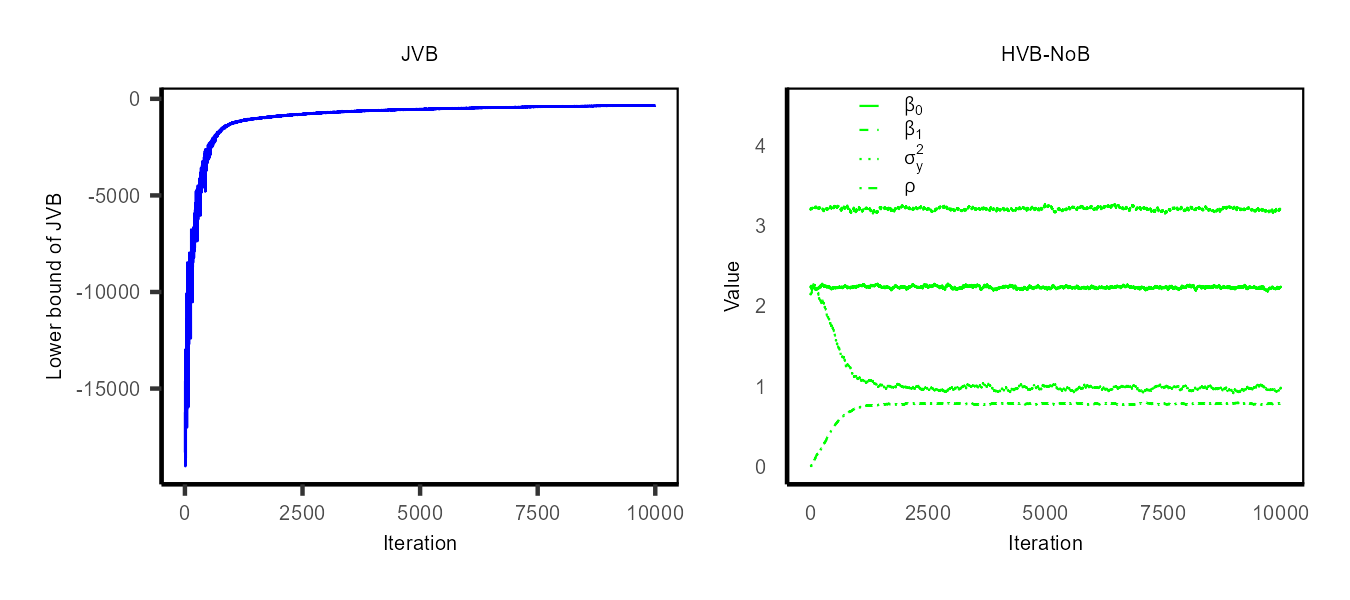}
    \caption{The lower bound for the JVB algorithm (left figure), and the trajectories of variational means of SEM parameters for the HVB-NoB algorithm over iterations (right figure), under MAR for the simulated dataset with $n=625$, $n_u$= 469 (i.e. missing percentage is 75\%)}
    \label{fig:con_vb_MAR_625_75}
\end{figure}

\begin{figure}[H]
    \centering
    \includegraphics[width=0.9\linewidth]{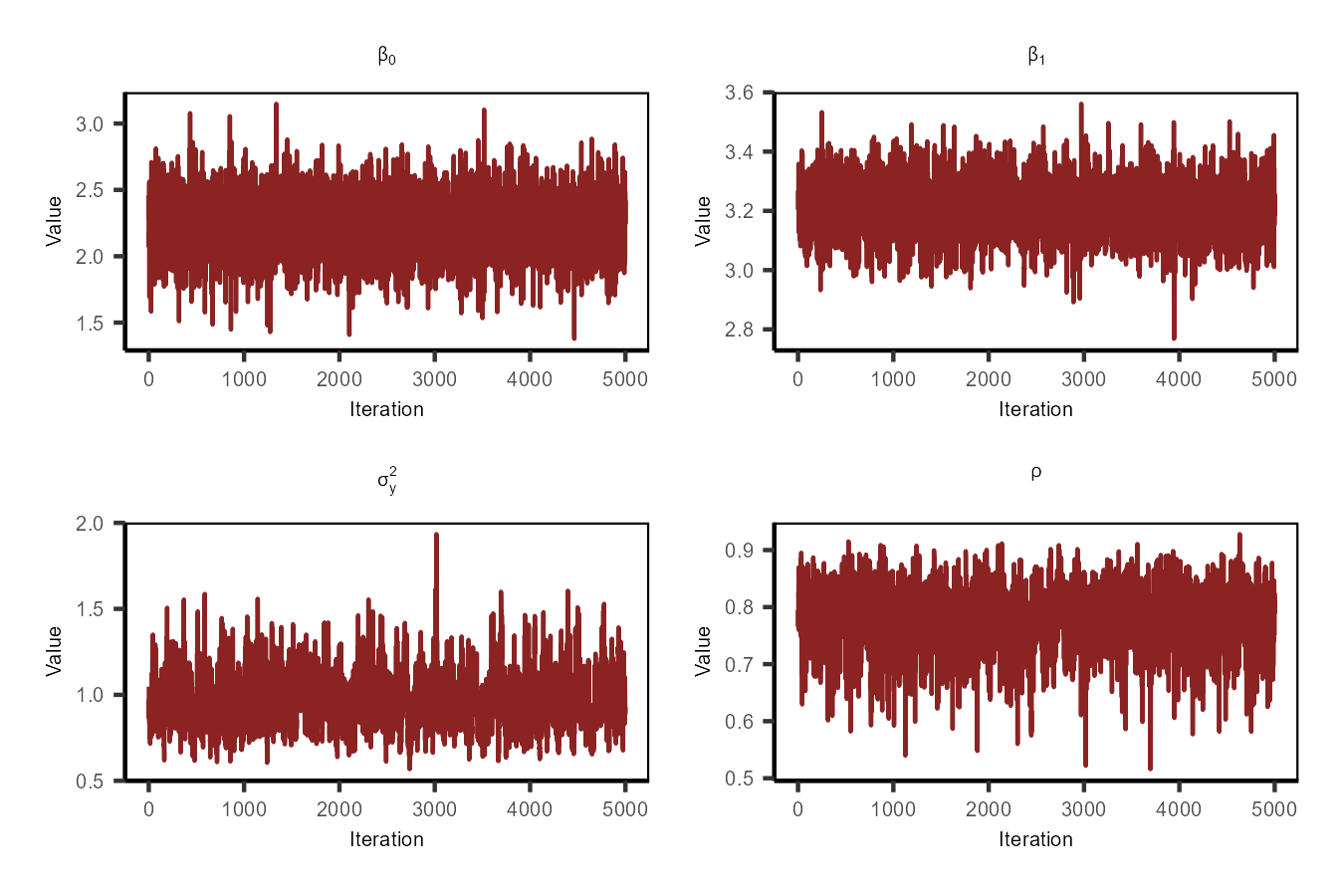}
    \caption{Trace plots of posterior samples of SEM parameters under MAR from the HMC method after excluding burn-in iterations for the simulated dataset with $n=625$ and $n_u=469$ (i.e., the missing percentage is 75\%)}
    \label{fig:HMC_MAR_75_625_TP}
\end{figure}

\begin{figure}[H]
    \centering\includegraphics[width=0.9\textwidth, keepaspectratio]{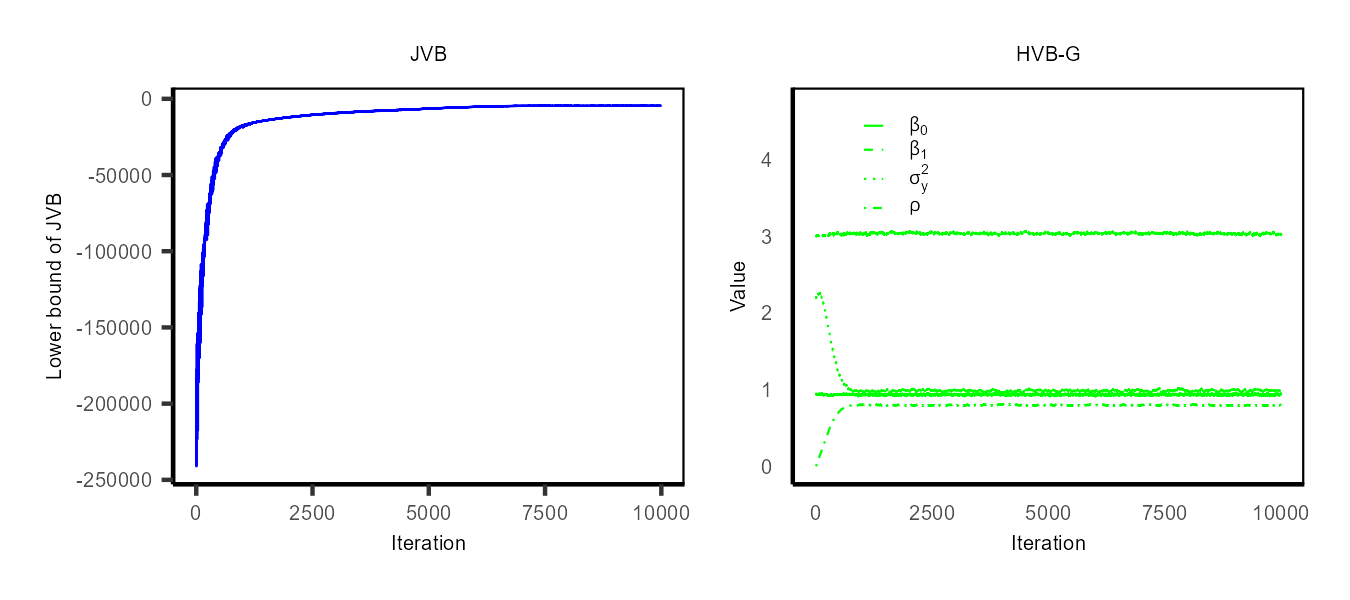}
    \caption{The lower bound for the JVB algorithm (left figure), and the trajectories of variational means of SEM parameters for the HVB-G algorithm over iterations (right figure), under MAR for the simulated dataset with $n=10,000$, $n_u=7,500$ (i.e. missing percentage is 75\%)}
    \label{fig:con_vb_MAR_10000_75}
\end{figure}

Figures~\ref{fig:HMC_MAR_25_625_TP} and \ref{fig:HMC_MAR_75_625_TP} display trace plots of posterior samples for SEM parameters under the MAR mechanism, obtained using the HMC method after discarding burn-in iterations for the simulated datasets with $n=625$ and different missing value percentages ($n_u$). 
The trace plots indicate good mixing for both cases.

\subsubsection{Convergence analysis for the simulation studies for the SEM under MNAR}
\begin{figure}[H]
    \centering
    \includegraphics[width=0.9\textwidth, keepaspectratio]{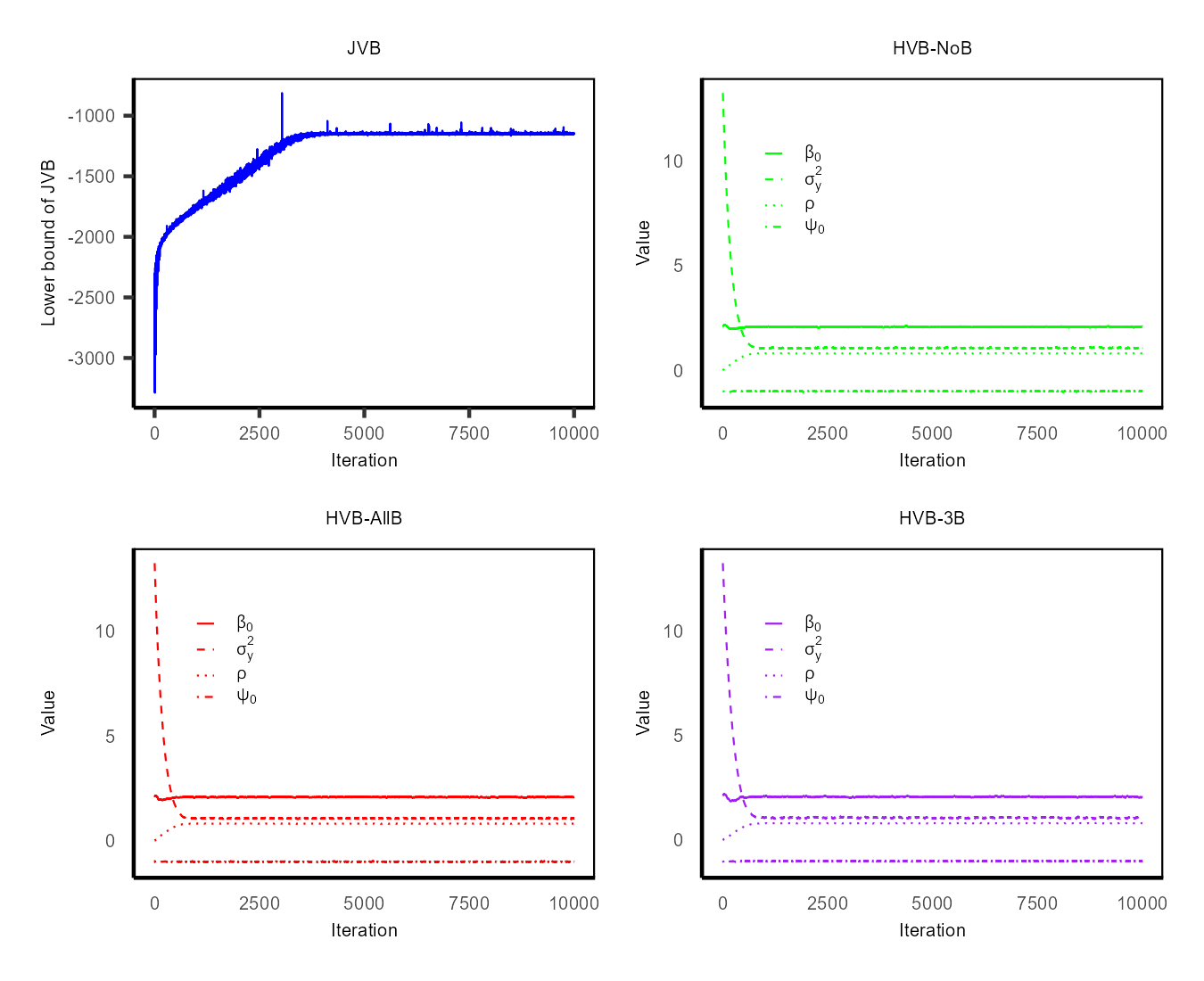}
    \caption{The lower bound for the JVB algorithm (top left figure), and the trajectories of variational means of SEM and missing data model parameters for the HVB algorithms over iterations (top right figure and bottom figures), under MNAR for the simulated dataset with $n=625$, $n_u$= 170 (i.e. missing percentage is 25\%)}
    \label{fig:con_vb_MNAR_625_25}
\end{figure}

Figures~\ref{fig:con_vb_MNAR_625_25}, \ref{fig:con_vb_MNAR_625_75}, and \ref{fig:con_vb_MNAR_10000_75} show lower bounds for the JVB algorithm (displayed in the top left subplot), and the trajectories of variational means of SEM and missing data model parameters for the HVB algorithms (shown in the remaining subplots) across VB iterations, under MNAR, for the simulated datasets with various combinations of $n$ and $n_u$. As observed in the simulation study under MAR, all VB algorithms achieve convergence well before the final iteration. Additionally, the HVB algorithms consistently achieve convergence in fewer iterations compared to the JVB algorithm.

\begin{figure}[H]
    \centering
    \includegraphics[width=0.9\linewidth]{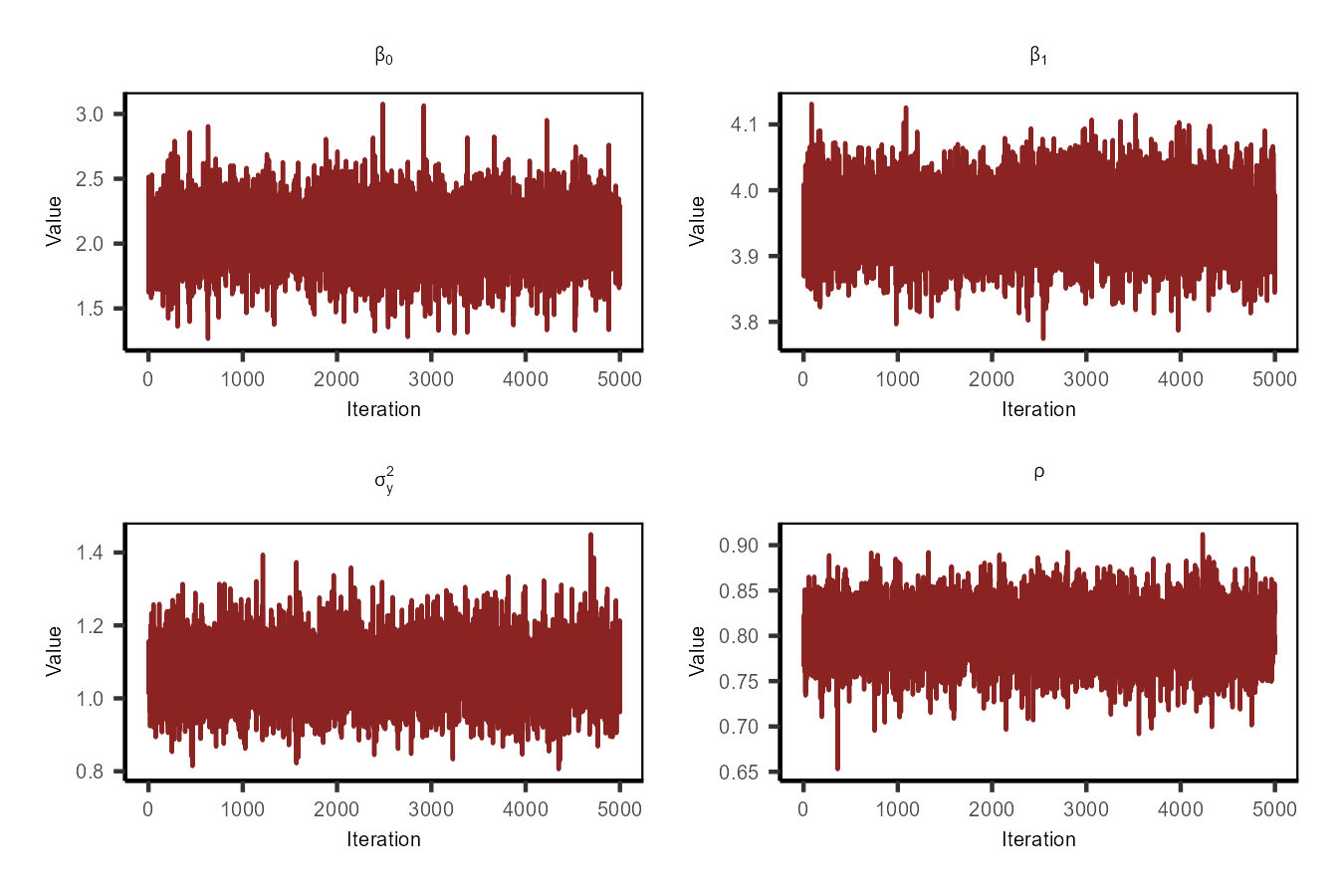}
    \caption{Trace plots of posterior samples of SEM parameters under MNAR from the HMC method after excluding burn-in iterations for the simulated dataset with $n=625$ and $n_u=170$ (i.e., the missing percentage is 25\%)}
    \label{fig:HMC_MNAR_25_625_TP}
\end{figure}

\begin{figure}[H]
    \centering
    \includegraphics[width=0.9\textwidth, keepaspectratio]{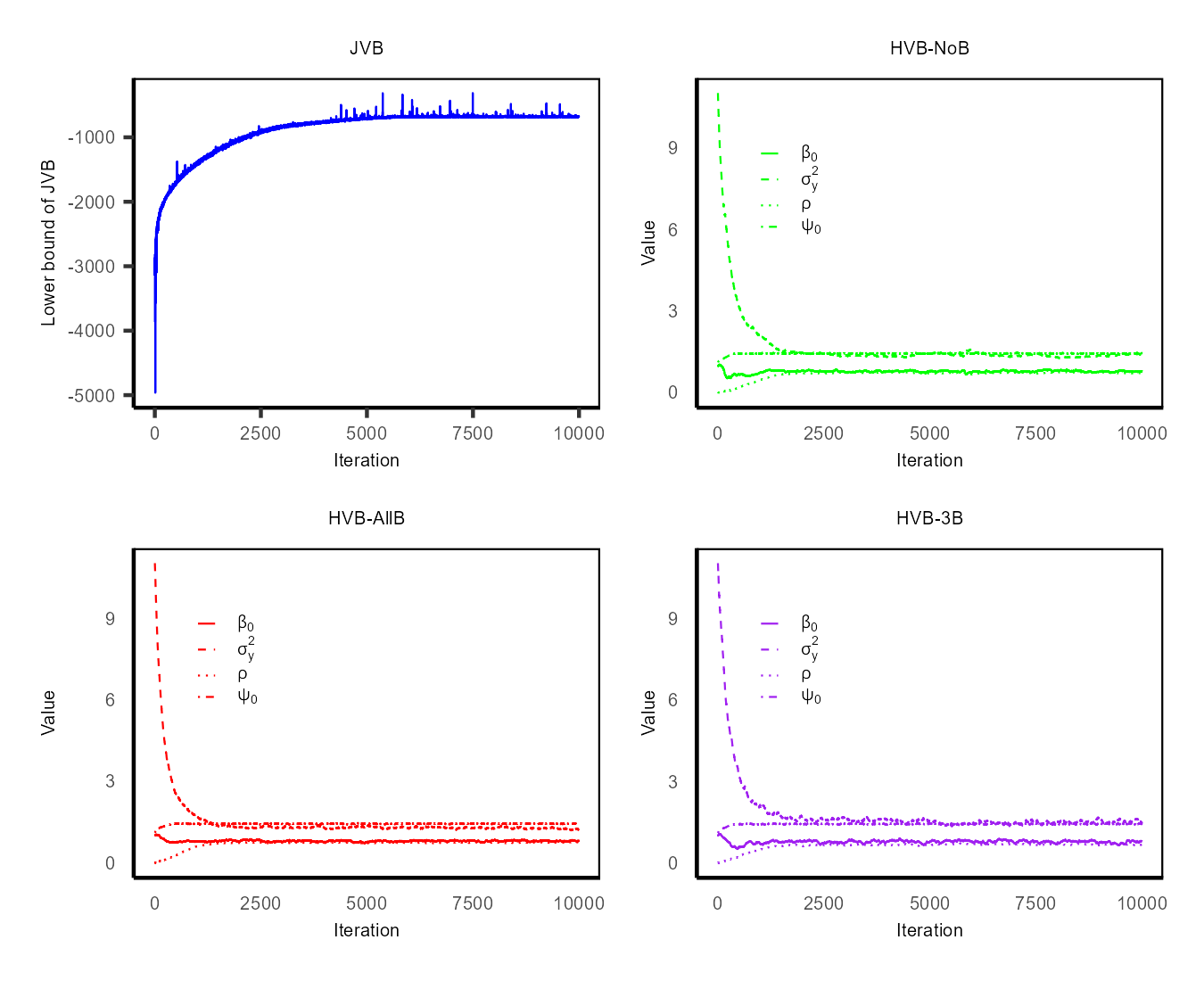}
    \caption{The lower bound for the JVB algorithm (top left figure), and the trajectories of variational means of SEM and missing data model  parameters for the HVB algorithms over iterations (top
right figure and bottom figures), under MNAR for the
simulated dataset with $n=625$, $n_u$= 469 (i.e. missing percentage is 75\%)}
    \label{fig:con_vb_MNAR_625_75}
\end{figure}

\begin{figure}[H]
    \centering
    \includegraphics[width=0.9\linewidth]{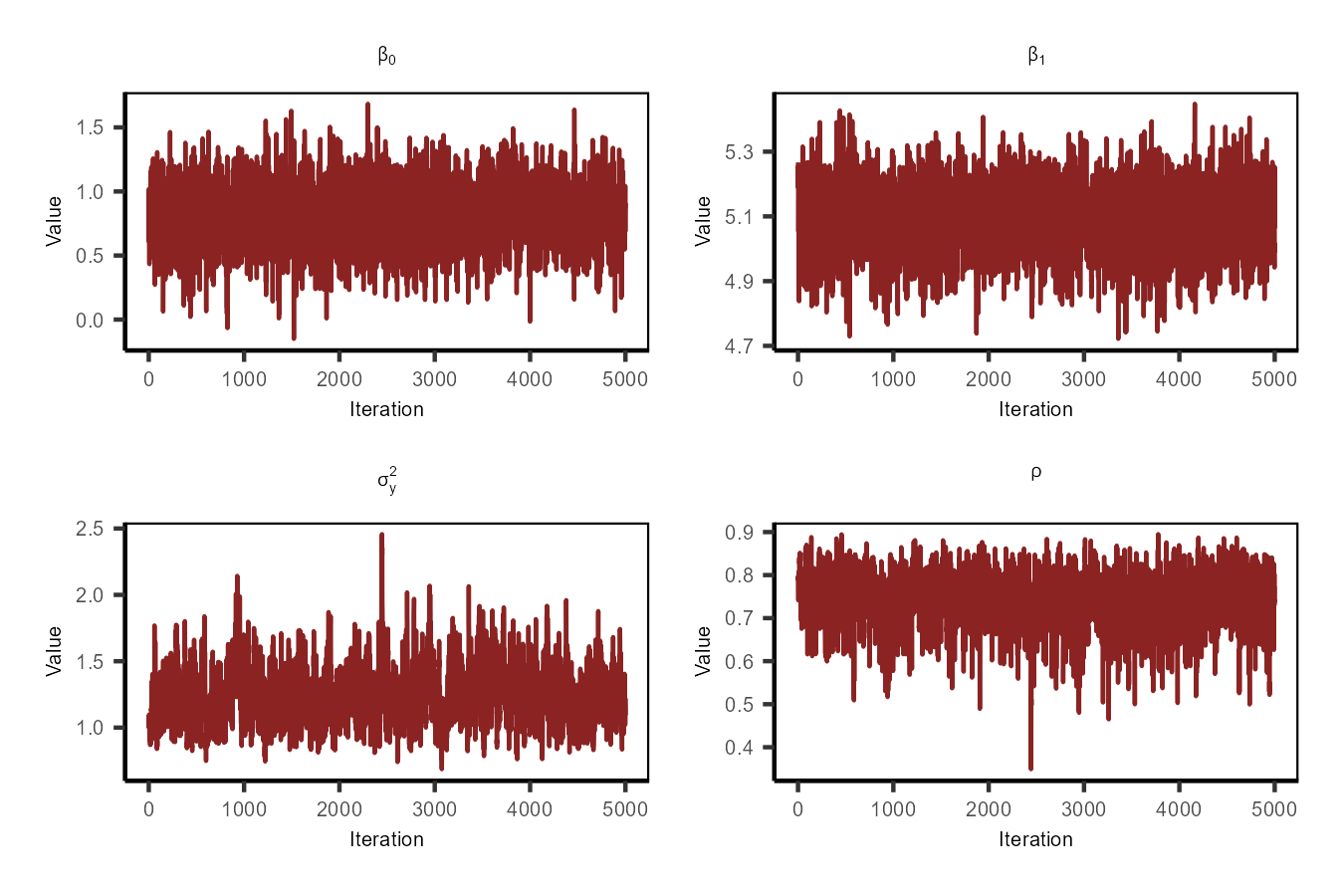}
    \caption{Trace plots of posterior samples of SEM parameters unde MNAR from the HMC method after excluding burn-in iterations for the simulated dataset with $n=625$ and $n_u=469$ (i.e., the missing percentage is 75\%)}
    \label{fig:HMC_MNAR_75_625_TP}
\end{figure}

\begin{figure}[H]
    \centering
    \includegraphics[width=0.9\textwidth, keepaspectratio]{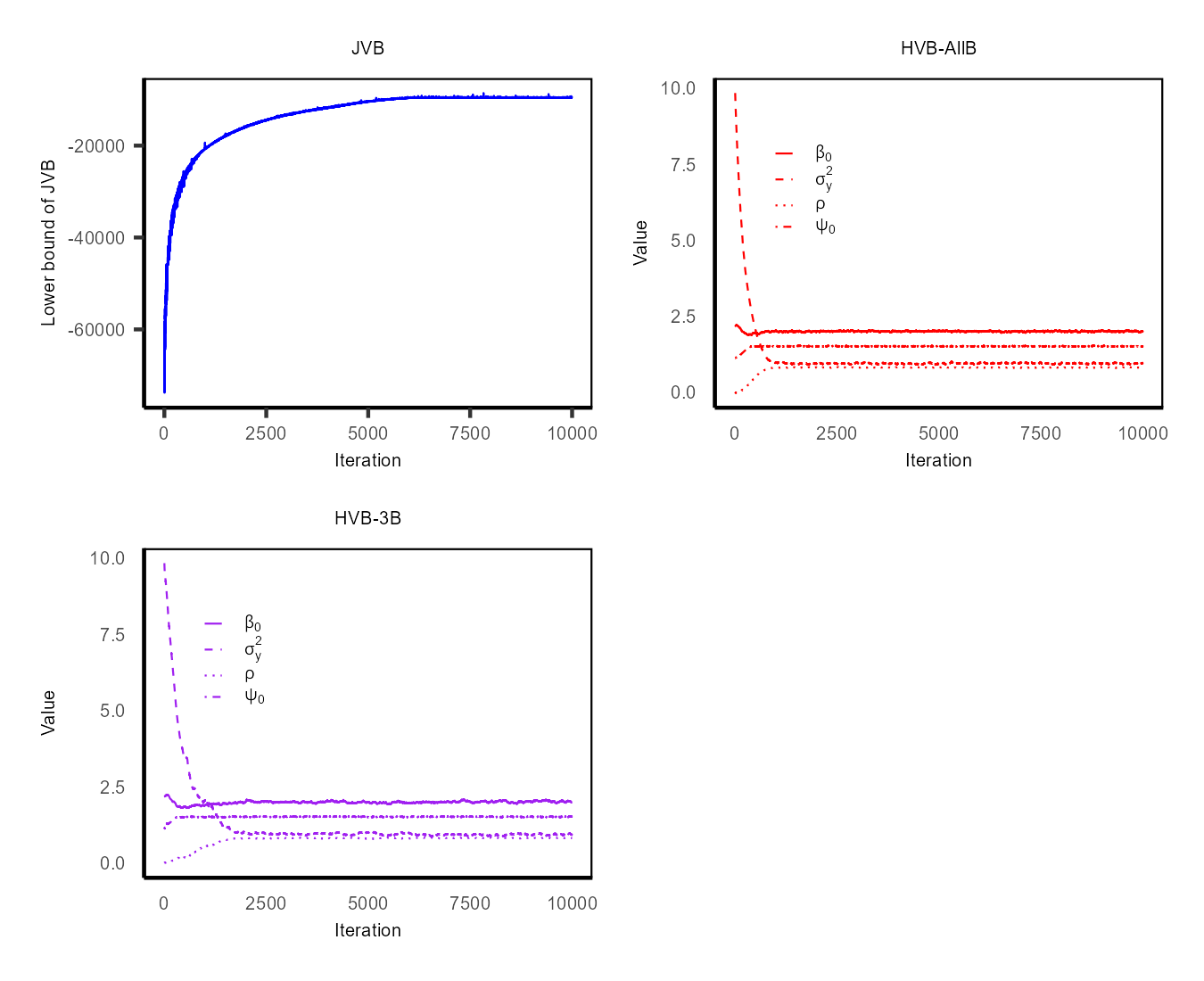}
    \caption{The lower bound for the JVB algorithm  (top left figure), and the trajectories of variational means of SEM and missing data model parameters for the HVB algorithms over iterations (top right and bottom left figures), under MNAR for the simulated dataset with $n=10,000$, $n_u= 7,542$ (i.e. missing percentage is approximately 75\%)}
    \label{fig:con_vb_MNAR_10000_75}
\end{figure}

Figures~\ref{fig:HMC_MNAR_25_625_TP} and \ref{fig:HMC_MNAR_75_625_TP} present trace plots of posterior samples of SEM parameters under the MNAR mechanism, obtained using the HMC method after discarding burn-in iterations, for the simulated datasets with a sample size of $n=625$ and different missing value percentages ($n_u$). These trace plots display stable, random-like patterns, suggesting good mixing for all SEM parameters.


\subsection{Convergence analysis for the Real data examples}

This subsection provides convergence analysis plots for the real world example presented in Section~\ref{sec:RealWorldAnalysis} of the main paper.

\subsubsection{Convergence analysis for Real data examples under MAR}
\label{sec:sup:con.real.MAR}

\begin{figure}[H]
    \centering\includegraphics[width=0.9\textwidth, keepaspectratio]{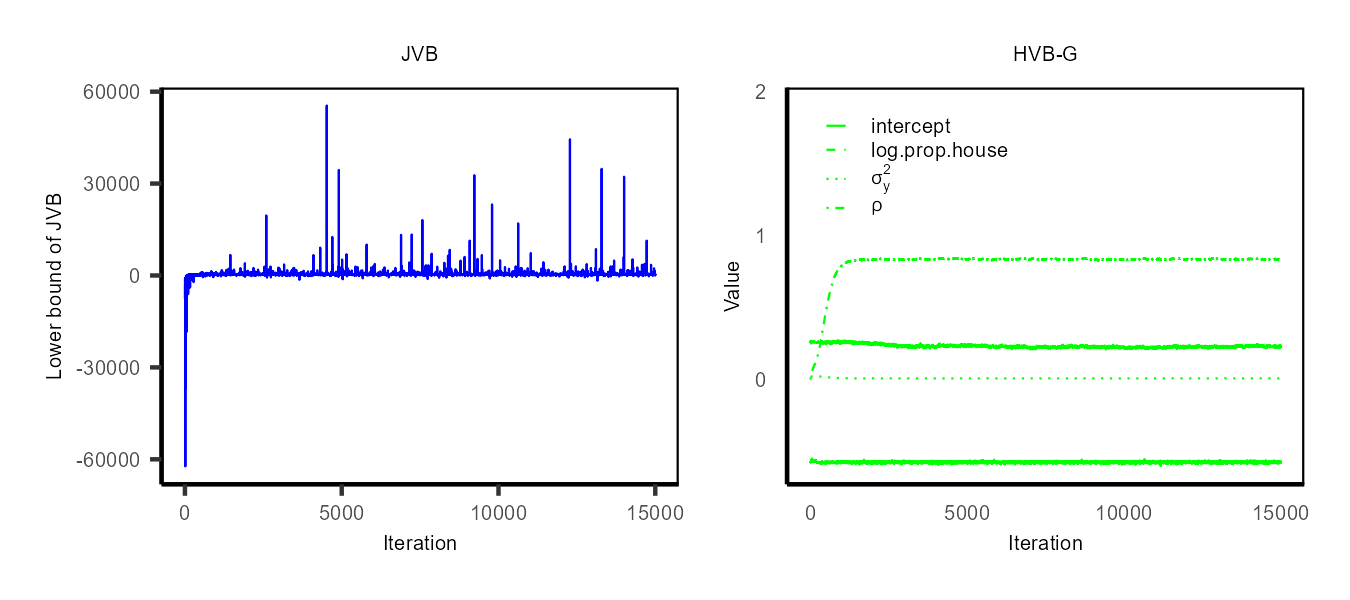}
    \caption{The lower bound for the JVB algorithm (left figure), and the trajectories of variational means of SEM parameters for the HVB-G algorithm over iterations (right figure) for the 1980 presidential election dataset, under MAR with $n_u=2,330$ (i.e. missing percentage is $75\%$)}
    \label{fig:con_vb_elec_MAR}
\end{figure}

Figure~\ref{fig:con_vb_elec_MAR} shows the lower bound for the JVB algorithm, and the trajectories of variational means of SEM parameters obtained using the HVB-G algorithm across iterations, for the 1980 presidential election dataset, under MAR with $n_u=2,330$. These plots clearly indicate convergence for both the HVB-G and JVB algorithms.

\subsubsection{Convergence analysis for the Real data example under MNAR}
\label{sec:sup:con.real.MNAR}
\begin{figure}[H]
    \centering\includegraphics[width=0.9\textwidth, keepaspectratio]{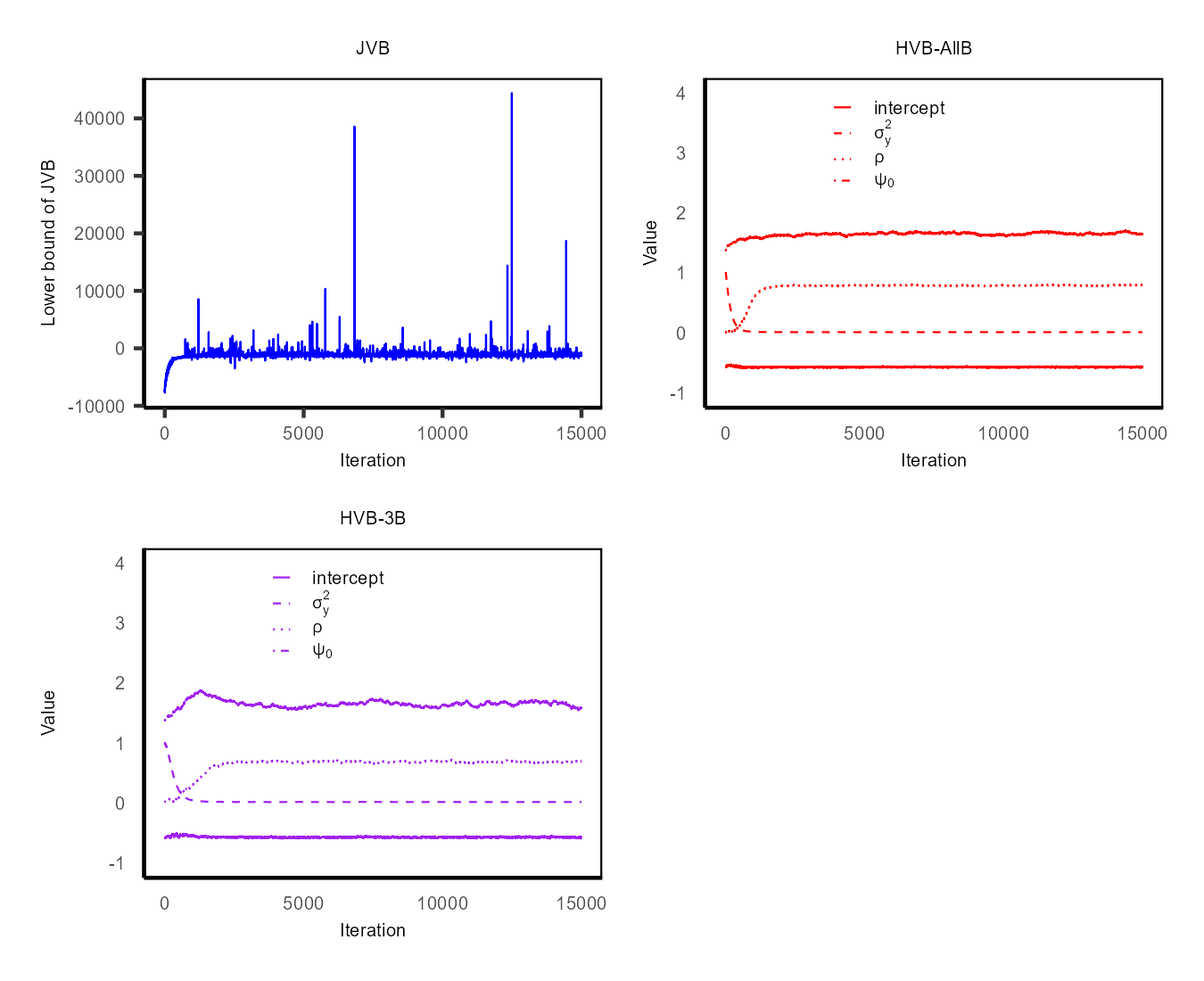}
    \caption{The lower bound for the JVB algorithm (top left figure), and the trajectories of variational means of SEM and missing data model parameters for the HVB-AllB and HVB-3B algorithms over iterations (top right and bottom left figures) for the 1980 presidential election dataset, under MNAR with $n_u=2,477$ (i.e. missing percentage is around $80\%$)}
    \label{fig:con_vb_MNAR_elec}
\end{figure}

Figure~\ref{fig:con_vb_MNAR_elec} illustrates the lower bound for the JVB algorithm, and the trajectories of the variational means of SEM and missing data model parameters for the HVB algorithms; HVB-AllB and HVB-3B, across iterations for the 1980 presidential election dataset under MNAR with $n_u=2,477$. Flat lines indicate that all algorithms have converged before the $15,000^{th}$ iteration.

\end{document}